\documentclass[aps,pra,twocolumn]{revtex4-2}
\usepackage{commath}
\usepackage{amsmath}
\usepackage{mathtools}
\usepackage{amssymb}
\usepackage{comment}
\usepackage{multirow}
\usepackage{xcolor}
\usepackage{soul}
\usepackage[colorlinks=true,linkcolor=blue,citecolor=blue,urlcolor=blue]{hyperref}

\usepackage{tikz}  

\usepackage{cleveref}
\crefname{fig}{Fig.}{Figs.}
\crefname{extendedfig}{Extended Data Fig.}{Extended Data Figs.}
\crefname{eq}{Eq.}{Eqs.}
\crefname{appdx}{Appendix}{Appendices}
\crefname{supp}{Supplementary Section}{Supplementary Sections}
\crefname{table}{Table}{Tables}
\crefname{main}{Main text Section}{Main text Sections}

\begin{document}

\title[Article Title]{A Linear Quantum Coupler for Clean Bosonic Control}

\author{Aniket Maiti}\email{aniket.maiti@yale.edu}

\author{John W. O. Garmon}
\author{Yao Lu}
\altaffiliation{Present address: 
Superconducting Quantum Materials and Systems Center, Fermi National Accelerator Laboratory (FNAL), Batavia, IL 60510, USA}
\author{Alessandro Miano}
\author{Luigi Frunzio}
\author{Robert J. Schoelkopf}\email{robert.schoelkopf@yale.edu}

\affiliation{Departments of Applied Physics and Physics, Yale University, New Haven, 06511, CT, USA}

\affiliation{Yale Quantum Institute, Yale University, New Haven, 06511, CT, USA}


\begin{abstract}
Quantum computing with superconducting circuits relies on high-fidelity driven nonlinear processes.
An ideal quantum nonlinearity would selectively activate desired coherent processes at high strength, without activating parasitic mixing products or introducing additional decoherence.
The wide bandwidth of the Josephson nonlinearity makes this difficult, with undesired drive-induced transitions and decoherence limiting qubit readout, gates, couplers, and amplifiers.
Significant strides have been recently made into building better `quantum mixers', with promise being shown by Kerr-free three-wave mixers that suppress driven frequency shifts, and balanced quantum mixers that explicitly forbid a significant fraction of parasitic processes.
We propose a novel mixer that combines both these strengths, with engineered selection rules that make it essentially linear (not just Kerr-free) when idle, and activate clean parametric processes even when driven at high strength.
Further, its ideal Hamiltonian is simple to analyze analytically, and we show that this ideal behavior is first-order insensitive to dominant experimental imperfections.
We expect this mixer to allow significant advances in high-Q control, readout, and amplification.
\end{abstract}

\newcommand{\red}[1]{\textcolor{red}{#1}}
\newcommand{\JWOG}[1]{\textcolor{cyan}{#1}}

\sethlcolor{yellow}
\newcommand{\reviewchanges}[1]{#1}

\newcommand{\reviewchangesAlt}[1]{#1}

\newcommand{\old}[1]{\textcolor{brown}{#1}}

\maketitle

\section{Introduction}
Driven superconducting circuits provide a platform for studying complex nonlinear quantum optics and manipulating quantum information in the microwave domain.
The intrinsic nonlinearity of Josephson junctions \cite{JunctionBasicsReview1979} has been utilized to design numerous circuit elements, ranging from quantum-limited amplifiers \cite{YurkeJPA,LehnertJPA,OliverJPA,JPC_Paper,JTWPA_2015,SNAILPaper,FPJA,AumentadoAmplifierReview}, novel non-reciprocal devices \cite{KatrinaCirculatorAmplifierJPC,ChapmanCirculator,GoogleCirculator,AumentadoCirculatorReview}, ultra-low-noise detectors \cite{FlurinSinglePhotonCounter,FlurinChargeSensor,HAYSTAC_Nature}, and, most notably, quantum information processing through the field of circuit Quantum Electrodynamics \reviewchanges{(circuit QED)} \cite{OriginalcQEDExpt2004,CircuitQEDReview}.
A significant portion of these applications are enabled by nonlinear quantum mixers that activate specific desired processes when parametrically driven with a microwave drive.
Such parametric mixers have found use as couplers \cite{Yvonne2018_PRX_ProgrammableInterference,ATS_Paper,SQUIDBeamsplitterPaper,ChapmanSNAILBeamsplitterPaper}, amplifiers \cite{AumentadoAmplifierReview}, and even nonlinear oscillators in their own right \cite{GrimmFrattiniKerrCat,BerkeleyCat,CubicPhaseExperiment}.

Ideally, one desires fast high-fidelity operations that manipulate sensitive quantum information without introducing errors.
This requires the driven quantum mixer to turn on desired interactions at high strength, without activating parasitic processes such as drive-induced frequency shifts and leakage to uncontrolled states.
A further complication is the finite coherence of the quantum mixer, where any unintended excitation of the mixer out of its ground state can lead to mixer decoherence dominating the infidelity of the parametric process \cite{Yaxing_PRA_BilinearModeCoupling}.
In particular, when the mixer is used as a coupler between high-Q modes, it must activate a parametric coupling between them on demand, while not reducing both their idle and driven coherence.
One thus aims to engineer the mixer such that it is minimally invasive when idling, and also remains in its ground state and prevents parasitic processes when driven.

A common approach to mitigate some parasitic effects is to use a `Kerr-free three-wave mixer' \cite{FlurinJPC,KerrFreeJRMHatridge,SNAILPaper,SNAIL_Analysis_Sivak,KerrFreeRFSQUID}, where one utilizes a third-order nonlinearity to generate the desired process with a single drive tone, while nulling the fourth-order (Kerr) nonlinearity.
Suppressing the Kerr suppresses spurious processes like the AC Stark shift, which can lead to reduced saturation power in the amplifier context \cite{GangJPCSaturation,SNAIL_Analysis_Frattini}, or frequency collisions with other sensitive modes in the coupler context \cite{SQUIDBeamsplitterPaper}.
However, for high-fidelity applications, this may still be insufficient -- in such fragile systems, the loss of even a single photon to some uncontrolled degree of freedom can ruin sensitive quantum information.
In particular, at strong drive powers, even Kerr-free mixers are inevitably spoiled by the onset of numerous transitions induced by nonlinearities beyond the fourth order.
The effect of these transitions is especially evident at drive powers where the mixer ionizes \cite{BlaisReminiscenceChaos,BlaisTransmonIonization,BlaisUnifiedIonization2024,Yaxing_PRA_BilinearModeCoupling,QuantumHeatingYaxingRecommendation} into higher-lying states or enters a chaotic regime, irreversibly spoiling the system's information.

\begin{figure*}
\includegraphics[width=0.98 \textwidth]{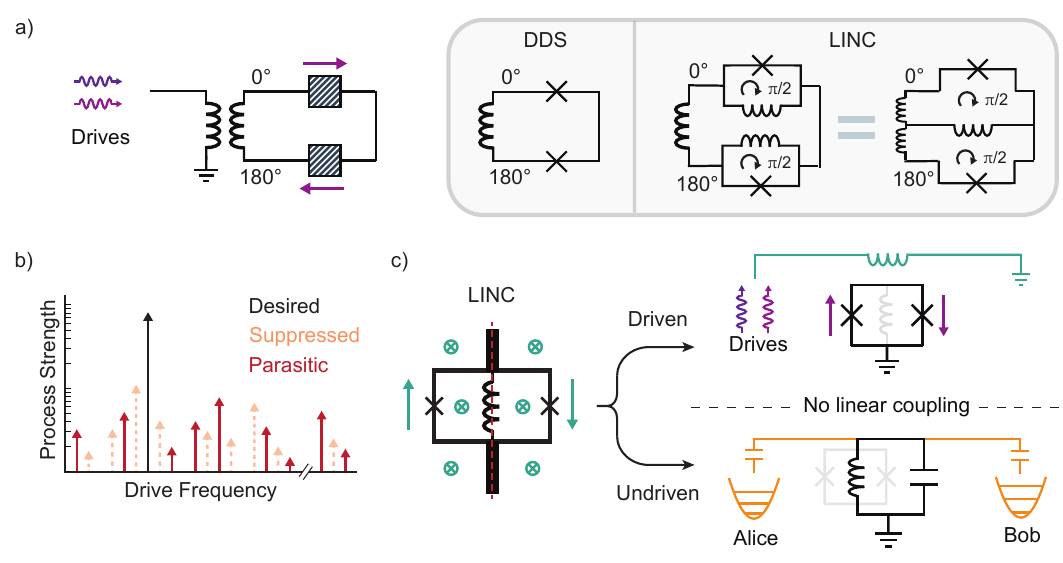}
\caption{
\textbf{The LINC, a quantum single-balanced mixer}
\textbf{a}, Fundamentals of mixer balancing. By taking two identical nonlinear elements and driving them $180^\circ$ out of phase (e.g., with a balun), we can create a single-balanced mixer. If the nonlinear element is a single junction, this circuit is the familiar differentially-driven SQUID (DDS,~\cite{SQUIDBeamsplitterPaper}). If the nonlinear element is instead a DC flux-biased RF-SQUID, this circuit results in the Linear Inductive Coupler (LINC). \reviewchanges{The $\Pi$-T transformation provides a simpler circuit representation.}
\textbf{b}, An illustration of the advantage of a balanced mixer. 
In general, a driven nonlinear element will permit numerous nonlinear processes, of which only a small subset is typically desired. Balancing the nonlinear element will suppress a significant fraction of spurious processes (light orange), but preserve the desired process (black) at high strength. Any remaining parasitic processes (red) still permitted by the symmetry must be avoided through other means, like careful selection of drive frequencies.
\textbf{c}, The LINC circuit, with a symmetric loop and capacitive pads, enables a separation of driven and undriven mixer behavior. At the operating point where the total DC flux in the outer SQUID loop is half a flux quantum, the loop junctions are biased to effectively infinite inductance, leaving only the linear shunt as the coupler's inductor.
This simultaneously nulls all nonlinearity when idle, which is beyond just being Kerr-free.
When threaded with AC flux, the outer loop activates and drives a balanced three-wave mixing process.
}\label[fig]{fig:fig1}
\end{figure*}

To suppress these higher-order nonlinear processes, we can leverage certain symmetries and intuitions in mixer design that further improve their performance by imposing an explicit selection rule on the types of allowed processes that the mixer can activate.
These design principles mimic mixer balancing in the classical electrical engineering context, where for example, a single-balanced mixer utilizes symmetries of both the circuit and the drive delivery to coherently suppress approximately half the undesired mixing products  (see \cref{fig:fig1}a-b).
This significantly increases the amplitude and frequencies at which the mixer can be driven without spoiling fidelity, especially in the presence of drive-induced shifts.
Such symmetries have been put to use before in circuit QED, both in the amplifier context (for e.g., with the JPC \cite{JPC_Paper,FlurinJPC}) as well as in high-Q control \cite{ATS_Paper}.
Particularly in the parametric coupler context, employing this symmetry in four-wave mixing with a Differentially Driven SQUID (DDS) has shown significant improvement over regular four-wave mixing with a single junction  \cite{SQUIDBeamsplitterPaper}.

\vspace{-0.03em}

Utilizing similar design principles to the DDS, here we introduce a mixer that has an identical Hamiltonian to a balanced flux-biased RF-SQUID pair (\cref{fig:fig1}a), but can manifest as a simple dipole element (\cref{fig:fig1}c).
When operated at a particular DC bias point, idle effects due to the Josephson junctions disappear, and the mixer is exceptionally linear.
When driven, the outer SQUID loop activates a three-wave mixing process that obeys a strict selection rule that we call parity protection, forbidding a significant fraction of the remaining parasitic processes.
This means that the mixer is not just Kerr-free when driven, but the Kerr-free bias point also coincides with the extinguishment of all even-order nonlinearity \textit{simultaneously}.

\section{The Linear Inductive Coupler}
We now introduce the new mixer and highlight the advantages that arise from its inherent symmetry.
We start by reminding the reader of the Differentially Driven SQUID (DDS), a symmetric superconducting loop that contains two junctions, such that the drive always causes an equal and opposite phase drop across the two junctions.
Imposing this symmetry on the driven circuit results in the coherent cancellation of half of the nonlinear resonances allowed in a single junction circuit, significantly cleaning up the parasitic processes that this mixer can activate, even when strongly driven.
However, this coupler still retains its idle Kerr nonlinearity, and has significant driven frequency shifts.

The improved circuit we propose shunts a DDS with a linear inductor, and biases the circuit at a constant external DC magnetic field of half a flux quantum threading the SQUID loop.
The inductive shunt is placed in a precise manner such that the drive across it exactly cancels when the outer SQUID loop is differentially driven (see~\cref{fig:fig1}c).
Biasing this circuit at half flux would generally set its frequency to zero in the absence of the linear inductor, since the junctions in its outer loop are biased to effectively infinite Josephson inductance.
However, the presence of the shunting inductor stabilizes the circuit and sets the mixer's frequency when undriven.

More precisely, the Hamiltonian of the mixer can be written in terms of its two available degrees of freedom under the above symmetry constraints, the flux-induced differential phase across each junction $\phi = \left(\phi_{\text{DC}} + \phi_{\text{AC}} \right) = \frac{\pi}{\Phi_0} \Phi_{\text{ext}}$, and the common mode phase across the circuit $\hat{\theta}$ that is conjugate to the charge $\hat{n}$ on its capacitive island:
\begin{equation}
    H_{\text{LINC}} = \underbrace{4E_C \hat{n}^2 + E_L \dfrac{\hat{\theta}^2}{2}}_{\text{static}} + \underbrace{2 E_J \sin{\left(\phi-\pi/2\right)} \cos{\hat{\theta}}}_{\text{driven}}.
    \label[eq]{eq:LincHamiltonian}
\end{equation}
Following standard notation in circuit quantum-electrodynamics, $E_C$ denotes the circuit's charging energy, $E_J$ the Josephson energy of each junction in the outer loop, $E_L$ the inductive energy of the shunting inductor, and $\hbar$ has been set to 1.
Note that any static charge offsets are redistributed by the shunting inductor, and hence absent in the Hamiltonian.
For the circuit's potential energy to have a single-valued solution at all values of DC flux, the shunt's inductive energy must dominate the circuit $E_L > 2E_J$, which is equivalent to its frequency never crossing zero (\cref{fig:fig2}a).

Let us now analyze the circuit's behavior, both when idle and when driven.
At the special flux point of $\phi_{\text{DC}} = \frac{\pi}{2}$, the nonlinearity of the circuit completely disappears when idle, resulting in an exactly linear static Hamiltonian:
\begin{equation}
    H_{\text{LINC}} \left[\dfrac{\pi}{2}\right] = 4 E_C \hat{n}^2 + E_L \dfrac{\hat{\theta}^2}{2}
\end{equation}
This is beyond a simple `Kerr-free' mixer -- at this flux point, the circuit's static nonlinearity at all orders vanishes.
We will show later that this property holds even in the presence of experimental imperfections like a parasitic series linear inductance, which is often present in realistic mixers due to geometric constraints.
The mixer hence provides an extremely linear environment to any modes it mediates a coupling between in the coupler context, which is particularly useful for bosonic control.
We thus name this mixer the `Linear INductive Coupler' (LINC).

When the LINC is driven, the drive exactly cancels across the shunting inductor by design and does not displace the common-mode phase.
The LINC's Hamiltonian is thus separable into two non-interacting parts, the undriven and the driven, as shown in \cref{eq:LincHamiltonian}.
The driven portion contains beneficial symmetries due to the balancing of the drive, enforcing a selection rule on the allowed nonlinear processes in a manner similar to the DDS.
Only processes where the number of coupler and resonator photons are even can be activated.
Even within these, processes that are even-order in the drive, like the drive-induced frequency shift, are suppressed.
This is in stark contrast to prevalent mixers like the SNAIL, where nominally every process can occur, but individual nonlinearities can be manually tuned to zero at specific flux bias points.
We will show that this selection rule significantly expands the frequencies and strengths at which the LINC can be driven without causing spurious transitions, and reduce inter-modulation products in the presence of multiple drive tones.
At leading order, the LINC functions as a three-wave mixer, activating either a beamsplitting (red sideband or conversion) or squeezing (blue sideband or gain) process when driven with a single tone of appropriate frequency.

We note that while the simultaneous balancing of the drive and the bias of this circuit is novel and crucial to its function, the circuit topology itself is fairly common in literature \cite{ATS_Paper,gSNAIL,gQuarton,BerkeleySTS_theory,QuantumFluxParametron}.
The LINC is a sibling to the Asymmetrically Threaded SQUID (ATS), with nominally identical circuit geometries but an exactly opposing notion of bias point.
This allows the LINC to be optimized as a three-wave mixing coupler, while the ATS is utilized for Kerr-free four-wave mixing applications, like parametric two-photon dissipation~\cite{ATS_Paper}.

The LINC also shares its topology with the gradiometric SNAIL~\cite{gSNAIL}, where the flux bias in the SQUID loops is used to tune the effective junction ratio of a regular charge-driven SNAIL, and symmetry is not a strict constraint.
One key physical feature where the LINC might differ from these circuits is in its shunting inductor, which for the SNAIL and ATS typically consists of an array of a few junctions.
\reviewchanges{While different circuits in the past have utilized inductive shunts for various improvements, here the shunt does not simply dilute nonlinearity. Due to the balanced nature of this circuit, this shunt entirely sets the linearity of its undriven Hamiltonian, and allows a suppression of unwanted interactions without reducing the strength of the desired process.
}
Since bosonic applications might require highly linear idle circuits, the LINC shunt could benefit from using high kinetic inductance materials such as NbN or granular Aluminum (grAl), or a meandering geometric inductor, to minimize its static nonlinearity.

\section{Idle nonlinearity and decoherence}

\begin{figure}
\includegraphics[width=0.48 \textwidth]{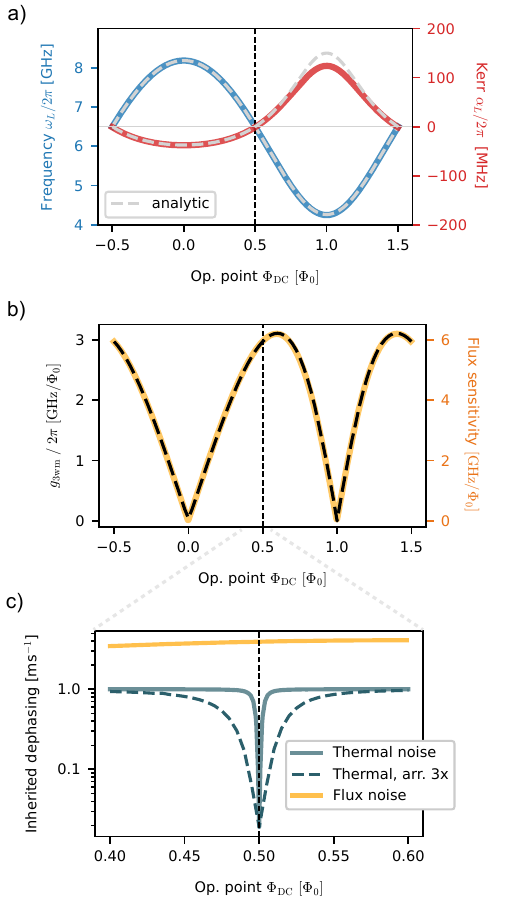}
\caption{
\textbf{Static Hamiltonian and flux noise sensitivity.}
\textbf{a}, The static LINC frequency and Kerr as a function of DC flux. Numerical values (solid lines) are calculated by exact diagonalization of a LINC Hamiltonian with $E_L/h=52.8$ GHz, $E_J/h=15.84$ GHz, $E_C/h=100$ MHz, with the center shunt composed of an array of 10 junctions, where $h$ is Planck's constant. Overlayed analytic curves follow~\cref{eq:analytic_Kerr,eq:analytic_freq}. At the operating point of $\phi_{\text{DC}}=\pi/2$, the coupler is linear.
\textbf{b}, Three-wave mixing strength (dashed black) and sensitivity to flux-noise (yellow) as a function of DC flux. The former is calculated from a Floquet simulation of a parametric squeezing process. The latter is a simple derivative of the static frequencies calculated in part \textbf{a} ($1/2\pi \;d\omega_L/d\Phi$). These quantities are equal, up to a scaling factor of $2$ (see~\cref{app:Circuit}).
\textbf{c}, Inherited dephasing for a coupled quantum memory, as a function of DC flux. Thermal noise-induced dephasing is calculated for \reviewchanges{$n_{\text{th}}^{\text{coupler}} = 2\%, T_1 = 20 \mu$s, and is minimized at the operating point, where the coupler is linear with no dispersive shift $\chi$ to the resonator. Arraying multiple LINCs together dilutes nonlinearity and therefore $\chi$, broadening the range over which $\chi \lesssim 1/T_1$ and dephasing is suppressed}. The low-frequency dephasing due to inherited flux noise is calculated through~\cref{eq:kappa_phi_cav}, with noise amplitude $A_\Phi = 1\mu\Phi_0/\sqrt{Hz}$. 
}\label[fig]{fig:fig2}
\end{figure}

While the well-controlled mixing properties of the LINC make it advantageous for a number of applications, the LINC particularly shines as a coupler for high-Q bosonic quantum control.
In this context, the LINC must activate a parametric coupling to one or more resonators (bosonic modes) when driven, and minimally spoil their quantum information when static.
Since the resonators in isolation are linear and only incur slow single-photon decay errors (at rate $\kappa_{\text{res}}$), an ideal bosonic coupler must neither limit this decay rate, nor introduce any additional nonlinearity or dephasing to the resonators.
Suppressing the decay inherited from the coupler limits the allowable energy participation~\cite{EPRPaper} of the resonator in the LINC mode, which we assume for the rest of this article to be $p_{\text{res}} = 0.01$.
Minimizing the resonator nonlinearity and dephasing bounds the acceptable static nonlinearity of the coupler.
It is important to note that, while a four-wave mixer like the transmon or DDS induces resonator dephasing through the cross-Kerr (dispersive) interaction, a three-wave mixer like the SNAIL or LINC suppresses such dephasing but can introduce low-frequency fluctuations from flux-noise.
We analyze the effect of both the LINC's linearity and propagated flux noise below.

The ideal LINC's nonlinearity goes to zero at the operating point, as shown in~\cref{fig:fig2}a.
The curious shape of this curve can be explained entirely analytically by considering the flux dependence of the fourth-order nonlinearity (see~\cref{app:Circuit}):
\begin{equation}
\begin{split}
    \alpha_{L}\left[\phi_{\text{DC}}\right] &= -2E_J \cos{\phi_{\text{DC}}} \; \left(\theta_{zpf} [ \phi_{\text{DC}} ] \right)^4 \\
    &= -\dfrac{2E_J \cos{\phi_{\text{DC}}}}{E_L+2E_J \cos{\phi_{\text{DC}}}} E_C.
\end{split}
\label[eq]{eq:analytic_Kerr}
\end{equation}
Note that the un-driven LINC has no third-order nonlinearity at any bias point ($\phi_{\text{DC}}$), and thus has no perturbative corrections to the coupler's idle Kerr.
This means that the LINC's self-Kerr and cross-Kerr to coupled modes always simultaneously go to zero at the same operating point.
The LINC's frequency can also be simply calculated through:
\begin{equation}
    \omega_{L}\left[\phi_{\text{DC}}\right] = \sqrt{8 E_C \left(E_L + 2E_J \cos{\phi_{\text{DC}}}\right)} - \alpha_L\left[\phi_{\text{DC}}\right].
\label[eq]{eq:analytic_freq}
\end{equation}
We derive these analytical results in~\cref{app:Circuit} and show them overlayed with exact numerical diagonalization in \cref{fig:fig2}a.
These analytics describe static physics near the operating point exceptionally well, and only deviate from numerics around $\Phi_{\text{DC}}=\Phi_0$, where the truncated Taylor expansion does not accurately represent the coupler's nonlinearity.

If instead of an ideal inductor, the LINC's shunt is composed of an array of $N$ Josephson Junctions, its static Hamiltonian at the operating point is given by:
\begin{equation}
\label[eq]{eq:JJ_array}
\begin{split}
    H_{\text{LINC}}\left[\dfrac{\pi}{2}\right] &= 4 E_C \hat{n}^2 - N E_{L,J} \cos{\dfrac{\hat{\theta}}{N}} \\
    \left. \alpha_{L} \right\vert_{\frac{\pi}{2}} &= E_C / N^2,
\end{split}
\end{equation}
where each junction in the shunting array has been scaled appropriately to preserve the same total inductance ($E_{L,J} = N E_L$).
To keep the effect of any inherited resonator nonlinearity smaller than the resonator's linewidth, the number of junctions in the LINC shunt must be greater than $\left(p_{res}^2 \bar{n} E_C / \kappa\right)^{\frac{1}{2}}$, for an intended resonator population of $\bar{n}$.
However, note that the LINC behaves as a protected three-wave mixer even if it is not this linear, including when its shunting inductor is just a single Josephson Junction \cite{BerkeleySTS_theory}.

While nulling the LINC's nonlinearity suppresses shot-noise induced dephasing, a major alternative source of dephasing in any LINC architecture will be low-frequency flux noise.
In fact, since the LINC is a flux-driven three-wave mixer, there is an exact trade-off in the sensitivity of the coupler to flux noise and the effective 3-wave mixing strength $g_{\text{3wm}}$ (see~\cref{fig:fig2}b).
This means that at the operating point of $\phi_{\text{DC}}=\pi/2$, both the $g_{\text{3wm}}$ and flux-noise sensitivity are close to maximum (anti-sweet spot).
However, since the LINC itself remains in the ground state, this dephasing is primarily harmful if it propagates errors into the desired parametric process, or to the modes that are statically hybridized with the LINC.
The effect of flux noise on the strength of the parametric process can be estimated by the simplified infidelity limit:
\begin{equation}
\begin{split}
    1-\mathcal{F}_p &\leq \left. \left(\dfrac{2\pi}{g_{\text{3wm}}} \dfrac{d g_{\text{3wm}}}{d \Phi} \right)^2 \right\vert_{\frac{\pi}{2}} \int_{0}^{\infty} S_{\Phi \Phi} [f]\; g_N[f \tau_{\text{g}}] \: df \\
    &\approx \left(\dfrac{2E_J}{E_L} \right)^2 \int_{1/\tau_{\text{exp}}}^{1/\tau_{\text{g}}} \frac{1}{\Phi_0^2} \;S_{\Phi \Phi} [f] \: df.
\end{split}
\label[eq]{eq:g3wm_flux_noise}
\end{equation}
Here $S_{\Phi \Phi}[f]$ is the spectral density of flux noise, $g_N[f \tau_{\text{g}}]$ is a characteristic function defined by the pulse sequence, and the flux-noise sensitivity is analytically derived (see~\cref{app:Circuit}). For an upper bound on the infidelity, we approximate $g_N$ as a rectangular window between the relevant timescales of a single gate ($\tau_{\text{g}} \sim 1 \text{ ns}$) and the total experiment ($\tau_\text{exp} \sim 1 \text{ s}$).
Note that the quadratic dependence on flux-noise sensitivity assumes that the variance in the strength of $g_{\text{3wm}}$ is slow and small, and therefore appears as a coherent offset to the intended pulse evolution.
For $1/f$ type noise $S_{\Phi \Phi}[f] \sim A_\Phi^2/f$, typical values of the noise amplitude $A_\Phi \sim 1\;\mu \Phi_0/\sqrt{Hz}$ at 1 Hz~\cite{,BylanderCPMG,FeiYanSpinLocking,MIT_SQUID_flux_noise_2020} lead to an estimate of  $1-\mathcal{F}_p\sim10^{-10}$, which means this mechanism should not be a limiting factor in the mixer's performance.

A more relevant effect might be the inherited flux noise in coupled information-storing modes, which we estimate analytically  in~\cref{fig:fig2}c.
This inherited dephasing scales $\propto p_{\text{res}}$\cite{Martinis_dephasing,Peter_dephasing}:
\begin{equation}
\begin{split}
    \kappa_\varphi &\propto \sqrt{\left(d\omega_{\text{res}}/d\Phi\right)^2 \int_0^\infty S_{\Phi \Phi}[f] \;g_N[f\tau]\: df} \\
    &\approx p_{\text{res}} \left\vert \frac{d\omega_L}{d\Phi} \right\vert_{\phi_{\text{DC}}=\pi/2} A_\Phi \; C,
\end{split}
\label[eq]{eq:kappa_phi_cav}
\end{equation}
where $C = \sqrt{2 \vert \ln{(2\pi \tau/ \tau_\text{exp})} \vert} \sim 3-5$ is a slow time dependence on when $\kappa_\phi$ is evaluated ($\tau$), and the total length of the experiment ($\tau_{\text{exp}}$).
The inherited dephasing can therefore be significant, but it is low-frequency, which means it could be mitigated with techniques like dynamical decoupling \cite{BylanderCPMG,FeiYanSpinLocking,MIT_SQUID_flux_noise_2020}, or stabilized bosonic codes\cite{GrimmFrattiniKerrCat,ATS_Paper}.
This makes it potentially still beneficial to operate near the anti-sweet spot, where the coupled resonator is sensitive to inherited flux noise but not to thermal noise-induced dephasing, since the latter requires non-trivial strategies for suppression \cite{YvonneMagneticHose,SergeShotNoiseFeedback2024,IoanFluxoniumTunableChi}.

\section{The LINC as a driven mixer}
We now focus on the driven behavior of the LINC and its performance as a balanced quantum mixer.
With ideal symmetry, the LINC's driven behavior is independent of its center shunt (up to a normalization of its impedance), and is given by:
\begin{equation}
    H_{\rm{driven}} = 2E_J \underbrace{\sin{\phi_{\rm{AC}}}}_{\rm{odd}} \underbrace{\cos{\hat{\theta}}}_{\rm{even}}.
\end{equation}
There are two important points to note about this driven Hamiltonian.
The first is that the drive, $\phi_{\text{AC}}$, acts on an entirely orthogonal degree of freedom to the LINC mode.
This means that the drive does not displace the mode, and the LINC in general remains in its undriven ground state, similar to the DDS \cite{SQUIDBeamsplitterPaper}.
In stark contrast to charge-driven mixers like the transmon or the SNAIL, this decouples the effective drive strength from the frequency of the LINC mode, allowing for independent optimization of the drive delivery and LINC frequency.

\begin{figure*}[ht]
\centering
\includegraphics[width=0.98 \textwidth]{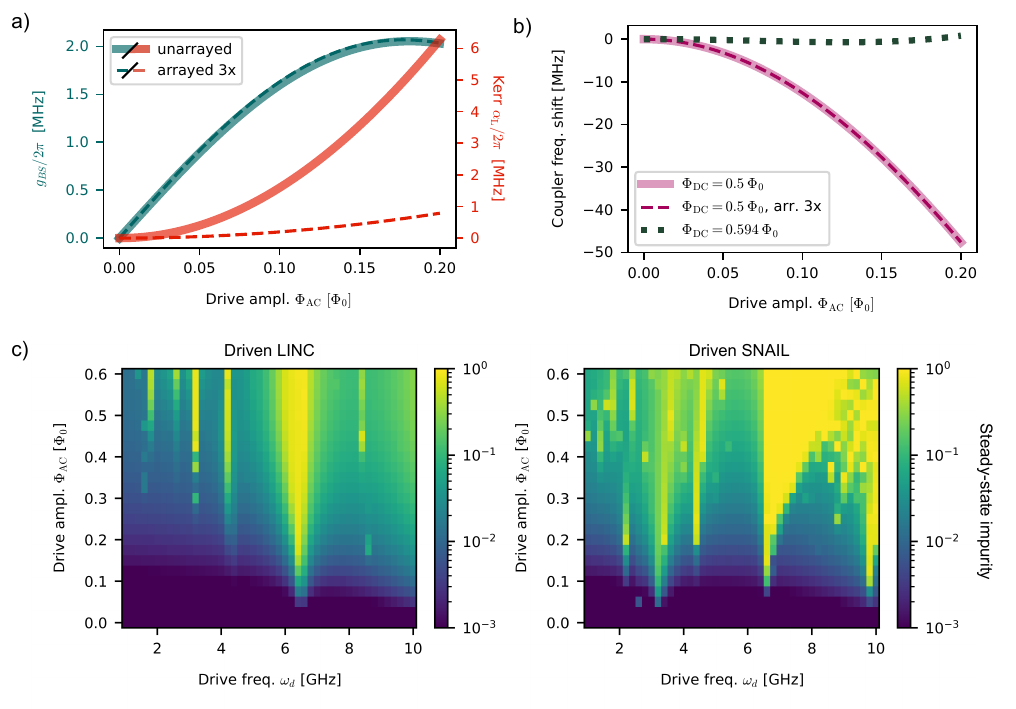}
\caption{
\textbf{The LINC as a driven mixer}
\textbf{a}, LINC beamsplitting strength ($g_{BS}$) and Kerr ($\alpha_{L}$) as a function of drive strength ($\phi_{\text{AC}}$), from an exact Floquet simulation. While the LINC is Kerr-free when idle, higher-order nonlinearities can induce a driven Kerr. This can be suppressed by arraying multiple LINCs while preserving the (driven) flux through each LINC loop (arrayed 3x, dashed).
\textbf{b}, Driven LINC frequency shift (solid purple) vs drive strength. Unlike in charge-driven mixers like the SNAIL, this frequency shift is independent of the coupler Kerr, and persists even if the coupler is arrayed (dashed purple). It is possible to minimize this shift by biasing to a flux point where the coupler frequency has an inflection point as a function of DC flux (dotted black).
\textbf{c}, Driven steady-state \reviewchanges{im}purity of the LINC (\reviewchanges{left}) and \reviewchanges{an equivalent} SNAIL (\reviewchanges{right}) as a function of drive frequency and amplitude.
Both couplers are operated at a bias point where they are Kerr-free, and their parameters are optimized to match their beamsplitting strength and frequency at this bias point. Each coupler is simulated with equal decay \reviewchanges{and low-frequency dephasing}.
The LINC displays a significantly cleaner driven spectrum due to its parity protection.
}\label[fig]{fig3}
\end{figure*}

Second, the order of the allowed parametric processes obeys a strict selection rule -- the only processes allowed are of the type $\phi_{\text{AC}}^m \hat{\theta}^n$, where $n$ is strictly even.
We call this selection rule `parity protection', analogous to the selection rule in the DDS.
Interestingly, the LINC has this parity protection at arbitrary operating points, even though it is only linear at $\phi_{\text{DC}}=\pi/2$.
Note that despite the drive amplitude appearing as an odd function in the Hamiltonian, there is only a weak parity protection in the number of drive photons $m$.
This is because modulating the drive parameter $\phi_{AC}$ can also modulate the spread of the wave-function $\theta_{zpf}$, causing non-trivial corrections to the strength of parametric processes.
Additionally, odd-order processes can combine at higher orders in perturbation theory to form even-order processes.
The simplest non-trivial effect where this is observable is in the driven frequency shift of the LINC, which is suppressed but non-zero at the operating point (see~\cref{fig3}b).
However, since the LINC is a `true' parametric coupler, all such effects are well-predicted by measuring static properties of the LINC as a function of the parameter ($\phi_{\text{DC}}$), and computing appropriate derivatives.
We explore this in more detail in~\cref{app:Circuit}.

The dominant mixing process in the driven LINC is an effective 3-wave mixing process, which for small drive strengths at $\phi_{\text{DC}}=\pi/2$ is given by:
\begin{equation}
\begin{split}
    H_{\text{driven}}^{(3)} &= \left( \dfrac{E_J}{2E_L} \right) \omega_L \; J_1 \left(\vert \phi_{\text{AC}} \vert \right) \cos{\omega_d t} \left(\hat{c} + \hat{c}^\dagger \right)^2 \\
    &\approx \left. g_{\text{3wm}} \right\vert_{\frac{\pi}{2}} \;\phi_{\text{AC}}(t) \left(\hat{c} + \hat{c}^\dagger \right)^2,
\end{split}
\end{equation}
where $\omega_L=\sqrt{8E_LE_C}$ is the frequency of the LINC mode at the operating point and $\hat{c}$ and $\hat{c}^\dagger$ are its ladder operators. \reviewchanges{$J_1(x)$ is the first-order Bessel function arising from the sinusoidal dependence of the drive in the LINC Hamiltonian, with} $\phi_{\text{AC}}(t) = \vert \phi_{\text{AC}} \vert \cos{\omega_d t}$ being the drive.
For two modes Alice ($\hat{a}$, $\omega_a$) and Bob ($\hat{b}$, $\omega_b$) that are coupled to the LINC (with energy participations $p_a$ and $p_b$), driving at select frequencies can activate various desired processes, such as:
\begin{equation}
    \begin{split}
        \omega_d = \omega_a - \omega_b &\implies g_{\text{BS}} \left(\hat{a}^\dagger \hat{b} + \hat{a} \hat{b}^\dagger \right) \\
        \omega_d = \omega_a + \omega_b  &\implies g_{\text{TS}} \left(\hat{a}^\dagger \hat{b}^\dagger + \hat{a} \hat{b} \right) \\
        \omega_d = 2\omega_a &\implies g_{\text{Sa}} \left(\hat{a}^{\dagger^2} + \hat{a}^2 \right) \\
        g_{\text{BS}} = g_{\text{TS}} &= 2g_{\text{Sa}} \approx g_{\text{3wm}} \vert \phi_{\text{AC}} \vert \sqrt{p_a p_b},
    \end{split}
\label[eq]{eq:3wmProcesses}
\end{equation}
with $p_a=p_b$ for the single-mode squeezing process.
The strength of the first of these (beamsplitting) has been shown in~\cref{fig3}a for illustrative purposes.

While the LINC is linear when un-driven, turning on a drive can induce some nonlinearity in the coupler due to higher-order effects (\cref{fig3}a).
This finite coupler Kerr is not directly harmful, as the LINC remains in its ground state when driven, but it can induce undesired inherited Kerr in coupled linear quantum modes.
Since the Kerr arises from higher-order nonlinear effects, it can be suppressed simply by arraying multiple LINCs while keeping the strength of the desired process constant.
This results in the following Hamiltonian, assuming each LINC loop is still threaded by the same flux (see~\cref{app:Circuit}):
\begin{equation}
\begin{split}
    H_{\text{LINC}}^{(M)} = \; &4E_C \hat{n}^2 + \frac{E_L}{2}  \hat{\theta}^2\\
    &- 2M^2 E_J \sin{\phi_{\text{AC}}} \; \cos{\frac{\hat{\theta}}{M}},
\end{split}
\label[eq]{eq:arrayedLINC}
\end{equation}
where the inductance of each LINC in the array has been scaled down to preserve the same total inductance.
This suppresses the driven Kerr of the LINC by a factor of $1/M^2$, as shown in~\cref{fig3}a, and all remaining undesired processes are similarly suppressed.
In fact, with a large array of LINCs, this circuit element approaches the ideal three-wave mixing of a modulated linear inductor:
\begin{equation}
    H_{\text{LINC}}^{(\infty)} = 4E_C \hat{n}^2 + E_L \left( 1 + \frac{2E_J}{E_L} \phi_{\text{AC}} \right) \frac{\hat{\theta}^2}{2}.
\label[eq]{eq:idealLINC}
\end{equation}
We analyze the beamsplitter performance of both the arrayed and un-arrayed LINC in more detail in~\cref{app:FullParametric}.

Unfortunately, even an ideal three wave mixer incurs drive-induced frequency shifts, as can be seen from the average frequency of the above element $\langle \omega_L^{(\infty)}\rangle \propto \langle \sqrt{\left( 1 + 2E_J/E_L \phi_{\text{AC}}\right)} \rangle \neq 0$.
In fact, in contrast to a charge-driven element like the SNAIL, the driven frequency shift of the LINC is unaffected by arraying (see~\cref{fig3}b), and is not related to the coupler's anharmonicity.
It is instead given by the susceptibility of the LINC frequency to change in the flux, $\Delta \omega_L \propto \left. d^2 \omega / d \phi^2 \right\vert_{\frac{\pi}{2}}$.
One can minimize this driven shift (for example, when using the LINC in a parametric amplifier) by finding the inflection point $d^2 \omega / d \phi^2\left[\phi_{\text{DC}}\right] = 0$, which for the perfectly symmetric LINC (with $E_J/E_L = 0.3$) occurs at $\phi_{\text{DC}} \sim 0.594 \pi$ (dotted line in~\cref{fig3}b).
However, we note that this operating point is not \reviewchanges{Kerr-free and it breaks the weaker parity protection with respect to the drive}, and therefore might suffer parasitic processes less desirable than the coupler frequency shift.

To demonstrate how parity protection helps suppress parasitic processes, one can examine the driven behavior of the LINC in the presence of one or more drive tones, and compare it to a Kerr-free SNAIL.
Ideally, both couplers must remain in their ground state even at strong drive amplitudes, which is a challenge similar to mitigating measurement-induced state transitions during qubit readout \cite{BlaisTransmonIonization,BlaisUnifiedIonization2024}.
For a realistic simulation, we can incorporate an environment that induces both coupler decay and dephasing, which in the presence of the drive can be transformed into nonlinear coupler heating \cite{Yaxing_PRA_BilinearModeCoupling,QuantumHeatingYaxingRecommendation}.
We also include a coupled, information-storing qubit mode, which must ideally remain unaffected by the coupler during any parametric process.
Assuming the coupler is not periodically reset, this coupler-qubit system will reach a driven steady state after a sufficient number of operations.
This steady state will be pure if the coupler remains in its ground state, i.e. is neither affected by parasitic transition, dressed decoherence, nor stray interactions with the qubit.
Any impurity in the coupler state will translate into an infidelity in the desired  parametric process, due to state-dependent shifts of the process' strength and resonance condition (see~\cref{app:FullParametric}).
We thus evaluate the purity of the driven coupler steady state as a measure for evaluating coupler performance.

In~\cref{fig3}c, we compute this driven purity for a 6.5 GHz LINC through Floquet-Markov simulations, with decay $\gamma_1=(26.7\mu \text{s})^{-1}$, flux-noise dephasing $S_{\Phi\Phi}[\omega]=[1 \mu \Phi_0 / \sqrt{Hz}]^2 / \omega$, and coupling to a qubit (4.9 GHz) in ~\cref{fig3}c.
The comparison to an equivalent SNAIL, with identical frequency, beamsplitting strength and decoherence at the Kerr-free point, makes the advantage of parity protection clear -- the LINC remains significantly more pure at all drive frequencies.
This advantage is even more apparent in the presence of multiple drive tones, where the SNAIL sees substantial spurious transitions over most of the frequency range, but a large fraction of these inter-modulation products are suppressed by parity protection in the LINC (\cref{fig3}d).
Overall, these simulations show that in realistic settings, the LINC  should offer important advantages over the SNAIL in high-Q and multi-tone applications.

\section{Effect of asymmetry and parasitic inductance}
Our analysis of the LINC so far has been restricted to a perfectly symmetric circuit under a purely differential DC flux and drive.
However, practical implementations of the coupler will come with experimental imperfections, and it is important to consider the effect of finite asymmetry on the LINC's static and driven performance.
Specifically, we want to evaluate whether the LINC remains quasi-linear when idle, and whether it retains the parity protection in its driven processes.

To quantify the asymmetry of the outer junctions, we use the ratio $(E_{J1}-E_{J2})/E_L \coloneqq \beta_\Delta$, which from fabrication imperfections we expect to be $\lesssim 2\%$.
The second imperfection is a finite difference in the DC flux in the two LINC sub-loops, arising from a gradient in the residual magnetic field in which the experimental package cooled down.
This imperfection is in principle possible to cancel in-situ with two dedicated flux lines per coupler, such that the relative currents in the flux lines can be tuned to achieve arbitrary flux biases in the two sub-loops.
However, the LINC ideally only requires a single fast flux line for its drive and DC flux, which always applies symmetric flux to both sub-loops.
In this scenario, a residual DC flux difference $\phi_{\Delta}$ will change both the LINC's driven and undriven behavior, which we will analyze below.
In the fully general case, one may also have an asymmetry in the applied drive due to imperfect drive line engineering, which might also affect performance.
However, this asymmetry can be largely avoided by appropriately designing the flux distribution and capacitance matrix of the device, aided by numerical simulation techniques~\cite{SQUIDBeamsplitterPaper,Lu2025}.

Let us first analyze the effect of asymmetries on the linearity of the idle LINC.
Since the drive is off, the asymmetries to consider are the junction and DC flux asymmetries, both of which result in an effective shift in the potential minima of the LINC to $\theta_c^{\text{min}} \neq 0$.
Specifically,
\begin{equation}
    \theta_c^{\text{min}} \approx \frac{2}{3}\phi_{\Delta} - \beta_\Delta
\end{equation}
to lowest order in $\beta_\Delta$ and $\phi_{\Delta}$.
This new minima results in a change in the undriven potential, meaning that Taylor expansions to compute parameters of interest should be expanded about this new point.
In light of this shift, the static Kerr at the operating point becomes (see \cref{app:Asymm})  
\begin{equation}
    \alpha_L^{\text{asym}} \approx E_c\beta_\Delta\sin\left(\phi_{\Delta} - (1 + \sqrt{2})\beta_\Delta\right)
\end{equation}
for small asymmetries.
Thus any static nonlinearity gained by the LINC in the presence of these asymmetries is only second-order in $\{\beta_\Delta,\phi_{\Delta}\}$.
Notably, the sign of the individual asymmetries is important, as the two effects can either constructively or destructively interfere. Additionally, in the absence of junction asymmetry, the static Kerr is nulled for any value of $\phi_{\Delta}$.
As a reference, for a $\beta_\Delta = 2\%$ and a $\phi_{\Delta} = -\frac{\pi}{2}\times1\% $, the static Kerr of the LINC is roughly $-130$ KHz for an $E_c = 100$ MHz. 
Additionally, the shifted Kerr-free point can be found by changing the symmetric flux bias by less than $0.005\pi$ (see \cref{app:Asymm}).

When a LINC with non-zero asymmetry is driven, the LINC may turn on processes that would have otherwise been parity-protected.
As an example, we consider the third-order mixing process involving two drive photons and a single LINC photon, corresponding to a parasitic sub-harmonic displacement of the coupler.
This process is always allowed in any charge-driven mixer (like the SNAIL), yet is forbidden in the ideal LINC.
The strength of this process, which we represent as $H_{\text{sub}} = g_{1,2} \phi_{\text{AC}}^2 (\hat{c}+\hat{c}^\dagger)$, is given by (\cref{app:Asymm}):
\begin{equation}
\begin{split}
        g_{12}  &\approx -\frac{E_L\theta_{zpf}}{2}\Big[
    \beta_\Delta
    + \frac{1}{2}\beta_\Sigma^2\sin(\phi_{\Delta} - \beta_\Delta) 
    \Big]
\end{split}
\end{equation}
 and is thus linearly sensitive to both junction and flux asymmetry, with their relative sensitivity depending on $\beta_\Sigma$. For modest asymmetries (again $\beta_\Delta = 2\%$ and $\phi_{\Delta} = -\frac{\pi}{2}\times1\% $), the relative suppression of such parasitic resonances is $ g_{12}\phi_{\text{AC}}^2/g_{21}\phi_{\text{AC}} \approx 5\%$ for $\phi_{\text{AC}} = 0.2\pi$, still achieving over an order of magnitude reduction compared to an ordinary 3-wave mixer.
 \reviewchanges{In effect, since the LINC is a balanced RF-SQUID, any asymmetry only brings its performance closer to the performance of a regular RF-SQUID or SNAIL.
 The full effect of asymmetries on parasitic resonances can be captured by plotting the available drive-space for the LINC, similar to }\cref{appx_fig3}\reviewchanges{, which add no significant spurious processes up to $\phi_{\text{AC}}=0.2\pi$ even at junction asymmetries up to $10\%$.}

Finally, for reasonable coupler geometries, one may expect the LINC to have a non-negligible parasitic linear inductance in series with the coupler.
Such a parasitic inductor often has an impact beyond diluting the driven nonlinearity of a general coupler -- for example, in the SNAIL, this can cause the Kerr-free operating point to shift significantly.
In the LINC, this linear inductance is far less harmful, since it does not directly interact with the flux (drive) degree of freedom.
In fact, any parasitic series inductance fully preserves the linearity of the idle LINC at the same operating point of $\phi_{\text{DC}}=\pi/2$, where every even order nonlinearity (including Kerr) is simultaneously suppressed.
On the other hand, parasitic inductances within the LINC loop, while relatively small in magnitude, may shift the linear operating point slightly.
We include a detailed analysis  in~\cref{app:LoopInductance,app:LinearInductance}.
Overall, our simulations predict that the LINC remains a robust and protected quantum mixer in the presence of realistic experimental imperfections.

\section{Conclusions and Outlook}
We have introduced a protected quantum mixer that combines the benefits of Kerr-free three-wave mixing and mixer balancing.
This results in a nonlinear element that is nearly linear when idling, and only activates its nonlinearity when it is driven to turn on gates.
Even when driven, the mixer balancing enforces selection rules that prevent a large fraction of the parasitic processes allowed by a general Josephson nonlinearity.
The benefits of such a mixer are significant, offering possible advantages in bosonic and qubit control, frequency conversion, and amplification.

In the bosonic context, the LINC promises to break the trade-off between fast nonlinear control and the idle errors introduced by a nonlinear ancillary mode (similar to~\cite{ECDControlPaper,CubicPhaseTheory,CubicPhaseExperiment,ding2024oscillatorkerrcat}).
This is particularly important for multi-photon encodings, where inherited Kerr and thermal-noise induced dephasing irreversibly spoil logical information.
Even in single-photon encodings like the dual-rail bosonic qubit, the LINC should reduces static dispersive interactions between neighboring oscillators, while enabling fast and clean universal control through parametric beamsplitting.
This advantage could even extend to non-Gaussian control, where the LINC could serve as an interface between the bosonic mode and its ancillary qubit, effectively shielding the nonlinearity when idling \cite{OndrejHighQMemories}.

The parity-protection in the LINC could also provide important advantages in its power handling and multi-tone operation.
This is useful in multiple contexts, with the simplest being the activation of a simultaneous parametric coupling between multiple neighboring elements.
If the LINC were used as an amplifier, an array of LINCs should perform equivalently to an array of balanced RF-SQUIDs, potentially outperforming both SNAIL and regular RF-SQUID-based amplifiers.
Such an implementation would potentially allow simultaneous \reviewchanges{ amplification of several signals without degradation from intermodulation products} \cite{Pozar_2012,SNAIL_Analysis_Frattini,IndermodulationDistortionTWPAWalraff}, easing the constraints on multiplexed readout \cite{IndermodulationDistortionTWPAWalraff,NakamuraMultiplexedReadout}.
Finally, the LINC could also simultaneously activate multiple types of parametric processes, giving rise to new bosonic control techniques.
For example by activating a resonant beamsplitting and two-mode squeezing between two oscillators in the high-Q regime, one could realize a direct parametric quadrature-quadrature coupling between them, enabling two-qubit gates for the Gottesman-Kitaev-Preskill (GKP) code \cite{OriginalGKP,GKPCiteFromShraddha}.

The advantages predicted in our simulations will require experimental proof, which is the subject of immediate future work.
Specifically, one must solve the engineering challenges of delivering the differential bias and drive in a compact architecture, and fabricating high-Q shunting inductors or junction-array.
\reviewchanges{
Thankfully, the LINC incorporates standard circuit elements and does not operate in an exotic parameter regime.
We do not foresee any major fabrication challenges, as the LINC shares the topology and parameter range of the ATS, which has been implemented in multiple experiments} \cite{ATS_Paper,A&B_100_sec_bit_flip_w_ATS,AmazonConcatenatedCats_w_ATS,Philippe_four_photon_diss_w_ATS}\reviewchanges{. 
As the LINC must be operated away from any sweet-spot (since $g_\text{3wm} \propto  \frac{\partial\omega}{\partial\Phi}$), it is vital to mitigate flux noise, but the LINC does not require unreasonably low flux noise to perform well.
For scaling up, the LINC is readily compatible with existing processor architectures comprising any mixture of qubits and resonators (bosonic modes), and can act as a parametric coupler between them.
As the control lines for the LINC are nearly identical to the DC-SQUID, any architecture capable of fast-flux control could utilize LINCs. While cross-talk between LINCs would be heavily implementation-specific, we note that the platform of flux-tunable transmons has seen great success in large-scale implementations} \cite{GoogleSurfaceCodeBelowThreshold,ZuchongzhiChinaProcessor}.\reviewchanges{
Additionally, our} analysis of the LINC's sensitivity to small asymmetries shows that its experimental implementation need not be impractically precise, and that the coupler should perform well even with realistic imperfections.
Overall, the LINC is a promising new element in the toolbox of superconducting quantum circuits, offering a unique combination of linearity, high-fidelity control, and practical design constraints.

\section{Acknowledgements}
We are grateful for discussions with P. Winkel on high-kinetic inductance shunting inductors and with S. J. de Graaf on the SNAIL as a three-wave mixer. We also thank M. Devoret and C. Zhou for helpful discussions about the LINC Hamiltonian and dephasing. We thank M. Hays and W. Oliver for general discussions about the coupler.

This research was sponsored by the Army Research Office (ARO) under grant no. W911NF-23-1-0051, and by the U.S. Department of Energy (DoE), Office of Science, National Quantum Information Science Research Centers, Co-design Center for Quantum Advantage ($\text{C}^{\text{2}}$QA) under contract number DE-SC0012704.The views and conclusions contained in this document are those of the authors and should not be interpreted as representing the official policies, either expressed or implied, of the ARO, DoE or the US Government. The US Government is authorized to reproduce and distribute reprints for Government purposes notwithstanding any copyright notation herein. L.F. and R.J.S. are founders and shareholders of Quantum Circuits Inc. (QCI).
\\
\indent A.M. and Y.L. conceptualized the coupler. A.M., J.W.O.G., Y.L., and Al.M. developed methodology and performed formal analysis. A.M., J.W.O.G., Y.L., and Al.M. wrote the manuscript with feedback and supervision from L.F. and R.J.S.

\begin{center}
\rule[0.5ex+0.25pt]{0.06\textwidth}{0.5pt}
\rule[0.5ex+0.pt]{0.06\textwidth}{1.pt}
\rule[0.5ex-0.25pt]{0.06\textwidth}{1.5pt}
\rule[0.5ex-0.5pt]{0.06\textwidth}{2.pt}
\rule[0.5ex-0.25pt]{0.06\textwidth}{1.5pt}
\rule[0.5ex]{0.06\textwidth}{1.0pt}
\rule[0.5ex+0.25pt]{0.06\textwidth}{0.5pt}
\end{center}


\appendix


\begin{figure*}
\centering
\includegraphics[width=0.98 \textwidth]{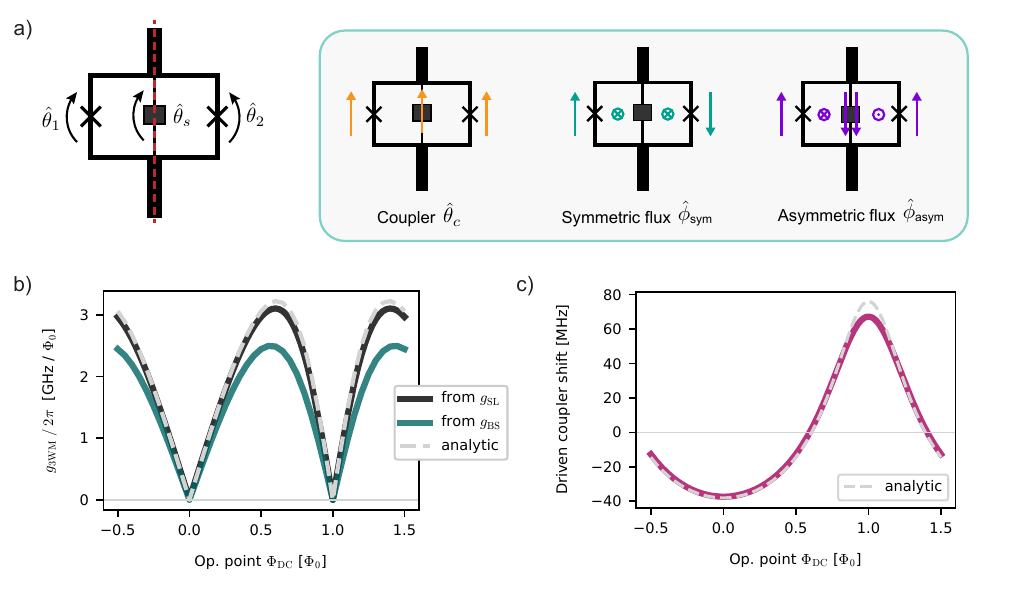}
\caption{
\textbf{Analyzing the LINC circuit}
\textbf{a}, Modes of operation of the circuit. The circuit, with an arbitrary inductive element as its center shunt, is defined by the three variables $\left\{\theta_1, \theta_2, \theta_s\right\}$.
These can be re-written in terms of the more operationally relevant charge-dipole ($\hat{\theta}_c$), symmetric flux ($\hat{\phi}_{sym}$), and anti-symmetric flux ($\hat{\phi}_{asym}$) modes.
\textbf{b}, LINC three-wave mixing strength as a function of the operating point, comparing analytic formula (grey dashed, \cref{eq:g3wm_derivation}) to exact time-domain and Floquet results from a squeezing operation within the LINC ($g_{\text{SL}}$, black) and a beamsplitter operation between two external resonators (teal, details in~\cref{app:FullParametric}) respectively. The discrepancy in the beamsplitting prediction may be due to driven changes in the effective resonator-LINC participations, which are captured in the Floquet simulation but not in the analytics.
\textbf{c}, Comparison of the analytic formulae (grey dashed, \cref{eq:Zeeman_derivation}) to exact Floquet simulation for driven coupler Zeeman shift as a function of DC flux, for fixed drive amplitude $\phi_{AC}=0.1\pi$.
}\label[fig]{appx_fig1}
\end{figure*}

\section{Analyzing the LINC circuit}
\label[appdx]{app:Circuit}
We derive here the Hamiltonian of the LINC circuit, along with analytical formulae that describe its dominant characteristics, as a function of its operating DC flux point.
The circuit in \cref{appx_fig1} contains three inductive branches, and two galvanic loops that can be independently threaded by DC magnetic flux, setting its operating point.
In the presence of non-zero field, each of these three branches can incur a different voltage drop across them, with corresponding superconducting phase drops across the three inductive elements, $\hat{\theta}_1$, $\hat{\theta}_s$ and $\hat{\theta}_2$.
We will rewrite these phase drops in terms of more convenient independent variables,
\begin{equation}
    \begin{split}
        \hat{\theta}_c &= \left(\hat{\theta}_1 + \hat{\theta}_2 + \hat{\theta}_s \right)/3 \\
        \hat{\phi}_{\text{sym}} &= \left(\hat{\theta}_1 - \hat{\theta}_2\right)/2 \\
        \hat{\phi}_{\text{asym}} &= \left(\hat{\theta}_1 + \hat{\theta}_2 - 2\hat{\theta}_s \right)/2.
    \end{split}
\end{equation}
$\hat{\theta}_c$ corresponds to the common mode of the circuit, and is the phase variable conjugate to the charge on the capacitive pads ($\hat{n}_c$).
In the absence of a magnetic field, this phase fully describes the circuit, which behaves similar to an inductively shunted transmon (IST, \cite{ClarkeISTQuantization}).
A symmetric flux in both loops displaces $\hat{\phi}_{\text{sym}}$, and forms the bias and the differential drive for the LINC.
An anti-symmetric flux on the other hand, displaces $\hat{\phi}_{\text{asym}}$ and forms the bias for the asymmetrically threaded SQUID (ATS) mode of operating the circuit.

The full Hamiltonian, with no assumptions on symmetry, is given by:
\begin{equation}
    \begin{split}
    H_{\text{full}} &= 4E_C \hat{n}^2_c + E_L \dfrac{(\hat{\theta}_c - 2\hat{\phi}_{\text{asym}}/3)^2}{2} \\
    &-  (E_{J_1}+E_{J_2})\cos (\hat{\theta}_c) \cos (\hat{\phi}_{\text{asym}}/3) \cos (\hat{\phi}_{\text{sym}})\\
    &+  (E_{J_1}+E_{J_2})\sin (\hat{\theta}_c) \sin (\hat{\phi}_{\text{asym}}/3) \cos (\hat{\phi}_{\text{sym}})\\
    &+  (E_{J_1}-E_{J_2})\cos (\hat{\theta}_c) \sin (\hat{\phi}_{\text{asym}}/3) \sin (\hat{\phi}_{\text{sym}})\\
    &+  (E_{J_1}-E_{J_2})\sin (\hat{\theta}_c) \cos (\hat{\phi}_{\text{asym}}/3) \sin (\hat{\phi}_{\text{sym}})
    \end{split}
\label[eq]{eq:Hfull}
\end{equation}
For the rest of this section, let us assume that all symmetry constraints in the circuit are satisfied, specifically that the outer junctions have equal Josephson energy $E_J$, and any DC flux and drive enter both loops symmetrically.
We will explore the effects of deviating from this ideal in~\cref{app:Asymm}.
This constraint lets us set $\hat{\phi}_{\text{asym}} = 0$, thus giving:
\begin{equation}
\begin{split}
    (\hat{\theta}_1 + \hat{\theta}_2) &= 2\hat{\theta}_s \\
    \implies \hat{\theta_c} &= \hat{\theta}_s, \\
    \hat{\theta}_1 &= \hat{\theta}_c + \hat{\phi}_{\text{sym}}, \\
    \hat{\theta}_2 &= \hat{\theta}_c - \hat{\phi}_{\text{sym}}.
\end{split}
\end{equation}
We additionally assume the symmetric flux degree of freedom $\hat{\phi}_{\text{sym}}$ is stiff, and can be replaced by a classical variable $\phi_d = \langle \hat{\phi}_{\text{sym}} \rangle = \pi \frac{\Phi_{\text{loop}}}{\Phi_0}$, where $\Phi_{\text{loop}}$ is the total flux threading the outer loop and $\Phi_0$ is the flux quantum.
Note that this assumption is only valid for a weakly coupled flux-drive port, and is similar to the usual treatment of the differential mode in a SQUID circuit  \cite{SQUIDBeamsplitterPaper}.
Under these assumptions, $H_{\text{full}}$ reduces to the LINC Hamiltonian in the main text:
\begin{equation}
    H_{\text{LINC}} = 4E_C \hat{n}^2_c + E_L \dfrac{\hat{\theta}^2_c}{2} - 2 E_J \cos{\phi_d} \cos{\hat{\theta}_c},
\end{equation}
where at the operating point, $\phi_d = \pi/2 + \phi_{AC}$.

Let us now analyze the LINC's frequency and nonlinearity analytically, through a Taylor expansion.
Since the two variables $\hat{\theta}_c$ and $\phi_d$ are independent, our strategy in general will be to study the bi-variate Taylor expansion of the LINC potential $U_{\text{LINC}}(\hat{\theta}_c, \phi_d)$ about its minima at $\hat{\theta}_c^{\text{min}} = 0$ and $\phi_d = \phi_{\text{DC}}$.
However, this expansion comes with a subtlety -- we aim to study the LINC's physics in a bosonic basis defined by ladder operators $\left\{\hat{c}, \; \hat{c}^\dagger \right\}$, where
\begin{equation}
\begin{split}
    \hat{\theta}_c &= \theta_{zpf} \left(\hat{c} + \hat{c}^\dagger\right) \\
    \theta_{zpf} &= \left( \dfrac{2 E_C}{\left. \partial^2  U_{\text{LINC}} / \partial \hat{\theta}^2_c \right\vert_{\hat{\theta}_c=0}} \right)^{1/4} \\
    &= \left(\dfrac{2 E_C}{E_L + 2E_J \cos{\phi_d}}\right)^{1/4}.
\end{split}
\end{equation}
The bosonic basis is thus itself a function of the LINC flux point, and modulating this flux modulates not just the LINC potential, but also the effective spread of the un-driven wavefunction $\theta_{zpf}(\phi_d)$.
This latter modulation results in an important re-normalization of the LINC physics, for example, resulting in a non-zero driven coupler shift at $\phi_{\text{DC}}=\pi/2$.
We can capture this physics by expanding the LINC potential as follows:
\begin{equation}
\begin{split}
    &U_{\text{LINC}} = \frac{E_L}{2} \theta^2_c - 2 E_J \cos{\phi_d} \cos{\hat{\theta}_c} \\
    &= \sum_{\substack{m \in \text{even}, \\ n }} \left. \dfrac{\partial^n}{\partial \phi_d^n} \left(\left. \dfrac{\partial^{m} \; U_{\text{LINC}}}{\partial \hat{\theta}^m_c} \right\vert_{\hat{\theta}_c=0}\dfrac{\hat{\theta}^m_c}{m!}\right) \right\vert_{\phi_{\text{DC}}} \dfrac{\phi^n_{AC}}{n!} \;  \\
    &= \sum_{\substack{m \in \text{even}, \\ n }} g_{mn}[\phi_{\text{DC}}] \; \phi^n_{AC} \; \left(\hat{c}+\hat{c}^\dagger \right)^m, 
\end{split}
\label[eq]{eq:Derivatives}
\end{equation}
with generalized parametric strengths $g_{mn}$ defined as:
\begin{equation}
\begin{gathered}
    g_{mn}[\phi_{\text{DC}}] := \dfrac{1}{m!\:n!} \sum_{k=0}^n \binom{n}{k} \left.u_{m,n-k} \; \frac{\partial^k \theta_{zpf}^m}{\partial \phi_d^k} \right\vert_{\phi_d = \phi_{\text{DC}}} , \\
    u_{m,k} := \left. \dfrac{\partial^{m+k} \; U_{\text{LINC}}}{\partial \hat{\theta}^m_c \; \partial \phi_d^k} \right\vert_{\hat{\theta}_c=0}.
\end{gathered}
\label[eq]{eq:g_derivatives}
\end{equation}
The expansion in \cref{eq:Derivatives} explicitly only contains the even terms in $m$, highlighting that the inherent parity-protection rule holds at arbitrary operating points.
Note that at $\phi_{\text{DC}} = \pi/2$, an additional protection is enforced which sets part of the LINC potential to zero, which makes the un-driven Hamiltonian linear:
\begin{equation}
    \left. u_{m > 2, \;0} \right\vert_{\phi_{\text{DC}}=\pi/2} = 0.
\end{equation}

We now derive the LINC's frequency, effective three-wave mixing strength, and anharmonicity explicitly.
We first use \cref{eq:Derivatives} to expand the undriven LINC Hamiltonian to fourth order, deriving its frequency ($\omega_L$) and Kerr nonlinearity ($\alpha_L$):
\begin{equation}
\begin{split}
    \left. H_{\text{LINC}}^{(4)} \right\vert_{\phi_{AC}=0} &= \omega_L \hat{c}^\dagger \hat{c} + \dfrac{\alpha_L}{2} \hat{c}^{\dagger^2} \hat{c}^2 \\
    \\
    \omega_L [\phi_{\text{DC}}] &= 4\; g_{2,0}[\phi_{\text{DC}}] + \alpha_L[\phi_{\text{DC}}]\;  \\
                         &= \sqrt{8 E_C (E_L + 2E_J \cos{\phi_{\text{DC}}}) } + \alpha_L \\
    \\
    \alpha_L [\phi_{\text{DC}}] &= 12 \; g_{4, 0}[\phi_{\text{DC}}] \;  \\
             &= -\dfrac{2E_Jcos{\phi_{\text{DC}}}}{E_L+ 2E_J cos{\phi_{\text{DC}}}} E_C
\end{split}
\end{equation}
At the operating point, $\omega_L[\pi/2]=\sqrt{8E_L E_C}$ and $\alpha_L[\pi/2]=0$.
This analytical prediction is overlaid on results from an exact diagonalization of the Hamiltonian of a LINC with a realistic center shunt comprised of an array of 10 junctions in~\cref{fig:fig2}a.

Similar to the static derivation, one can also find the effective three-wave mixing strength when driven:
\begin{equation}
\begin{split}
    g_{\text{3wm}}  [\phi_{\text{DC}}] &= 2\;g_{21}[\phi_{\text{DC}}] \\
    &= \frac{E_J}{E_L}\;\sqrt{\dfrac{2E_L E_C}{1 + 2 \frac{E_J}{E_L}\cos{\phi_{\text{DC}}}}} \sin{\phi_{\text{DC}}} \\
    \Rightarrow g_{\text{3wm}}  \left[\frac{\pi}{2}\right] &= \frac{E_J}{2E_L} \; \omega_L.
\end{split}
\label[eq]{eq:g3wm_derivation}
\end{equation}
Note that this is almost exactly half the flux-noise sensitivity $d\omega_L / d\phi_{\text{DC}} = 4 g_{2,1} + d\alpha_L/d\phi_{\text{DC}}$, due to conventions for the definition of $g_{3wm}$ chosen in~\cref{eq:3wmProcesses}.
This analytic formula matches the numerically derived three-wave mixing strength reasonably well (\cref{appx_fig1}b).
The latter is calculated from a Floquet simulation of either a self-squeezing of the LINC (with an added Kerr to induce $|0\rangle \leftrightarrow |2\rangle$ oscillations), or a full beamsplitting process (see~\cref{app:FullParametric}).
It is also simple to derive the relative sensitivity of this process to flux noise, for~\cref{eq:g3wm_flux_noise} in the main text:
\begin{equation}
\begin{split}
    \left. \frac{1}{g_{\text{3wm}}} \frac{dg_{\text{3wm}}}{d\phi} \right\vert_{\phi_{\text{DC}}} &= \frac{1 + \cos^2{\phi_\text{DC}}+\frac{E_L}{E_J}\cos{\phi_\text{DC}}}{\sin{\phi_{\text{DC}}}\left(\frac{E_L}{E_J} + 2\cos{\phi_\text{DC}}\right)} \\
    \implies\left. \frac{2\pi}{g_{\text{3wm}}} \frac{dg_{\text{3wm}}}{d\Phi} \right\vert_{\Phi_0/2} &\approx  \frac{1}{\Phi_0} \frac{2E_J}{E_L},
\end{split}
\end{equation}
where we have used $\phi_{\text{DC}}=\frac{\pi}{\Phi_0} \Phi_{\text{DC}}$.

Finally, the driven Zeeman shift, which is independent of the coupler's Kerr nonlinearity, is given by:
\begin{equation}
    \begin{gathered}
    \Delta \omega_L [\phi_{\text{DC}}, \phi_{AC}] = 2\; g_{2,2}[\phi_{\text{DC}}] \;|\phi_{AC}|^2 \\ \\
    g_{2,2} = - \frac{E_J}{4} \sqrt{\frac{2 E_C}{E_L + 2E_J\cos{\phi_{\text{DC}}}}} \times \\\;\;\frac{ E_J(1+\cos^2{\phi_{\text{DC}}}) +E_L\cos{\phi_{\text{DC}}}}{E_L + 2E_J\cos{\phi_{\text{DC}}}} \\ \\
    \implies \Delta\omega_L [\pi/2, \phi_{\text{AC}}] = -\left(\frac{E_J}{2E_L} \right)^2 \omega_L \; |\phi_{\text{AC}}|^2
\end{gathered}
\label[eq]{eq:Zeeman_derivation}
\end{equation}
This matches realistic Floquet simulations of the Zeeman shift with $\phi_{AC}=0.1\pi \sin{\omega_d t}$ well, as seen in~\cref{appx_fig1}c.
Note that this Zeeman shift can also be used to analytically account for changes in participations during driven operations.

In pursuit of additional linearity, one could also choose to array the LINC with $M$ LINC loops, shunted by a single capacitor.
To preserve the same frequency and parametric process strength as the single-loop LINC, the inductive energy $E_L$ and Josephson energy $E_J$ of each loop must be scaled by $M$, and the drive and bias in each loop must be nominally identical to the single-loop device.
 This simple extension of the circuit can be analyzed by the transformation $\hat{\theta}_c \rightarrow \hat{\theta}_c / N$, $(E_L, E_J) \rightarrow (M E_L, M E_J)$, resulting in~\cref{eq:arrayedLINC} in the main text.
 In effect, this preserves the inductance and three-wave mixing of the LINC, but suppresses all higher processes $\propto \hat{c}^{2n}$ by a factor of $1/M^{2(n-1)}$.
 As $M\rightarrow \infty$, this approaches the ideal three-wave mixing properties of a linear resonator with a modulated inductance (\cref{eq:idealLINC}).

 \begin{figure}[ht]
    \centering
    \includegraphics[width=0.49 \textwidth]{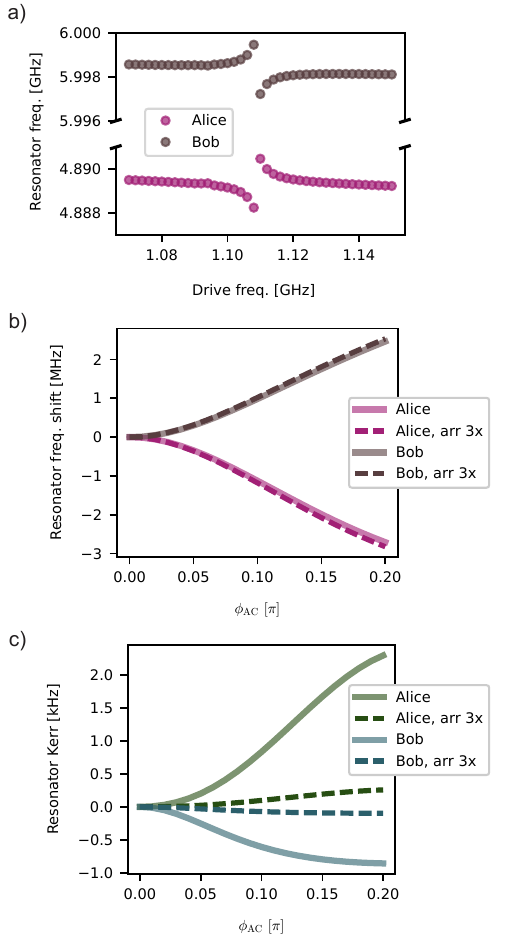}
    \caption{
    \textbf{Parametric beamsplitting with the LINC}
    \textbf{a}, Driven avoided crossing due to the beamsplitter interaction between two storage modes, Alice (4.9 GHz) and Bob (6.0 GHz), as a function of drive detuning from beamsplitter resonance, at a fixed drive amplitude of $\phi_{AC}=0.2\pi$.
    \textbf{b}, Driven frequency shift of each mode for an unarrayed (solid lines) vs arrayed (dashed lines) LINC. Arraying does not make a noticeable difference.
    \textbf{c}, Driven Kerr of each mode for an unarrayed (solid lines) vs arrayed (dashed lines) LINC. Arraying 3 LINCs reduces the inherited Kerr by a factor of $\sim 9$.
    }\label[fig]{appx_fig2}
    \end{figure}

\vspace{-1em}
\section{Driven operations with the LINC}
\label[appdx]{app:FullParametric}

We now analyze the LINC as a parametric coupler and evaluate its performance in a realistic frequency-stack.
Consider the LINC coupled to two resonators (Alice and Bob)  with linear coupling strengths and frequencies $g_a, \omega_a$ and $g_b, \omega_b$.
One can analyze the resulting dressed Hamiltonian through a simple transformation:
\begin{equation}
    \begin{split}
        \hat{c} &= \hat{\tilde{c}} + \sqrt{p_a} \hat{\tilde{a}} + \sqrt{p_b} \hat{\tilde{b}} \\
        p_a, p_b &= \left(\frac{g_a}{\Delta_{La}} \right)^2, \left(\frac{g_b}{\Delta_{Lb}}\right)^2,
    \end{split}
\end{equation}
where $\Delta_{L\;a,b} = \omega_L - \omega_{a,b}$.
In a frame rotating at $\omega_a$, the Hamiltonian of the system up to the third order is then:
\begin{equation}
    \begin{split}
        H_{\text{3wm}} &= \tilde{\omega}_a \hat{\tilde{a}}^\dagger \hat{\tilde{a}} + \tilde{\omega}_b \hat{\tilde{b}}^\dagger \hat{\tilde{b}} + \tilde{\omega}_L \hat{\tilde{c}}^\dagger \hat{\tilde{c}} \\
        &+ g_{\text{3wm}} \frac{\vert \phi_{\text{AC}} \vert}{2} \left(e^{i \omega_d t} + e^{-i \omega_d t}\right) \times \\
         &\left(\sqrt{p_a} \;\hat{\tilde{a}} + \sqrt{p_b} \;\hat{\tilde{b}} + \hat{\tilde{c}} + \text{h.c.} \right)^2
    \end{split}
\end{equation}
It is then simple to derive the strength of each parametric process when the drive frequency $\omega_d$ is set to the appropriate resonance condition, as in~\cref{eq:3wmProcesses}.

While it is convenient to follow analytic derivations from $H_{\text{3wm}}$, from an experimental perspective, it is difficult to directly measure $g_{\text{3wm}}$ and $p_a, p_b$.
Since these processes are truly a result of modulating the parameter $\phi$, one can instead just measure the frequency dependence of the Alice and Bob modes as a function of operating flux $\phi_{\text{DC}}$, and then take appropriate derivatives of these flux curves as in~\cref{eq:g_derivatives}.
For the beamsplitting process, this results in a simplified relation (ignoring driven frequency shifts):
\begin{equation}
    g_{\text{BS}} = \sqrt{\frac{\partial \omega_a}{\partial \phi_{\text{DC}}} \frac{\partial \omega_b}{\partial \phi_{\text{DC}}}} \left\vert \frac{\phi_{\text{AC}}}{2} \right\vert.
\end{equation}
which is similar to other flux-driven couplers~\cite{Zakka2011,ParametricCQEDPaper}.

To numerically demonstrate such a parametric beamsplitting process, we consider the specific frequency stack of $\omega_a = 2\pi\times4.9$GHz, $\omega_b = 2\pi\times6.0$GHz, and $\omega_c = 2\pi\times6.5$GHz, with Rabi coupling strengths of $g_{ac} = 2\pi\times 120$MHz and $g_{bc} = 2\pi\times 50$MHz. We choose the LINC circuit parameters as $E_C = 2\pi\times100$MHz, $E_J = 2\pi\times15.84$GHz, and $E_L = 2\pi\times52.8$GHz. With M LINC loops in an array, the system Hamiltonian is given by 
\begin{equation}
    \begin{split}
        H_\text{LINC} &=\omega_a \hat{a}^\dagger \hat{a} +\omega_a \hat{b}^\dagger \hat{b} +\omega_c \hat{c}^\dagger \hat{c}\\
        &- g_{ac}(\hat{a}^\dagger-\hat{a})(\hat{c}^\dagger-\hat{c})- g_{bc}(\hat{b}^\dagger-\hat{b})(\hat{c}^\dagger-\hat{c})\\&-2M^2E_J\sin{\phi_\text{AC}}\cos{\frac{\hat{\theta}_c}{M}},
    \end{split}
\end{equation}
assuming that each LINC loop is modulated by $\phi_\text{ext}=\pi/2 + \phi_\text{AC}$.

Using standard Floquet theory, we can numerically compute the quasi-energies of the Floquet modes, and extract drive-induced frequency shifts and Kerr shifts as a function of the drive frequency and amplitude, which are shown in~\cref{appx_fig2}.
As in the case of the LINC itself, it is clear that the resonator driven frequency shifts do not change upon arraying, but their driven Kerr can be significantly suppressed.

\section{\reviewchangesAlt{Evaluating driven performance}}
\label[appdx]{app:Floquet}
\reviewchanges{We now focus on quantifying the driven coupler's performance, in the presence of environmental noise and the resonator modes, Alice and Bob. Specifically, we show that the LINC's is resilient to drive-induced parasitic processes and dressed decoherence, which should lead to an improvement in parametric process fidelity.
As a benchmark for comparison, we compare it to a SNAIL at its Kerr-free flux point.}

\reviewchanges{We model the driven SNAIL Hamiltonian in the displaced frame as}
\begin{equation}
    \begin{split}
        H_\text{SNAIL} &=\omega_a \hat{a}^\dagger \hat{a} +\omega_a \hat{b}^\dagger \hat{b} +\omega_c \hat{c}^\dagger \hat{c}\\
        &- g_{ac}(\hat{a}^\dagger-\hat{a})(\hat{c}^\dagger-\hat{c})- g_{bc}(\hat{b}^\dagger-\hat{b})(\hat{c}^\dagger-\hat{c})
        \\&-\alpha M^2E_J\cos_\text{nl}{\left(\theta_\text{eq,1}+\phi_\text{AC}+\frac{\hat{\theta}}{M}\right)}
        \\&-NM^2E_J\cos_\text{nl}{\left(\frac{1}{N}\left(\theta_\text{eq,2}+\phi_\text{AC}+\frac{\hat{\theta}}{M}\right)\right)},
    \end{split}
\end{equation}
\reviewchanges{where $M$ is the number of SNAIL loops in an array, $N$ is the number of Josephson junctions in the array of each SNAIL loop, and $\theta_\text{eq,i}$ is the frustration phase across the i-th branch of the SNAIL loop, as a result of the dc flux penetrating the loop. For consistency, we denoted the phase displacement of the SNAIL as $\phi_{AC}$, even though it has a different physical origin than flux modulation in the LINC circuit. We carefully choose the SNAIL parameters as 
$E_C = 2\pi\times100$ MHz, $E_J = 2\pi\times276$ GHz, and $\alpha = 0.193$.
Under these parameters, the SNAIL circuit yields the same frequency of $\omega_c = 2\pi\times6.5$GHz as the LINC, as well as the same $g_{3wm}$, at its Kerr-free point of $\Phi_\text{ext} = 0.442\Phi_0$.}

\begin{figure*}[ht]
\centering
\includegraphics[width=0.95 \textwidth]{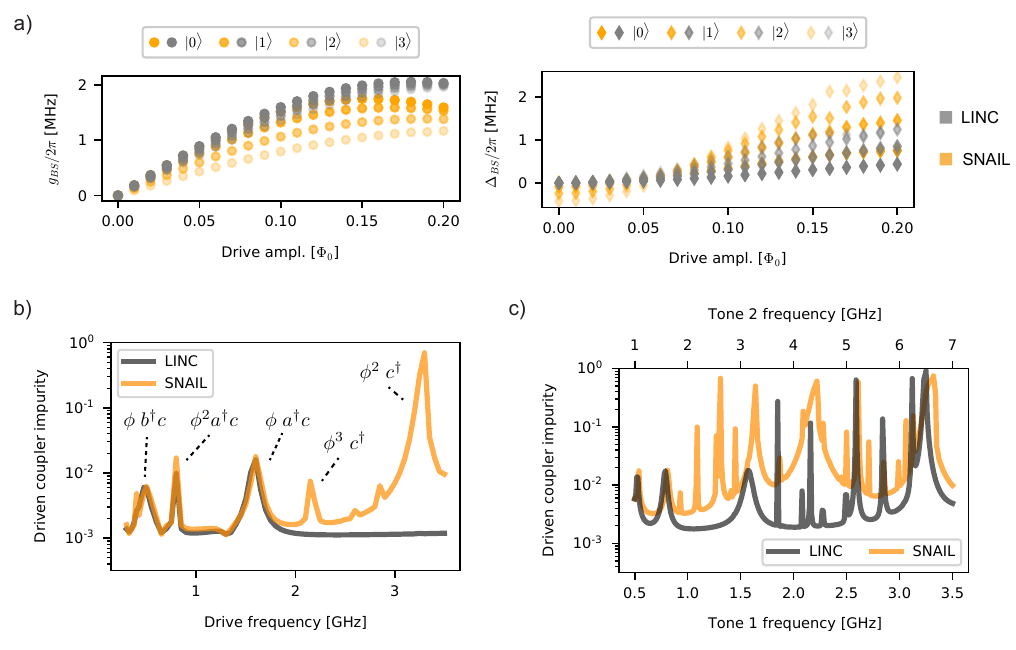}
\caption{
\textbf{\reviewchanges{Driven coupler impurity}}
\textbf{a}, Dispersion in beamsplitting strength (left) and resonance condition (right) as a function of coupler (undriven) state $|0\rangle, \dots, |3\rangle$. The LINC (grey) has less dispersion than the SNAIL (yellow), implying that for similar excited state populations, it will have higher process fidelity.
\textbf{b}, Steady-state impurity of the driven LINC and the (charge) driven SNAIL, in the presence of decay and flux-noise dephasing, as a function of drive frequency at fixed amplitude ($\phi_{\text{AC}}=0.2\pi$). The simulation also includes additional coupled modes Alice (4.9 GHz) and Bob (6.0 GHz). The LINC remains significantly more pure at all drive frequencies. Dominant processes are explicitly labeled.
\textbf{c}, Driven coupler impurity in the presence of two drive tones, each with amplitude $\phi_{AC}=0.1\pi$, and coupled Alice and Bob modes, displaying the effects of parasitic intermodulation products. The two drive frequencies are scaled simultaneously maintaining a fixed ratio of 1:2, to ease Floquet-Markov simulations.
}\label[fig]{appx_fig3}
\end{figure*}

\reviewchanges{To characterize their performance as microwave beamsplitters, we first examine the rate and resonance condition of the Alice-Bob beamsplitter interaction activated by driven the LINC and SNAIL respectively, as a function of drive amplitude.
We employ time-dependent perturbation theory for both couplers in the Floquet mode basis}~\cite{Yaxing_PRA_BilinearModeCoupling} \reviewchanges{to calculate the beamsplitter rate as a function of coupler state, $\vert m \rangle$.
This amounts to finding the coefficient of the effective Hamiltonian,}
\begin{equation}
    H_\text{eff} = g_\text{BS,m}\hat{a}^\dagger \hat{b} \vert m \rangle\langle m\vert+\text{h.c.}
\end{equation}
\reviewchanges{When the coupler is in its ground state ($m=0$), this calculation provides the beamsplitter rate reported in the main text.
When the coupler is excited to higher levels, we observe state-dependent shifts of the beamsplitter rate and resonance condition (}\cref{appx_fig3}\reviewchanges{a,b).
This dispersion makes it clear that if the coupler leaves its ground state and enters a mixed state due to the parametric drive, it can dephase the beamsplitting process.
As a result, the drive-induced excitation of the coupler state can lead to process infidelity, with magnitude set by the amount of state-dependent dispersion in $g_{\text{BS}}$ and $\Delta_{\text{BS}}$.}

\reviewchanges{To study such drive-induced excitation in the driven LINC and SNAIL as well as spurious coupler-resonator interactions, we employ the Floquet-Markov approach.
We consider an environment that contains a coupler thermal relaxation channel and a dephasing channel for both circuits.
To efficiently capture the effects of additional modes in the system, we model Alice and Bob as two Lorentzians set by their frequencies and linewidths, resulting in a total coupler relaxation spectrum of}
\begin{align}
\Gamma_c(\omega) &= \frac{\omega}{\omega_q}(1+n_B(\omega,T_c))\gamma_c \\ \nonumber &+ \sum_{i=A,B}(1+n_B(\omega,T_i))\frac{g_i^2}{(\omega-\omega_i)^2+(\frac{\kappa_i}{2})^2}\kappa_i.
\end{align}
\reviewchanges{Here, $\gamma_c = (26.7\mu \text{s})^{-1}$ is the coupler decay rate, $n_B(\omega, T) = 1/(e^{\hbar\omega/k_B T} -1)$ is the Bose-Einstein occupation factor.
We set Alice's and Bob's temperature and decay rates to 50mK and $(10\mu s)^{-1}$, higher than the coupler's, to efficiently capture and visualize the parasitic transitions. We also assume $1/f$ dephasing spectrum for the coupler, $S_{\Phi\Phi}[\omega]=\frac{(1\mu\Phi_0)^2}{\omega}$, with a cut-off frequency of 1 Hz.

Using the Floquet-Markov approach, we examine the impurity of the steady state over different drive frequencies and amplitudes, as a proxy for process infidelity caused by spurious coupler excitation, as well as by other parasitic coupler-resonator interactions.}
\reviewchanges{When the two circuits are driven by a single tone (}\cref{appx_fig3}\reviewchanges{b) across the frequency range of 0.5- 2 GHz, both impurities remain small, except for a few peaks that represent coupler-resonator transitions.
At higher frequencies above 2 GHz, LINC has substantially lower impurity, benefiting from its parity selection rule that forbids the spurious sub-harmonic excitations that SNAIL encounters at 2.17 GHz and 3.25 GHz. 
Further, when two drives are simultaneously applied to the circuits (}\cref{appx_fig3}\reviewchanges{c), we observe a significant difference between the LINC and SNAIL, where the former possesses much lower impurities and fewer parasitic transitions than the latter.
Finally, when sweeping drive amplitude, such as one might enact in the duration of a pulse sequence, the LINC has negligible drive-induced shifts in its parasitic resonances, as seen in}~\cref{fig3}\reviewchanges{c.
This is in stark contrast to Kerr-full couplers like a regular transmon, or the DDS.

Overall, LINC's low driven process dispersion and significantly cleaner, more stable driven spectrum promise an expanded drive space for high-fidelity parametric operations, including when dealing with multiple modes or pumps.}

\section{Effect of junction and DC flux asymmetries}
\label[appdx]{app:Asymm}
In the main text, we have seen how the LINC can leverage symmetries of the circuit, of the bias point, and of the drive to realize a parity-protected three-wave mixer. Here, we examine the LINC in the presence of some asymmetries inherent to any experimental realization, and predict the effect of these asymmetries on device performance.

To begin, similar to our treatment of $\hat{\phi}_{\text{sym}}$ in \ref{app:Circuit}, we assume that the asymmetric flux degree of freedom $\hat{\phi}_{\text{asym}}$ is stiff and can be replaced by the classical variable $\phi_{\Delta} = \langle \hat{\phi}_{\text{asym}} \rangle = \pi \frac{\Phi_{\text{L}} - \Phi_{\text{R}}}{\Phi_0}$. 
Here $\Phi_{\text{L,R}}$ is the flux threading the left and right loops respectively, with $\phi_{\Delta}$ represents an asymmetry in the flux threading the two loops, which may arise due to residual local gradients in the ambient magnetic field.
Further, to model the effect of imperfect junction fabrication of the outer arms, we introduce $\beta_\Delta = (E_{J_1} - E_{J_2})/E_L$ and $\beta_\Sigma = (E_{J_1} + E_{J_2})/E_L$, with $E_L$ setting the energy scale of the system.

With these definitions, we can express the static inductive potential in the presence of asymmetries as
\begin{equation}
\label{eq:LINC_asymm_U}
\begin{split}
    U_{\text{full}}/E_L = \frac{1}{2}&(\hat{\theta}_c - 2\phi_{\Delta}/3)^2 \\
    - \beta_\Sigma\cos(\phi_{d})&\left[\cos(\phi_{\Delta}/3)\cos(\hat{\theta}_c) - \sin(\phi_{\Delta}/3)\sin(\hat{\theta}_c)\right] \\
    +\beta_\Delta\sin(\phi_{d})&\left[\cos(\phi_{\Delta}/3)\sin(\hat{\theta}_c) + \sin(\phi_{\Delta}/3)\cos(\hat{\theta}_c)\right]
\end{split}
\end{equation}
\\
~
\\
To study the effects that these asymmetries will have on static properties of the LINC (such as the static Kerr), we can examine
the LINC's potential at the $\phi_d=\pi/2$ operating point, where the potential simplifies to
\begin{equation}
\label[eq]{eq:static_asymm_potential}
    \begin{split}
    &\left. U_{\text{full}} \right\vert_{\phi_{\text{d}} = \pi/2} = E_L\Bigg[ \dfrac{(\hat{\theta}_c - 2\phi_{\Delta}/3)^2}{2} \\
    &+  \beta_\Delta\left(\cos (\hat{\theta}_c) \sin (\phi_{\Delta}/3)+  \sin (\hat{\theta}_c) \cos (\phi_{\Delta}/3)\right) \Bigg]\\
    &= E_L \Bigg[\dfrac{(\hat{\theta}_c - 2\phi_{\Delta}/3)^2}{2}
    +  \beta_\Delta\sin (\hat{\theta}_c + \phi_{\Delta}/3)\Bigg].
    \end{split}
\end{equation}
Notably, the presence of asymmetry shifts the the potential minimum with respect to $\hat{\theta}_c$ from $\hat{\theta}_c^{\text{min}} = 0$ to 
\begin{equation}
    \theta_c^{\text{min}} \approx \frac{2}{3}\phi_{\Delta} - \beta_\Delta,
\end{equation}
to lowest order in $\beta_\Delta$ and $\phi_{\Delta}$
Due to this shift, we must re-compute parameters of interest by evaluating derivatives at this new $\theta_c^{\text{min}}$, following the general framework given in \cref{app:Circuit}. From this, we find that the LINC frequency and Kerr nonlinearity are modified to be
\begin{widetext}
\begin{equation}
    \omega_L[\phi_{\text{DC}}] = \sqrt{8E_cE_L(1 + \beta_\Sigma\cos(\phi_{d})\cos(\phi_\Delta -\beta_\Delta) - 
    \beta_\Delta\sin(\phi_{d})\sin(\phi_\Delta -\beta_\Delta))} + \alpha_L[\phi_{\text{DC}}],
\end{equation}
\begin{equation}
    \alpha_L[\phi_{\text{DC}}] = -E_c\left(\frac{\beta_\Sigma\cos(\phi_{d})\cos(\phi_\Delta -\beta_\Delta) - \beta_\Delta\sin(\phi_{d})\sin(\phi_\Delta -\beta_\Delta)}
{1 + \beta_\Sigma\cos(\phi_{d})\cos(\phi_\Delta -\beta_\Delta) - \beta_\Delta\sin(\phi_{d})\sin(\phi_\Delta -\beta_\Delta)}\right).
\end{equation}
\end{widetext}

 \begin{figure}
    \centering
    \includegraphics[width=0.49 \textwidth]{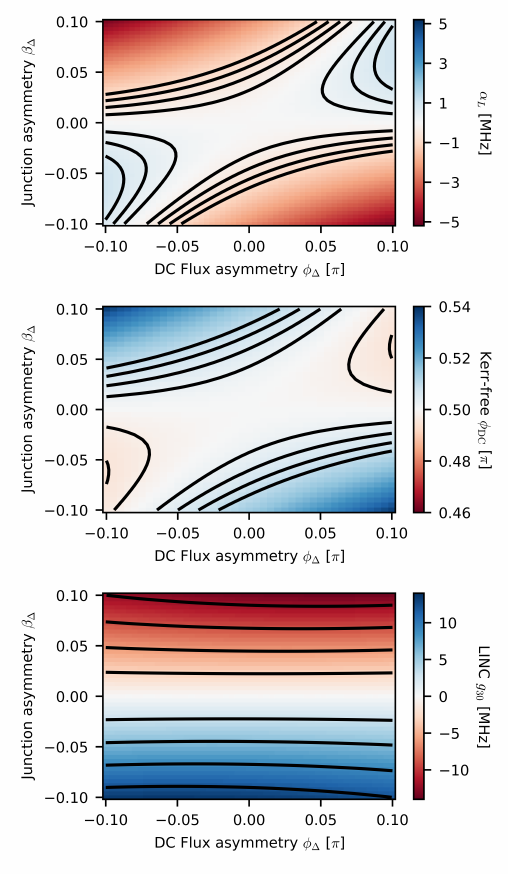}
    \caption{
    \textbf{LINC Kerr in the presence of asymmetries}
    \textbf{a}, Kerr of the LINC mode at $\phi_{\text{DC}}=\pi/2$, computed by direct diagonalization. Black contour lines correspond to $\vert\alpha_L\vert = 250, 500, 750, \text{and } 1000$ KHz respectively.
    \textbf{b}, Operating point $\phi_d$ where the LINC is Kerr-free, as a function of asymmetry. With a minor change in the symmetric DC flux, the Kerr induced by small asymmetry can be nulled. Black contour lines correspond to flux values from $98\%\times\pi/2$ to $102\%\times\pi/2$ in $0.5\%$ increments around the asymmetry-free DC operating point. 
    \textbf{c}, third-order nonlinearity of the LINC dipole mode, $g_{30}$. With finite junction asymmetry, the self-Kerr and cross-Kerr-free points will differ. Contour lines correspond to $\vert g_{30}\vert = 3,6,9, \text{and } 12$ MHz respectively. 
    }\label[fig]{fig:asymm_Kerr}
    \end{figure}

When focusing on the LINC's behavior at the operating point, this framework provides us with an approximate expression of the induced Kerr, given as
\begin{equation}
\label{eq:LINC_asymm_K_op_point}
\begin{split}
    \alpha_L^{\text{asym}} &\approx \frac{1}{2}\frac{\partial^2 U}{\partial \theta_c^4} \theta_{zpf}^4\Bigg\vert_{\theta_c = \theta_c^{\text{min}}, \phi_d = \pi/2} \\
 &\approx E_c\beta_\Delta\sin\left(\phi_{\Delta} - \beta_\Delta\right)
\end{split}
\end{equation}
to lowest order in asymmetry.
However, when comparing with exact diagonalization of the LINC Hamiltonian in the presence of asymmetries, we find that an additional factor must be added to agree with numerics.
These non-idealities, to leading order, therefore induce a static LINC Kerr nonlinearity of 
\begin{equation}
    \alpha_L^{\text{asym}} \approx E_c\beta_\Delta\sin\left(\phi_{\Delta} - (1 + \sqrt{2})\beta_\Delta \right).
\end{equation}
While it is not entirely clear where this additional factor comes from (possibly higher-order corrections to \cref{eq:LINC_asymm_K_op_point}), it is clear that the LINC's Kerr is only second-order sensitive to asymmetry.
To get a sense of the Kerr landscape, \cref{fig:asymm_Kerr} shows the LINC Kerr computed by direct diagonalization of the LINC Hamiltonian as a function of asymmetry.
Notably, in the absence of any junction asymmetry, the LINC is Kerr-free for arbitrary asymmetric DC flux threading the loops, so long as $\phi_d = \pi/2$, interpolating the device between the LINC ($\phi_d = 0$) and the ATS ($\phi_d = \pi/2$).
Intuitively, when the two outer junctions are identical, their potentials can entirely destructively interfere, leaving only the potential of the linear shunt as seen in \cref{eq:static_asymm_potential} for $\beta_\Delta = 0$.

Importantly, for small asymmetries, the LINC Kerr always has a zero-crossing.
Thus for applications requiring truly zero residual Kerr, the symmetric DC flux can be tuned slightly away from $\phi_d = 0.5\pi$ to arrive at a Kerr-free point, as shown in \cref{fig:asymm_Kerr}b.
However, straying from this bias point may compromise some of the protected driven qualities of the system.

In addition to an ideal LINC's Kerr going to zero at the operating point, any cross-Kerr to other modes are also zero in the absence of asymmetry due to the simultaneous extinguishment of all $g_{m>2,0}$. However, in the presence of asymmetry, these other nonlinearities re-appear and can cause the self-Kerr- and cross-Kerr-free points to no longer coincide. Specifically, the nonlinear term 
\begin{equation}
\label[eq]{eq:asymm_g30}
\begin{split}
        g_{30} &= \frac{1}{3!}\left[\frac{\partial^3U}{\partial \theta_c^3} \theta_{zpf}^3\right] \Bigg\vert_{\theta_c = \theta_c^{\text{min}}, \phi_d = \pi/2} \\
        &\approx  -E_L\beta_\Delta\cos(\phi_\Delta  -\beta_\Delta)\theta_{zpf}^3
\end{split}
\end{equation}
will enter into the induced cross-Kerr as $\chi_{\text{res,LINC}} = 24p_{\text{res}}^2(g_{40} + 6g_{30}^2\omega_L/(\omega_{\text{res}}^2-4\omega_L^2))$, but enter into the self-Kerr as $K = 12(g_4 - 5g_{30}^2/\omega_L)$ \cite{StijnThesis}. In effect, these two quantities now rely on an interplay between $g_{30}$ and $g_{40}$, meaning that they can no longer be eliminated simultaneously\cite{ChapmanSNAILBeamsplitterPaper}.

Now that we have studied the implications of asymmetry on the static behavior of the LINC, we move to examining the driven behavior.
To do this, we can look at the driven Hamiltonian at the operating point in the presence of asymmetries
\begin{equation}
\label[eq]{eq:asymm_driven}
\begin{split}
        U/E_L = &\frac{1}{2}(\hat{\theta}_c - 2\phi_{\Delta}/3)^2 \\
        + \beta_\Sigma\sin(\phi_{AC})&\left[\cos(\phi_{\Delta}/3)\cos(\hat{\theta}_c) - \sin(\phi_{\Delta}/3)\sin(\hat{\theta}_c)\right] \\
        +\beta_\Delta\cos(\phi_{AC})&\left[\cos(\phi_{\Delta}/3)\sin(\hat{\theta}_c) + \sin(\phi_{\Delta}/3)\cos(\hat{\theta}_c)\right].
\end{split}
\end{equation}
Utilizing \cref{eq:Derivatives,eq:g_derivatives} but evaluating at $\theta_c = \theta_c^{\text{min}}$, we can compute the strength of terms of interest, including those typically forbidden by the LINC's parity protection. In general one may also have an asymmetry in the applied drive, but
this asymmetry can be nearly entirely nulled through careful microwave design \cite{SQUIDBeamsplitterPaper}, so we focus only on junction and DC flux asymmetry here.

\begin{figure}
    \centering\includegraphics[width=0.49 \textwidth]{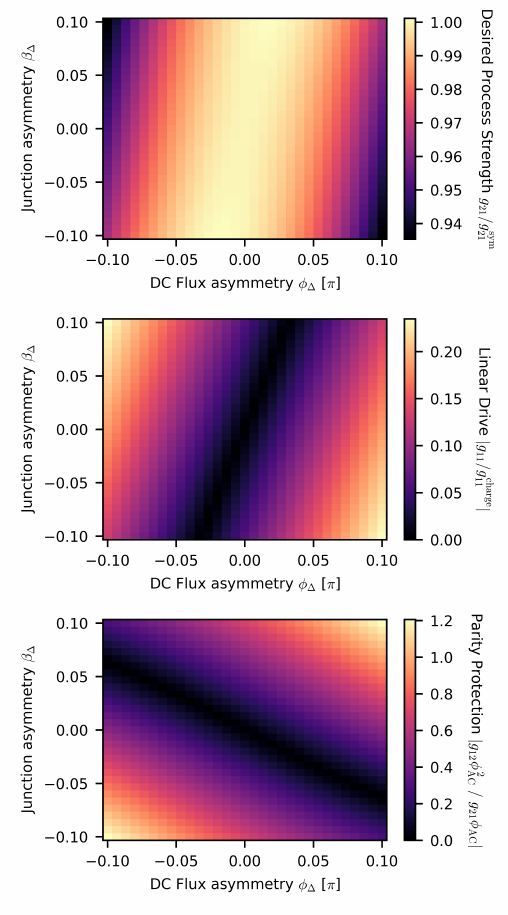}
    \caption{
    \textbf{Driven process strength in the presence of asymmetries}
    \textbf{a}, Fractional strength of the desired $g_{21}$ process in the presence of asymmetries. We see that the strength is primarily dependent upon the asymmetric flux $\phi_\Delta$, and is second order with respect to both asymmetries.
    \textbf{b}, strength of $g_{11}$, the linear coupling of the LINC mode to a symmetric flux drive, as compared to the LINC's linear coupling to a charge drive $g_{11}^{\text{charge}} \approx E_L\theta_{zpf}$.
    \textbf{c}, Parity protection within the 3rd order nonlinearity. Strength of $g_{12}\phi_{\text{AC}}^2$, that would permit subharmonic-driving-like processes, as compared to the desired $g_{21}\phi_{\text{AC}}$, is shown as a function of asymmetry for the drive value $\vert \phi_{\text{AC}} \vert = 0.2\pi$.     The above simulations are performed by numerically computing appropriate derivatives of the potential and $\theta_{zpf}$ as detailed in \cref{eq:asymm_g21,eq:asymm_g11,eq:asymm_g12}. The behavior of these driven process strengths with respect to asymmetry has been separately verified by floquet and/or time-domain simulations.
    }\label[fig]{fig:asymm_driven}
    \end{figure}

We first examine the strength of our desired three-wave mixing process, $g_{21}(\hat{c} + \hat{c}^\dagger)^2\phi_{\text{AC}}$, to see how the beamsplitting strength degrades. 
We find that the strength of our desired process suffers only quadratically with asymmetry,
\begin{equation}
\label[eq]{eq:asymm_g21}
\begin{split}
        g_{21} &= \frac{1}{2}\left[\frac{\partial^3U}{\partial \phi_d\partial \theta_c^2} \theta_{zpf}^2 + \frac{\partial^2U}{\partial \theta_c^2} \frac{\partial\theta_{zpf}^2}{\partial \phi_d}\right] \Bigg\vert_{\theta_c = \theta_c^{\text{min}}, \phi_d = \pi/2} \\
        &\approx  -\frac{E_L\beta_\Sigma\theta_{zpf}^2}{4}\cos(\phi_\Delta -\beta_\Delta).
\end{split}
\end{equation}
To show the fractional change in beamsplitting strength, we plot the relative change in $g_{21}$ as a function of asymmetry in \cref{fig:asymm_driven}a and find only a moderate reduction for reasonable asymmetries.
    
Next, we examine the leading-order undesired process, the linear drive $g_{11}(\hat{c} + \hat{c}^\dagger)\phi_{\text{AC}} $. 
The emergence of this linear coupling term scales as 
\begin{equation}
\label[eq]{eq:asymm_g11}
\begin{split}
    g_{11} &= \theta_{zpf}\frac{\partial^2 U}{\partial \phi_d \partial \theta_c}\Bigg\vert_{\theta_c = \theta_c^{\text{min}}, \phi_d = \pi/2}   \\
    &\approx -E_L\beta_\Sigma\theta_{zpf}\sin(\phi_\Delta - \beta_\Delta )
\end{split}
\end{equation}
for small asymmetries. 
As a benchmark to compare against, we can consider the alternative case of charge-driving a LINC which would result in a linear coupling of $g_{11}^{\text{charge}} \approx E_L\theta_{zpf}$. From \cref{fig:asymm_driven}b, we see that the linear drive strength from the flux drive is still only a fraction of the linear drive from a charge displacement even for moderate asymmetry.

From the simulations in figure \cref{fig:asymm_driven}a-b, we find that both $g_{21}$ and $g_{11}$ have a rather strong dependence on $\phi_\Delta$. This can be well understood through the relationship between the LINC and the ATS. In the absence of junction and DC flux asymmetry, the circuit operates as a LINC and generates a pure $g_{21}$ nonlinearity with no $g_{11}$ from the parity protection. In the other extreme, with $\phi_{\Delta} = \frac{\pi}{2}$, we recover the behavior of an ATS and have precisely the opposite parity protection, meaning that $g_{21}$ becomes forbidden while $g_{11}$ is amplified. So, as $\phi_\Delta$ is changed between the two regimes, we see the corresponding reduction in $g_{21}$ and enhancement of $g_{11}$. 

We now turn to specifically checking parity protection within the three-wave mixing processes, by looking at the term $g_{12} (\hat{c} + \hat{c}^\dagger)\phi_{\text{AC}}^2 $.
For context, note that for an unprotected mixer like the SNAIL, this driven term would always only be off by a factor of $\theta_{zpf}$ from the desired mixing process ($g_{21}$).
We find that the protected term, in the presence of asymmetry, becomes 
\begin{equation}
\label[eq]{eq:asymm_g12}
\begin{split}
    g_{12} &= \left[\frac{1}{2}
      \frac{\partial^3 U}{\partial \phi_d^2 \partial \theta_c} \theta_{zpf} 
    + \frac{\partial^2 U}{\partial \phi_d \partial \theta_c} \frac{\partial \theta_{zpf}}{\partial \phi_d}
    \right] \Bigg\vert_{\theta_c = \theta_c^{min}, \phi_d = \pi/2} \\
        &\approx -\frac{E_L\theta_{zpf}}{2}\Big[
    \beta_\Delta
    + \frac{1}{2}\beta_\Sigma^2\sin(\phi_{\Delta} - \beta_\Delta)
    \Big],
\end{split}
\end{equation}
which is again linearly sensitive to asymmetry.
This linear dependence on $\beta_\Delta$ can be seen from the form of \cref{eq:asymm_driven} as the term explicitly appears in the Hamiltonian. Interestingly, the linear dependence in $\phi_\Delta$ originates from a modulation of $\theta_{zpf}$, similar to the LINC's AC Zeeman shift. To get a better understanding of how much these forbidden processes are suppressed as compared to permitted processes of the same order, \cref{fig:asymm_driven}c shows the ratio of process strengths ($g_{12}\phi_{\text{AC}}^2$ vs. $g_{21}\phi_{\text{AC}}$) for a chosen drive strength of $\phi_{\text{AC}} = 0.2\pi$. As seen below, the LINC still offers substantial suppression for reasonable asymmetries. Notably, for a charge-driven mixer, the $g_{21}$ process would be a factor of $1/\theta_{zpf} \approx 4$ times stronger than $g_{12}$.

Overall, while asymmetries can spoil the perfect cancellation of many undesired quantities, the LINC is still rather robust.
The LINC's static properties are preserved quite well in the presence of small asymmetry, with the LINC Kerr depending only quadratically on asymmetry and always having a zero crossing at a slightly offset symmetric flux.
Additionally, while the strength of some undesired driven processes are first-order sensitive to asymmetry, a slightly unbalanced LINC will still suppress these processes significantly more than an otherwise unbalanced mixer like the SNAIL.

\begin{figure*}
\centering
\includegraphics[width=0.98 \textwidth]{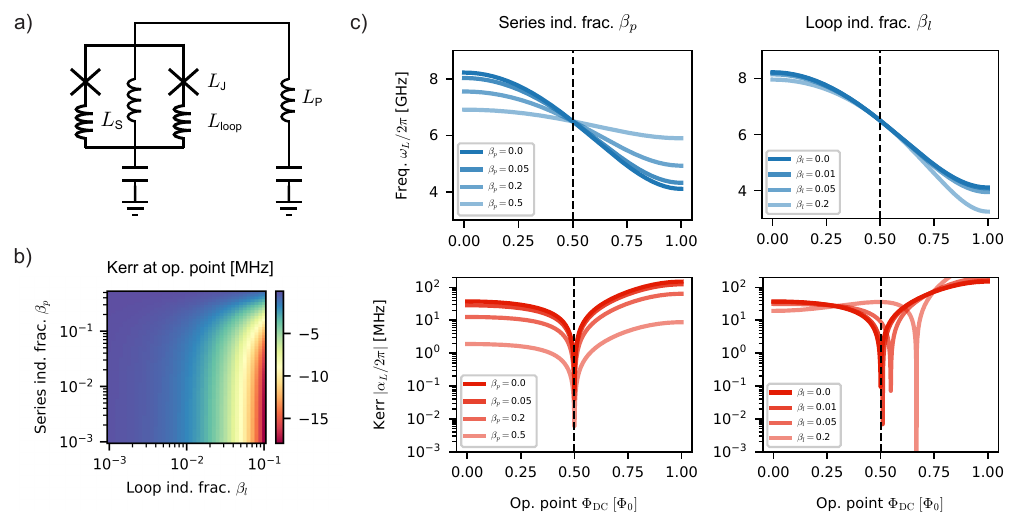}
\caption{
\textbf{Effects of parasitic inductance}
\textbf{a}, Circuit diagram for the LINC with parasitic loop inductance ($L_{\text{loop}}$) and series inductance ($L_{\text{P}}$). We analyze their effects by quantifying their fractional branch participations $\beta_l = L_{\text{loop}}/L_{\text{J}}$ and $\beta_p = L_{\text{P}}/L_{\text{S}}$.
\textbf{b}, Kerr at the operating point $\phi_{\text{DC}}=\pi/2$ as a function of inductance fractions $\beta_l$ and $\beta_p$. The series inductance has a minimal effect, but a large loop inductance can introduce significant Kerr.
\textbf{c}, Frequency and Kerr as a function of operating point $\phi_{\text{DC}}$, in the presence of parasitic inductances. A series inductor changes the participation of the loop and therefore the frequency tuning range, but does not change the Kerr-free point. A loop inductance has minimal changes to the frequency, but can change the LINC's Kerr. For reasonable loop sizes ($\leq 100\;\mu$m, $\beta_p\sim0.01$), the Kerr can be nulled by only a small shift in the operating point.
}\label[fig]{appx_fig4}
\end{figure*}

\section{Effects of parasitic loop inductance}
\label[appdx]{app:LoopInductance}
We now analyze the effect that a loop parasitic inductance has on the static nonlinearity of the LINC potential energy.
In particular, we assume each Josephson junction in the outer SQUID loop has a stray series inductance ($L_{\text{loop}}$ in~\cref{appx_fig4}a), and apply nonlinear current conservation methods developed in \cite{NinaPaper}.
We give intuition for and setup the equations for the analysis below, and then evaluate the equations numerically for actual comparison.
Note that because the loop inductance always shares current with an outer junction, it is difficult to directly give analytic solutions that are non-parametric.
The case of the series inductance in the coupler arms ($L_\text{P}$ in~\cref{appx_fig4}a) does not face this issue, and we treat it fully analytically in the next section.

In the absence of the central linear inductance of the LINC, e.g. in the simple DC-SQUID case, a small loop inductance would already bring the circuit's potential energy to a multistable configuration \cite{Clarkecos2phiTheory,ClarkeKITEcos2phiExpt}.
Thankfully, the linear shunt protects the LINC from becoming multi-stable in realistic geometries, as long as the ratio of loop inductance to shunt inductance is small.
To compute the potential energy expansion coefficients in the presence of these inductors, we start from a generic expression for the potential energy of a two-loop superconducting dipole with a symmetric external flux $\phi_d$
\begin{equation}
\label[eq]{eq:GENERIC_TWO_LOOP_POTENTIAL}
U(\theta, \phi) = U_\mathrm{L}\left(\theta_c - \frac{\phi_d}{2}\right) + U_\mathrm{C}(\theta_c) + U_\mathrm{R}\left(\theta_c + \frac{\phi}{2}\right),
\end{equation}
where $U_\mathrm{L,R}$ are the potential energies of the left and right branch, respectively, while $U_\mathrm{C}$ is the potential energy of the center branch, which is nominally just the shunting inductor's energy.
The m-th order derivative of the LINC with respect to $\theta_c$ is then
\begin{equation}
\label[eq]{eq:expansion_coefficients_loop_inductance}
\frac{\partial^m U}{\partial\theta_c^m} = \frac{\partial^m U_L}{\partial\theta_c^m} + \frac{\partial^m U_C}{\partial\theta_c^m} + \frac{\partial^m U_R}{\partial\theta_c^m}.
\end{equation}
Assuming perfect symmetry, these derivatives are evaluated  $\theta_{c,\text{ min}} = 0$.
We can rewrite these expansion coefficients as
\begin{equation}
U_{m0}(\phi) = l_m\left(\frac{\phi}{2}\right) + c_m(0) + r_m\left(-\frac{\phi}{2}\right)
\end{equation}
where $l_m$, $c_m$ and $r_m$ are the m-th order derivatives of the potential energy functions of the left, shunting and right LINC branches, respectively, evaluated at $\pm\frac{\phi}{2}$.
Finally, for driven behavior, we can also compute the n-th order derivative with respect to external flux, giving the general expansion coefficient
\begin{multline}
U_{mn}(\phi) = \\
\left\{
\begin{aligned}
&l_m\left(\frac{\phi}{2}\right) + c_m(0) + r_m\left(-\frac{\phi}{2}\right) & \mathrm{n=0} \\
&\frac{1}{2^n}\left[l_{m+n}\left(\frac{\phi}{2}\right) + (-1)^n r_{m+n}\left(-\frac{\phi}{2}\right)\right] & \mathrm{n>0}.
\end{aligned}\right.
\end{multline}

For a symmetric LINC with identical junctions and loop inductances, the potential energies $U_L$ and $U_R$ have the same functional expression, expressed parametrically as:
\begin{equation}
\left\{
\begin{aligned}
U_{L,R} & = \frac{E_\text{loop}}{2}\left(\beta_l\sin\varphi_{J_{L,R}}\right)^2 - E_J\cos\varphi_{J_{L,R}}\\
\phi_{L,R} & = \varphi_{J_{L,R}} + \beta_l\sin\varphi_{J_{L,R}}
\end{aligned}\right.
\end{equation}
where $E_{\text{loop}}$ is the energy of the stray loop inductance, $\beta_l$ is the ratio between the stray loop inductance and the Josephson inductance ($L_{\text{loop}}/L_J$), and $\varphi_J$ is the phase across the Josephson junction.
The effect of the symmetric flux bias is captured in the junction phases, forcing $\varphi_{J_L} = - \varphi_{J_R} = \varphi_{J}$.
Higher order derivatives of $U_{L,R}$ can be expressed in the same form \cite{NinaPaper}.

Combining these ingredients, the potential energy expansion coefficients $U_{mn}$ can be related to external flux via a parametric relation, noticing that the external flux $\phi_d$ can be expressed as
\begin{equation}
\phi_d = 2\left(\varphi_J + \beta_J\sin\varphi_{J}\right).
\end{equation}
It is then possible to graphically evaluate the exact expressions of the LINC Hamiltonian expansion coefficients.
We perform this numerically to evaluate effects on the static LINC in~\cref{appx_fig4}b, c.
We expect the loop inductance of the LINC to be geometrically limited to $\sim1$ pH/ $\mu$m, which for a $100\mu$m loop side implies $L_{\text{loop}}=0.2$nH $\implies \beta_l \sim 0.02$ (with $L_S \sim 3$nH, $L_J\sim 10$nH).
We observe that the static frequency of the LINC as a function of flux is barely affected for in this range of loop inductance.
This also implies that the parametric mixing strength $g_{3wm}\sim 0.5 \;d\omega_L/d\phi_d$ remains similar.
The static Kerr of the LINC is more significantly affected, with $\alpha_L=$1-10 MHz at $\phi_{DC}=\pi/2$, but the Kerr still has a zero-crossing very close to half flux, and can therefore always be nulled out for sensitive operations.

\section{Effect of a series parasitic inductance}
\label[appdx]{app:LinearInductance}
We now consider the effect of a parasitic inductance in series with the LINC loop, which may arise from the geometric inductance of the metallic arms of the coupler that form the electric dipole necessary for self-capacitance and coupling.
As in~\cref{app:Circuit}, we label the phase across the LINC as $\theta_c$ (with the flux threading it as $\phi_d$), with potential energy $U_\mathrm{LINC}$.
We then consider the phase across the whole effective dipole (LINC $+$ parasitic inductor) as $\theta$, with corresponding potential energy $U_\text{tot}$.
Each nonlinear process activated by the coupler is then described through derivatives of this total potential energy, and can be analytically derived.
We will see that for a symmetric LINC, this parasitic inductance will primarily deform the potential and dilute the nonlinearity, but not shift the potential minima -- thus preserving the LINC's linearity at $\phi_{\text{DC}}=\pi/2$.
We outline our derivation in a pedagogical manner below, with the actual analytic results for the LINC Hamiltonian detailed in~\cref{app:sub:LINCParasiticResults}.

Since much of the LINC's advantageous properties, including its static linearity, come from its symmetry-enforced selection rules, we want to quantify how well its parity-protection is preserved in the presence of a parasitic series inductance, i.e. whether
\begin{equation}
\label[eq]{eq:PP_STRAY_INDUCTANCE}
\frac{\partial^{m+n} \; U_\mathrm{LINC}(\theta_c,\phi_d)}{\partial\theta_c^m \; \partial\phi_d^n} = 0 \Rightarrow 
\frac{\partial^{m+n} \; U_{\text{tot}}(\theta,\phi_d)}{\partial\theta^m \; \partial\phi^n} \approx 0.
\end{equation}
To evaluate the correspondence in Eq. \eqref{eq:PP_STRAY_INDUCTANCE}, we appeal to the current conservation condition $i = i_\mathrm{LINC}$ to relate first-order partial derivatives of $U_\text{tot}$ and $U_\mathrm{LINC}$, following methodology similar to \cite{SNAIL_Analysis_Sivak,SNAIL_Analysis_Frattini,NinaPaper}. 
As $i = \frac{2\pi}{\Phi_0}\frac{\partial U_\text{tot}}{\partial\theta}$, and $i_\mathrm{LINC} = \frac{2\pi}{\Phi_0}\frac{\partial U_\mathrm{LINC}}{\partial\theta_c}$,
the following relation always holds:
\begin{equation}
\label[eq]{eq:CC_STRAY_INDUCTANCE}
\frac{\partial U_{\text{tot}}}{\partial\theta}(\theta,\phi_d) = 
\frac{\partial U_\mathrm{LINC}}{\partial\theta_c}[\theta_c(\theta,\phi_d),\phi_d].
\end{equation}
Higher order partial derivatives can then be computed from this identity, making it possible to relate partial derivatives of $U_{\text{tot}}$ as linear combinations of partial derivatives of $U_\mathrm{LINC}$, weighted by partial derivatives of $\theta_c(\theta, \phi_d)$.
This requires a change of variables from $(\theta_c,\phi_d)$ to the effective dipole variables $(\theta, \phi_d)$, which we outline below.

\subsection{\footnotesize{Representing the total potential energy and phase}}
In this section, we analyze the change of variables from the LINC variables $(\theta_c,\phi_d)$ to the effective dipole variables $(\theta, \phi_d)$ and study its invertibility.
We define the bi-variate function implementing the change of variables as $\mathbf{F}$, such that $(\theta, \phi_d) = \mathbf{F}(\theta_c,\phi_d)$.
 Let us define the superconducting phase drop across the parasitic inductor to be $\theta_p$, satisfying $\theta = \theta_c + \theta_p$, and its current to be  $i_p$.
 Our derivation will primarily use the current conservation relation $i_\mathrm{LINC} = i_p$.
 
 From the individual potential energies of the LINC and the parasitic inductor, we know the LINC current to be
\begin{equation}
\label[eq]{eq:LINC_CURRENT}
i_\mathrm{LINC}(\theta_c,\phi_d) = \frac{\Phi_0}{2\pi L}\theta_c + 2I_J\cos{\phi_d}\sin{\theta_c},
\end{equation}
and the current through the parasitic inductor to be
\begin{equation}
\label[eq]{eq:LP_CURRENT}
i_p = \frac{\Phi_0}{2\pi L_p}\theta_p.
\end{equation}
Applying current conservation and phase addition to \cref{eq:LINC_CURRENT,eq:LP_CURRENT}, the effective dipole phase reads,
\begin{equation}
\theta = 
(1 + \beta_p)\theta_c + 2\beta_J\cos{\phi_d}\sin{\theta_c}
\label[eq]{eq:LP_PHASE}
\end{equation}
where we have defined $\beta_p = L_p/L$ and $\beta_J = 2\pi L_p I_J/\Phi_0 = L_p / L_J$.
\cref{eq:LP_PHASE} thus defines the change of variable function $\mathbf{F}$.
To compute our desired derivatives, we actually require the partial derivatives of the inverse of this function instead, implementing the change of variable $(\theta_c,\phi_d) = \mathbf{F}^{-1}(\theta, \phi_d)$.

We now compute the conditions under which $\boldsymbol{F}$ is invertible. A necessary and sufficient condition is that the determinant of $\boldsymbol{J} = \nabla\boldsymbol{F}$ (the Jacobian matrix) does not change sign for any value of the independent variables. By computing partial derivatives of $\boldsymbol{F}$ and combining them into the determinant, we obtain
\begin{equation}
\label[eq]{eq:JACOBIAN_F}
\det(\boldsymbol{J})(\theta_c,\phi_d) = 1 + \beta_p + 2\beta_J\cos{\phi_d}\cos{\theta_c}.
\end{equation}
As $\det{\boldsymbol{J}}(0,0) > 0$, and by noticing that it is minimized when $\cos{\phi_d}\cos{\theta_c} = -1$, the condition for invertibility of $\boldsymbol{F}$ is
\begin{equation}
\label[eq]{eq:INVERTIBILITY_OF_F}
1 + \beta_p - 2\beta_J > 0,
\end{equation}
which is automatically satisfied if the LINC potential $U_\mathrm{LINC}$ is single-valued, i.e. 
\begin{equation}
    2E_J < E_L \Rightarrow 2\beta_J < \beta_p \Rightarrow \text{det}(\nabla\boldsymbol{F}) \neq 0
\end{equation} 

\subsection{\footnotesize{Computing derivatives of the total potential}}
We can now proceed to compute the partial derivatives of $\boldsymbol{F}^{-1}$ by first composing it with $\boldsymbol{F}$ to form the identity 
\begin{equation}
\label[eq]{eq:INVERT_F_IDENTITY}
\theta
=
\theta\left[\theta_c(\theta, \phi_d), \phi_d(\theta, \phi_d)\right]
\end{equation}
Defining the generalized participation of the LINC as
\begin{equation}
\label[eq]{eq:LINC_PARTICIPATION}
p_{mn} = \frac{\partial^{m+n}{\theta_c}}{\partial\theta^m\partial\phi_d^n},
\end{equation}
and computing first-order partial derivatives with respect to $\theta$ and $\phi_d$ on both sides of \eqref{eq:INVERT_F_IDENTITY}, we obtain the set of equations
\begin{equation}
\label[eq]{eq:PARTICIPATIONS_EQUATIONS}
\left\{
\begin{aligned}
&1 = \frac{\partial\theta}{\partial\theta_c}p_{10}\\
&0 = \frac{\partial\theta}{\partial\theta_c}p_{01} + \frac{\partial\theta}{\partial\phi_d}.
\end{aligned}
\right.
\end{equation}
Note that the implicit dependence of $\theta$ on $\phi_d$ via $\theta_c$ is included in the $p_{01}$ term. As a consequence, the partial derivative of $\theta$ with respect to $\phi_d$ has to only account for the explicit dependence on $\phi_d$ (consistently with the definition of the partial derivative).

We will find it useful to define the dimensionless Josephson part of the LINC potential, and its partial derivatives, as
\begin{equation}
\label[eq]{eq:LINC_JOSEPHSON_DERIVATIVES}
\begin{split}
    u(\theta_c, \phi_d) &= \dfrac{U_L + U_R}{E_J} = -2\cos{\phi_d}\cos{\theta_c} \\
    u_{mn} &= \frac{\partial^{m+n}u}{\partial\theta_c^m\phi_d^n}.
\end{split}
\end{equation}
The first-order participation ratios of the LINC are then obtained from \eqref{eq:PARTICIPATIONS_EQUATIONS} as
\begin{equation}
\label[eq]{eq:PARTICIPATIONS_EQUATIONS_EXPLICIT}
\left\{
\begin{aligned}
p_{10} & = \frac{1}{1 + \beta_p + \beta_J u_{20}}\\
p_{01} & = -\frac{\beta_J u_{11}}{1 + \beta_p + \beta_J u_{20}} = -\beta_J p_{10} u_{11}.
\end{aligned}
\right.
\end{equation}
Additionally, the following algebraic rules hold for Eq. \eqref{eq:CC_STRAY_INDUCTANCE}:
\begin{equation}
\begin{aligned}
\frac{\partial p_{mn}}{\partial\theta} & = p_{m+1\,n}\\
\frac{\partial p_{mn}}{\partial\phi_d} & = p_{m\,n+1}\\
\frac{\partial u_{mn}}{\partial\theta} & = p_{10}u_{m+1\,n}\\
\frac{\partial u_{mn}}{\partial\phi_d} & = p_{01}u_{m+1\,n} + u_{m\,n+1},
\label[eq]{eq:ALGEBRAIC_RULES}
\end{aligned}
\end{equation}
which ease the computation of higher-order coefficients.
One can now apply these rules to compute arbitrary partial derivatives of the effective dipole potential energy function, noting that the right hand side term of \eqref{eq:CC_STRAY_INDUCTANCE} can be written as
\begin{equation}
\begin{aligned}
\frac{\partial U_\mathrm{LINC}}{\partial\theta_c} & = E_L \theta_c + 2E_J\cos{\phi_d}\sin{\theta_c} \\
& = E_L p_{00} + E_Ju_{10}.
\label[eq]{eq:U_START}
\end{aligned}
\end{equation}

Defining the partial derivatives of $U_{\text{tot}}$ as
\begin{equation}
U_{mn} := \frac{\partial^{m+n}U_{\text{tot}}}{\partial\theta^m\phi_d^n}
\end{equation}
we have from \eqref{eq:CC_STRAY_INDUCTANCE}
\begin{equation}
U_{m+1n} = \frac{\partial^{m+n}}{\partial\theta^m\phi_d^n}(E_L p_{00} + E_Ju_{10})
\end{equation}
Importantly, $U_{mn}$ can be expressed as a function of $p_{10}$ and $\{u_{ij}\}_{^{i = 1 \dots m}_{k = 1 \dots n}}$ only.
For instance, $U_{20}$ can be computed applying the algebraic rules to \eqref{eq:U_START}, obtaining
\begin{equation}
U_{20} = E_Lp_{10} + E_Jp_{10}u_{20}
\end{equation}
and, deriving $U_{20}$ with respect to $\theta$ we obtain
\begin{equation}
U_{30} = E_Lp_{20} + E_J(p_{10}^2u_{30} + p_{20}u_{20}).
\end{equation}
The expression for the second order participation ratio $p_{20}$ can also be obtained by applying the algebraic rules to the first line of \eqref{eq:PARTICIPATIONS_EQUATIONS_EXPLICIT}, resulting in
\begin{equation}
p_{20} = -\beta_J p_{10}^3 u_{30}.
\end{equation}
Using this, we then obtain
\begin{equation}
U_{30} = \left[E_Jp_{10}^2 -\beta_J p_{10}^3(E_L + E_Ju_{20})\right]u_{30}
\end{equation}
which can be further simplified to
\begin{equation}
U_{30} = E_Jp_{10}^3u_{30}
\end{equation}
by performing the substitution
\begin{equation}
\label{eq:participation_substitution}
\beta_J(E_L + E_Ju_{20}) = E_J\left(\frac{1}{p_{10}} - 1\right)
\end{equation}
which can be demonstrated from the definitions of $\beta_J$, $\beta_L$ and $p_{10}$.
Importantly, for the LINC, $u_{30}=0$, which implies $g_{30} \propto U_{30} = 0$.
For higher order nonlinearities, performing an additional derivative with respect to $\theta$ will generate a $p_{20}$ term, which can be again substituted with the previously obtained expression.
It is thus clear that $U_{m0}$ can be expressed as a multinomial function of $p_{10}$ and $\{u_{20}, u_{30}, \dots u_{m0}\}$.

We can also take the same approach to compute the arbitrary derivative $U_{mn}$ starting from $U_{11}$, which reads
\begin{equation}
U_{11} = E_Lp_{01} + E_Jp_{01}u_{20} + E_Ju_{11}.
\end{equation}
Here, $p_{01}$ can be substituted by the expression obtained in \eqref{eq:PARTICIPATIONS_EQUATIONS_EXPLICIT}, giving
\begin{equation}
U_{11} = \left[E_J -\beta_J p_{10} (E_L + E_J u_{20})\right]u_{11}.
\end{equation}
This expression can be further simplified by performing the substitution in Eq. \eqref{eq:participation_substitution}, resulting in
\begin{equation}
U_{11} = E_Jp_{10}u_{11}.
\end{equation}
A further derivative with respect to $\phi_d$ will generate $U_{12}$, which will contain the participation $p_{11}$.
However, $p_{11}$ can be obtained by applying the algebraic rules to  \eqref{eq:PARTICIPATIONS_EQUATIONS_EXPLICIT}, as
\begin{equation}
p_{11} = - \beta_J (p_{20}u_{11} + p_{10} ^ 2 u_{21}).
\end{equation}
In combination with the expression for $p_{20}$ previously obtained, this gives an expression of $U_{12}$ that again only containes $p_{10}$ and $u_{mn}$ terms.
Thus by recursively substituting the expressions for $p_{01}$, $p_{11}$ and $p_{20}$, any partial derivative of $U_\text{tot}$ can be expressed as a multinomial function of $p_{10}$ and the set of nonlinear coefficients $u_{mn}$.

Finally, after the desired order of derivative is obtained, all the quantities have to be evaluated at the generic flux operating point $(\theta_{\text{min}},\phi_{\text{min}}) = (0,\phi_d)$.
Importantly, the invertibility of $\boldsymbol{F}$, along with \eqref{eq:LP_PHASE}, imply that this expansion point corresponds to $(\theta_c=0,\phi_d)$.
This means that the potential minima is unchanged, and one can then evaluate arbitrary $U_{mn}$ at this minima.

\subsection{\footnotesize{The LINC Hamiltonian with parasitic inductance}}
\label[appdx]{app:sub:LINCParasiticResults}
Given the derivatives of the total potential $U_{\text{tot}}$ with respect to $\hat{\theta}$ and $\phi_d$, we can find the relevant coupler Hamiltonian in the presence of linear inductance, which we derive here.
Similar to \cref{eq:Derivatives}, the Hamiltonian we derive describes the equations of motion corresponding to the bosonic operator $\hat{\theta} = \theta_{zpf} \left[ \phi_{\text{DC}} \right] (\hat{c} + \hat{c}^\dagger)$, which can be expanded in terms of $U_{mn}$.
Note that the parasitic inductance also affects $U_{2,0}$ and hence the spread of the wavefunction $\theta_{zpf}$, whose intrinsic flux dependence affects the every flux driven process.

First, let us consider the static LINC potential, given by:
\begin{equation*}
\begin{split}
    U_{\text{LINC}} &= \sum_{m} \tilde{g}_{m, 0}  \left(\hat{c}+\hat{c}^\dagger \right)^m \\
    \tilde{g}_{m, 0} &= U_{m,0} \left. \frac{\theta_{zpf}^m}{m!} \right\vert_{\phi_{\text{DC}}}.
\end{split}
\end{equation*}
We know that the frequency and impedance of the LINC shifts due to the parasitic inductance, specifically
\begin{equation*}
    \begin{gathered}
        U_{2,0} = p_{10} \left(E_L + 2E_J \cos{\phi_{\text{DC}}}\right) \\
        p_{10} = \left(1 + \beta_p +2\beta_J \cos{\phi_{\text{DC}}} \right)^{-1}\\
        \Rightarrow \omega_L(\beta_p, \phi_{\text{DC}}) = \sqrt{p_{10}}\; \omega_L(0, \phi_{\text{DC}})
    \end{gathered}
\end{equation*}
Crucially, this change is just a renormalization of the LINC's inductive energy by its linear participation in the parasitic inductor.

Higher-order nonlinearities are also similarly renormalized by the parasitic inductor, which means they can be recursively proven to be \textit{zero} at the operating point, for the ideal LINC.
As an example, the LINC Kerr in the presence of this parasitic inductance is given by
\begin{equation}
    \alpha_L(\beta_p, \phi_{\text{DC}}) = p_{10}^{3} \; \alpha_L(0, \phi_{\text{DC}}).
\end{equation}
Thus the undriven LINC remains linear at the operating point of $\phi_{\text{DC}}=\pi/2$, and we compute its static behavior in~\cref{appx_fig4}b, c.
The driven LINC terms are also similarly renormalized, for eg with the three-wave mixing strength changing to:
\begin{equation}
    g_{\text{3wm}}(\beta_p, \phi_{\text{DC}}) = p_{10}^{3/2} \; g_{\text{3wm}}(0, \phi_{\text{DC}})
\end{equation}
Overall the LINC with parasitic series inductance retains its static linearity at the same operating point, and has analytically predictable changes in its driven properties, which make its behavior significantly simpler than charge-driven mixers like the transmon or the SNAIL.

\bibliography{Linc_Arxiv}

\begin{thebibliography}{71}%
\makeatletter
\providecommand \@ifxundefined [1]{%
 \@ifx{#1\undefined}
}%
\providecommand \@ifnum [1]{%
 \ifnum #1\expandafter \@firstoftwo
 \else \expandafter \@secondoftwo
 \fi
}%
\providecommand \@ifx [1]{%
 \ifx #1\expandafter \@firstoftwo
 \else \expandafter \@secondoftwo
 \fi
}%
\providecommand \natexlab [1]{#1}%
\providecommand \enquote  [1]{``#1''}%
\providecommand \bibnamefont  [1]{#1}%
\providecommand \bibfnamefont [1]{#1}%
\providecommand \citenamefont [1]{#1}%
\providecommand \href@noop [0]{\@secondoftwo}%
\providecommand \href [0]{\begingroup \@sanitize@url \@href}%
\providecommand \@href[1]{\@@startlink{#1}\@@href}%
\providecommand \@@href[1]{\endgroup#1\@@endlink}%
\providecommand \@sanitize@url [0]{\catcode `\\12\catcode `\$12\catcode `\&12\catcode `\#12\catcode `\^12\catcode `\_12\catcode `\%12\relax}%
\providecommand \@@startlink[1]{}%
\providecommand \@@endlink[0]{}%
\providecommand \url  [0]{\begingroup\@sanitize@url \@url }%
\providecommand \@url [1]{\endgroup\@href {#1}{\urlprefix }}%
\providecommand \urlprefix  [0]{URL }%
\providecommand \Eprint [0]{\href }%
\providecommand \doibase [0]{https://doi.org/}%
\providecommand \selectlanguage [0]{\@gobble}%
\providecommand \bibinfo  [0]{\@secondoftwo}%
\providecommand \bibfield  [0]{\@secondoftwo}%
\providecommand \translation [1]{[#1]}%
\providecommand \BibitemOpen [0]{}%
\providecommand \bibitemStop [0]{}%
\providecommand \bibitemNoStop [0]{.\EOS\space}%
\providecommand \EOS [0]{\spacefactor3000\relax}%
\providecommand \BibitemShut  [1]{\csname bibitem#1\endcsname}%
\let\auto@bib@innerbib\@empty
\bibitem [{\citenamefont {Waldram}(1976)}]{JunctionBasicsReview1979}%
  \BibitemOpen
  \bibfield  {author} {\bibinfo {author} {\bibfnamefont {J.~R.}\ \bibnamefont {Waldram}},\ }\bibfield  {title} {\bibinfo {title} {The josephson effects in weakly coupled superconductors},\ }\href {https://doi.org/10.1088/0034-4885/39/8/002} {\bibfield  {journal} {\bibinfo  {journal} {Reports on Progress in Physics}\ }\textbf {\bibinfo {volume} {39}},\ \bibinfo {pages} {751} (\bibinfo {year} {1976})}\BibitemShut {NoStop}%
\bibitem [{\citenamefont {Yurke}\ \emph {et~al.}(1989)\citenamefont {Yurke}, \citenamefont {Corruccini}, \citenamefont {Kaminsky}, \citenamefont {Rupp}, \citenamefont {Smith}, \citenamefont {Silver}, \citenamefont {Simon},\ and\ \citenamefont {Whittaker}}]{YurkeJPA}%
  \BibitemOpen
  \bibfield  {author} {\bibinfo {author} {\bibfnamefont {B.}~\bibnamefont {Yurke}}, \bibinfo {author} {\bibfnamefont {L.~R.}\ \bibnamefont {Corruccini}}, \bibinfo {author} {\bibfnamefont {P.~G.}\ \bibnamefont {Kaminsky}}, \bibinfo {author} {\bibfnamefont {L.~W.}\ \bibnamefont {Rupp}}, \bibinfo {author} {\bibfnamefont {A.~D.}\ \bibnamefont {Smith}}, \bibinfo {author} {\bibfnamefont {A.~H.}\ \bibnamefont {Silver}}, \bibinfo {author} {\bibfnamefont {R.~W.}\ \bibnamefont {Simon}},\ and\ \bibinfo {author} {\bibfnamefont {E.~A.}\ \bibnamefont {Whittaker}},\ }\bibfield  {title} {\bibinfo {title} {Observation of parametric amplification and deamplification in a josephson parametric amplifier},\ }\href {https://doi.org/10.1103/PhysRevA.39.2519} {\bibfield  {journal} {\bibinfo  {journal} {Phys. Rev. A}\ }\textbf {\bibinfo {volume} {39}},\ \bibinfo {pages} {2519} (\bibinfo {year} {1989})}\BibitemShut {NoStop}%
\bibitem [{\citenamefont {Castellanos-Beltran}\ and\ \citenamefont {Lehnert}(2007)}]{LehnertJPA}%
  \BibitemOpen
  \bibfield  {author} {\bibinfo {author} {\bibfnamefont {M.~A.}\ \bibnamefont {Castellanos-Beltran}}\ and\ \bibinfo {author} {\bibfnamefont {K.~W.}\ \bibnamefont {Lehnert}},\ }\bibfield  {title} {\bibinfo {title} {Widely tunable parametric amplifier based on a superconducting quantum interference device array resonator},\ }\bibfield  {booktitle} {\emph {\bibinfo {booktitle} {Applied Physics Letters}},\ }\href {https://doi.org/10.1063/1.2773988} {\bibfield  {journal} {\bibinfo  {journal} {Applied Physics Letters}\ }\textbf {\bibinfo {volume} {91}},\ \bibinfo {pages} {083509} (\bibinfo {year} {2007})}\BibitemShut {NoStop}%
\bibitem [{\citenamefont {Yamamoto}\ \emph {et~al.}(2008)\citenamefont {Yamamoto}, \citenamefont {Inomata}, \citenamefont {Watanabe}, \citenamefont {Matsuba}, \citenamefont {Miyazaki}, \citenamefont {Oliver}, \citenamefont {Nakamura},\ and\ \citenamefont {Tsai}}]{OliverJPA}%
  \BibitemOpen
  \bibfield  {author} {\bibinfo {author} {\bibfnamefont {T.}~\bibnamefont {Yamamoto}}, \bibinfo {author} {\bibfnamefont {K.}~\bibnamefont {Inomata}}, \bibinfo {author} {\bibfnamefont {M.}~\bibnamefont {Watanabe}}, \bibinfo {author} {\bibfnamefont {K.}~\bibnamefont {Matsuba}}, \bibinfo {author} {\bibfnamefont {T.}~\bibnamefont {Miyazaki}}, \bibinfo {author} {\bibfnamefont {W.~D.}\ \bibnamefont {Oliver}}, \bibinfo {author} {\bibfnamefont {Y.}~\bibnamefont {Nakamura}},\ and\ \bibinfo {author} {\bibfnamefont {J.~S.}\ \bibnamefont {Tsai}},\ }\bibfield  {title} {\bibinfo {title} {Flux-driven josephson parametric amplifier},\ }\bibfield  {booktitle} {\emph {\bibinfo {booktitle} {Applied Physics Letters}},\ }\href {https://doi.org/10.1063/1.2964182} {\bibfield  {journal} {\bibinfo  {journal} {Applied Physics Letters}\ }\textbf {\bibinfo {volume} {93}},\ \bibinfo {pages} {042510} (\bibinfo {year} {2008})}\BibitemShut {NoStop}%
\bibitem [{\citenamefont {Bergeal}\ \emph {et~al.}(2010)\citenamefont {Bergeal}, \citenamefont {Vijay}, \citenamefont {Manucharyan}, \citenamefont {Siddiqi}, \citenamefont {Schoelkopf}, \citenamefont {Girvin},\ and\ \citenamefont {Devoret}}]{JPC_Paper}%
  \BibitemOpen
  \bibfield  {author} {\bibinfo {author} {\bibfnamefont {N.}~\bibnamefont {Bergeal}}, \bibinfo {author} {\bibfnamefont {R.}~\bibnamefont {Vijay}}, \bibinfo {author} {\bibfnamefont {V.~E.}\ \bibnamefont {Manucharyan}}, \bibinfo {author} {\bibfnamefont {I.}~\bibnamefont {Siddiqi}}, \bibinfo {author} {\bibfnamefont {R.~J.}\ \bibnamefont {Schoelkopf}}, \bibinfo {author} {\bibfnamefont {S.~M.}\ \bibnamefont {Girvin}},\ and\ \bibinfo {author} {\bibfnamefont {M.~H.}\ \bibnamefont {Devoret}},\ }\bibfield  {title} {\bibinfo {title} {Analog information processing at the quantum limit with a josephson ring modulator},\ }\href {https://doi.org/10.1038/nphys1516} {\bibfield  {journal} {\bibinfo  {journal} {Nature Physics}\ }\textbf {\bibinfo {volume} {6}},\ \bibinfo {pages} {296} (\bibinfo {year} {2010})}\BibitemShut {NoStop}%
\bibitem [{\citenamefont {Macklin}\ \emph {et~al.}(2015)\citenamefont {Macklin}, \citenamefont {O'Brien}, \citenamefont {Hover}, \citenamefont {Schwartz}, \citenamefont {Bolkhovsky}, \citenamefont {Zhang}, \citenamefont {Oliver},\ and\ \citenamefont {Siddiqi}}]{JTWPA_2015}%
  \BibitemOpen
  \bibfield  {author} {\bibinfo {author} {\bibfnamefont {C.}~\bibnamefont {Macklin}}, \bibinfo {author} {\bibfnamefont {K.}~\bibnamefont {O'Brien}}, \bibinfo {author} {\bibfnamefont {D.}~\bibnamefont {Hover}}, \bibinfo {author} {\bibfnamefont {M.~E.}\ \bibnamefont {Schwartz}}, \bibinfo {author} {\bibfnamefont {V.}~\bibnamefont {Bolkhovsky}}, \bibinfo {author} {\bibfnamefont {X.}~\bibnamefont {Zhang}}, \bibinfo {author} {\bibfnamefont {W.~D.}\ \bibnamefont {Oliver}},\ and\ \bibinfo {author} {\bibfnamefont {I.}~\bibnamefont {Siddiqi}},\ }\bibfield  {title} {\bibinfo {title} {A near--quantum-limited josephson traveling-wave parametric amplifier},\ }\href@noop {} {\bibfield  {journal} {\bibinfo  {journal} {Science}\ }\textbf {\bibinfo {volume} {350}},\ \bibinfo {pages} {307} (\bibinfo {year} {2015})}\BibitemShut {NoStop}%
\bibitem [{\citenamefont {Frattini}\ \emph {et~al.}(2017)\citenamefont {Frattini}, \citenamefont {Vool}, \citenamefont {Shankar}, \citenamefont {Narla}, \citenamefont {Sliwa},\ and\ \citenamefont {Devoret}}]{SNAILPaper}%
  \BibitemOpen
  \bibfield  {author} {\bibinfo {author} {\bibfnamefont {N.~E.}\ \bibnamefont {Frattini}}, \bibinfo {author} {\bibfnamefont {U.}~\bibnamefont {Vool}}, \bibinfo {author} {\bibfnamefont {S.}~\bibnamefont {Shankar}}, \bibinfo {author} {\bibfnamefont {A.}~\bibnamefont {Narla}}, \bibinfo {author} {\bibfnamefont {K.~M.}\ \bibnamefont {Sliwa}},\ and\ \bibinfo {author} {\bibfnamefont {M.~H.}\ \bibnamefont {Devoret}},\ }\bibfield  {title} {\bibinfo {title} {3-wave mixing josephson dipole element},\ }\bibfield  {booktitle} {\emph {\bibinfo {booktitle} {Applied Physics Letters}},\ }\href {https://doi.org/10.1063/1.4984142} {\bibfield  {journal} {\bibinfo  {journal} {Applied Physics Letters}\ }\textbf {\bibinfo {volume} {110}},\ \bibinfo {pages} {222603} (\bibinfo {year} {2017})}\BibitemShut {NoStop}%
\bibitem [{\citenamefont {Lecocq}\ \emph {et~al.}(2017)\citenamefont {Lecocq}, \citenamefont {Ranzani}, \citenamefont {Peterson}, \citenamefont {Cicak}, \citenamefont {Simmonds}, \citenamefont {Teufel},\ and\ \citenamefont {Aumentado}}]{FPJA}%
  \BibitemOpen
  \bibfield  {author} {\bibinfo {author} {\bibfnamefont {F.}~\bibnamefont {Lecocq}}, \bibinfo {author} {\bibfnamefont {L.}~\bibnamefont {Ranzani}}, \bibinfo {author} {\bibfnamefont {G.~A.}\ \bibnamefont {Peterson}}, \bibinfo {author} {\bibfnamefont {K.}~\bibnamefont {Cicak}}, \bibinfo {author} {\bibfnamefont {R.~W.}\ \bibnamefont {Simmonds}}, \bibinfo {author} {\bibfnamefont {J.~D.}\ \bibnamefont {Teufel}},\ and\ \bibinfo {author} {\bibfnamefont {J.}~\bibnamefont {Aumentado}},\ }\bibfield  {title} {\bibinfo {title} {Nonreciprocal microwave signal processing with a field-programmable josephson amplifier},\ }\href {https://doi.org/10.1103/PhysRevApplied.7.024028} {\bibfield  {journal} {\bibinfo  {journal} {Phys. Rev. Applied}\ }\textbf {\bibinfo {volume} {7}},\ \bibinfo {pages} {024028} (\bibinfo {year} {2017})}\BibitemShut {NoStop}%
\bibitem [{\citenamefont {Aumentado}(2020)}]{AumentadoAmplifierReview}%
  \BibitemOpen
  \bibfield  {author} {\bibinfo {author} {\bibfnamefont {J.}~\bibnamefont {Aumentado}},\ }\bibfield  {title} {\bibinfo {title} {Superconducting parametric amplifiers: The state of the art in josephson parametric amplifiers},\ }\href {https://doi.org/10.1109/MMM.2020.2993476} {\bibfield  {journal} {\bibinfo  {journal} {IEEE Microwave Magazine}\ }\textbf {\bibinfo {volume} {21}},\ \bibinfo {pages} {45} (\bibinfo {year} {2020})}\BibitemShut {NoStop}%
\bibitem [{\citenamefont {Sliwa}\ \emph {et~al.}(2015)\citenamefont {Sliwa}, \citenamefont {Hatridge}, \citenamefont {Narla}, \citenamefont {Shankar}, \citenamefont {Frunzio}, \citenamefont {Schoelkopf},\ and\ \citenamefont {Devoret}}]{KatrinaCirculatorAmplifierJPC}%
  \BibitemOpen
  \bibfield  {author} {\bibinfo {author} {\bibfnamefont {K.~M.}\ \bibnamefont {Sliwa}}, \bibinfo {author} {\bibfnamefont {M.}~\bibnamefont {Hatridge}}, \bibinfo {author} {\bibfnamefont {A.}~\bibnamefont {Narla}}, \bibinfo {author} {\bibfnamefont {S.}~\bibnamefont {Shankar}}, \bibinfo {author} {\bibfnamefont {L.}~\bibnamefont {Frunzio}}, \bibinfo {author} {\bibfnamefont {R.~J.}\ \bibnamefont {Schoelkopf}},\ and\ \bibinfo {author} {\bibfnamefont {M.~H.}\ \bibnamefont {Devoret}},\ }\bibfield  {title} {\bibinfo {title} {Reconfigurable josephson circulator/directional amplifier},\ }\href {https://doi.org/10.1103/PhysRevX.5.041020} {\bibfield  {journal} {\bibinfo  {journal} {Phys. Rev. X}\ }\textbf {\bibinfo {volume} {5}},\ \bibinfo {pages} {041020} (\bibinfo {year} {2015})}\BibitemShut {NoStop}%
\bibitem [{\citenamefont {Chapman}\ \emph {et~al.}(2017)\citenamefont {Chapman}, \citenamefont {Rosenthal}, \citenamefont {Kerckhoff}, \citenamefont {Moores}, \citenamefont {Vale}, \citenamefont {Mates}, \citenamefont {Hilton}, \citenamefont {Lalumi\`ere}, \citenamefont {Blais},\ and\ \citenamefont {Lehnert}}]{ChapmanCirculator}%
  \BibitemOpen
  \bibfield  {author} {\bibinfo {author} {\bibfnamefont {B.~J.}\ \bibnamefont {Chapman}}, \bibinfo {author} {\bibfnamefont {E.~I.}\ \bibnamefont {Rosenthal}}, \bibinfo {author} {\bibfnamefont {J.}~\bibnamefont {Kerckhoff}}, \bibinfo {author} {\bibfnamefont {B.~A.}\ \bibnamefont {Moores}}, \bibinfo {author} {\bibfnamefont {L.~R.}\ \bibnamefont {Vale}}, \bibinfo {author} {\bibfnamefont {J.~A.~B.}\ \bibnamefont {Mates}}, \bibinfo {author} {\bibfnamefont {G.~C.}\ \bibnamefont {Hilton}}, \bibinfo {author} {\bibfnamefont {K.}~\bibnamefont {Lalumi\`ere}}, \bibinfo {author} {\bibfnamefont {A.}~\bibnamefont {Blais}},\ and\ \bibinfo {author} {\bibfnamefont {K.~W.}\ \bibnamefont {Lehnert}},\ }\bibfield  {title} {\bibinfo {title} {Widely tunable on-chip microwave circulator for superconducting quantum circuits},\ }\href {https://doi.org/10.1103/PhysRevX.7.041043} {\bibfield  {journal} {\bibinfo  {journal} {Phys. Rev. X}\ }\textbf {\bibinfo {volume} {7}},\ \bibinfo {pages} {041043} (\bibinfo {year} {2017})}\BibitemShut
  {NoStop}%
\bibitem [{\citenamefont {Kwende}\ \emph {et~al.}(2023)\citenamefont {Kwende}, \citenamefont {White},\ and\ \citenamefont {Naaman}}]{GoogleCirculator}%
  \BibitemOpen
  \bibfield  {author} {\bibinfo {author} {\bibfnamefont {R.}~\bibnamefont {Kwende}}, \bibinfo {author} {\bibfnamefont {T.}~\bibnamefont {White}},\ and\ \bibinfo {author} {\bibfnamefont {O.}~\bibnamefont {Naaman}},\ }\bibfield  {title} {\bibinfo {title} {Josephson parametric circulator with same-frequency signal ports, 200 mhz bandwidth, and high dynamic range},\ }\href {https://doi.org/10.1063/5.0150427} {\bibfield  {journal} {\bibinfo  {journal} {Applied Physics Letters}\ }\textbf {\bibinfo {volume} {122}},\ \bibinfo {pages} {224001} (\bibinfo {year} {2023})}\BibitemShut {NoStop}%
\bibitem [{\citenamefont {Ranzani}\ and\ \citenamefont {Aumentado}(2019)}]{AumentadoCirculatorReview}%
  \BibitemOpen
  \bibfield  {author} {\bibinfo {author} {\bibfnamefont {L.}~\bibnamefont {Ranzani}}\ and\ \bibinfo {author} {\bibfnamefont {J.}~\bibnamefont {Aumentado}},\ }\bibfield  {title} {\bibinfo {title} {Circulators at the quantum limit: Recent realizations of quantum-limited superconducting circulators and related approaches},\ }\href {https://doi.org/10.1109/MMM.2019.2891381} {\bibfield  {journal} {\bibinfo  {journal} {IEEE Microwave Magazine}\ }\textbf {\bibinfo {volume} {20}},\ \bibinfo {pages} {112} (\bibinfo {year} {2019})}\BibitemShut {NoStop}%
\bibitem [{\citenamefont {Balembois}\ \emph {et~al.}(2024)\citenamefont {Balembois}, \citenamefont {Travesedo}, \citenamefont {Pallegoix}, \citenamefont {May}, \citenamefont {Billaud}, \citenamefont {Villiers}, \citenamefont {Est\`eve}, \citenamefont {Vion}, \citenamefont {Bertet},\ and\ \citenamefont {Flurin}}]{FlurinSinglePhotonCounter}%
  \BibitemOpen
  \bibfield  {author} {\bibinfo {author} {\bibfnamefont {L.}~\bibnamefont {Balembois}}, \bibinfo {author} {\bibfnamefont {J.}~\bibnamefont {Travesedo}}, \bibinfo {author} {\bibfnamefont {L.}~\bibnamefont {Pallegoix}}, \bibinfo {author} {\bibfnamefont {A.}~\bibnamefont {May}}, \bibinfo {author} {\bibfnamefont {E.}~\bibnamefont {Billaud}}, \bibinfo {author} {\bibfnamefont {M.}~\bibnamefont {Villiers}}, \bibinfo {author} {\bibfnamefont {D.}~\bibnamefont {Est\`eve}}, \bibinfo {author} {\bibfnamefont {D.}~\bibnamefont {Vion}}, \bibinfo {author} {\bibfnamefont {P.}~\bibnamefont {Bertet}},\ and\ \bibinfo {author} {\bibfnamefont {E.}~\bibnamefont {Flurin}},\ }\bibfield  {title} {\bibinfo {title} {Cyclically operated microwave single-photon counter with sensitivity of ${10}^{\ensuremath{-}22}\phantom{\rule{0.2em}{0ex}}\mathrm{W}/\sqrt{\mathrm{hz}}$},\ }\href {https://doi.org/10.1103/PhysRevApplied.21.014043} {\bibfield  {journal} {\bibinfo  {journal} {Phys. Rev. Appl.}\ }\textbf {\bibinfo {volume} {21}},\ \bibinfo
  {pages} {014043} (\bibinfo {year} {2024})}\BibitemShut {NoStop}%
\bibitem [{\citenamefont {Najera-Santos}\ \emph {et~al.}(2024)\citenamefont {Najera-Santos}, \citenamefont {Rousseau}, \citenamefont {Gerashchenko}, \citenamefont {Patange}, \citenamefont {Riva}, \citenamefont {Villiers}, \citenamefont {Briant}, \citenamefont {Cohadon}, \citenamefont {Heidmann}, \citenamefont {Palomo}, \citenamefont {Rosticher}, \citenamefont {le~Sueur}, \citenamefont {Sarlette}, \citenamefont {Smith}, \citenamefont {Leghtas}, \citenamefont {Flurin}, \citenamefont {Jacqmin},\ and\ \citenamefont {Del\'eglise}}]{FlurinChargeSensor}%
  \BibitemOpen
  \bibfield  {author} {\bibinfo {author} {\bibfnamefont {B.-L.}\ \bibnamefont {Najera-Santos}}, \bibinfo {author} {\bibfnamefont {R.}~\bibnamefont {Rousseau}}, \bibinfo {author} {\bibfnamefont {K.}~\bibnamefont {Gerashchenko}}, \bibinfo {author} {\bibfnamefont {H.}~\bibnamefont {Patange}}, \bibinfo {author} {\bibfnamefont {A.}~\bibnamefont {Riva}}, \bibinfo {author} {\bibfnamefont {M.}~\bibnamefont {Villiers}}, \bibinfo {author} {\bibfnamefont {T.}~\bibnamefont {Briant}}, \bibinfo {author} {\bibfnamefont {P.-F.}\ \bibnamefont {Cohadon}}, \bibinfo {author} {\bibfnamefont {A.}~\bibnamefont {Heidmann}}, \bibinfo {author} {\bibfnamefont {J.}~\bibnamefont {Palomo}}, \bibinfo {author} {\bibfnamefont {M.}~\bibnamefont {Rosticher}}, \bibinfo {author} {\bibfnamefont {H.}~\bibnamefont {le~Sueur}}, \bibinfo {author} {\bibfnamefont {A.}~\bibnamefont {Sarlette}}, \bibinfo {author} {\bibfnamefont {W.~C.}\ \bibnamefont {Smith}}, \bibinfo {author} {\bibfnamefont {Z.}~\bibnamefont {Leghtas}}, \bibinfo {author} {\bibfnamefont
  {E.}~\bibnamefont {Flurin}}, \bibinfo {author} {\bibfnamefont {T.}~\bibnamefont {Jacqmin}},\ and\ \bibinfo {author} {\bibfnamefont {S.}~\bibnamefont {Del\'eglise}},\ }\bibfield  {title} {\bibinfo {title} {High-sensitivity ac-charge detection with a mhz-frequency fluxonium qubit},\ }\href {https://doi.org/10.1103/PhysRevX.14.011007} {\bibfield  {journal} {\bibinfo  {journal} {Phys. Rev. X}\ }\textbf {\bibinfo {volume} {14}},\ \bibinfo {pages} {011007} (\bibinfo {year} {2024})}\BibitemShut {NoStop}%
\bibitem [{\citenamefont {Backes}\ \emph {et~al.}(2021)\citenamefont {Backes}, \citenamefont {Palken}, \citenamefont {Kenany}, \citenamefont {Brubaker}, \citenamefont {Cahn}, \citenamefont {Droster}, \citenamefont {Hilton}, \citenamefont {Ghosh}, \citenamefont {Jackson}, \citenamefont {Lamoreaux}, \citenamefont {Leder}, \citenamefont {Lehnert}, \citenamefont {Lewis}, \citenamefont {Malnou}, \citenamefont {Maruyama}, \citenamefont {Rapidis}, \citenamefont {Simanovskaia}, \citenamefont {Singh}, \citenamefont {Speller}, \citenamefont {Urdinaran}, \citenamefont {Vale}, \citenamefont {van Assendelft}, \citenamefont {van Bibber},\ and\ \citenamefont {Wang}}]{HAYSTAC_Nature}%
  \BibitemOpen
  \bibfield  {author} {\bibinfo {author} {\bibfnamefont {K.~M.}\ \bibnamefont {Backes}}, \bibinfo {author} {\bibfnamefont {D.~A.}\ \bibnamefont {Palken}}, \bibinfo {author} {\bibfnamefont {S.~A.}\ \bibnamefont {Kenany}}, \bibinfo {author} {\bibfnamefont {B.~M.}\ \bibnamefont {Brubaker}}, \bibinfo {author} {\bibfnamefont {S.~B.}\ \bibnamefont {Cahn}}, \bibinfo {author} {\bibfnamefont {A.}~\bibnamefont {Droster}}, \bibinfo {author} {\bibfnamefont {G.~C.}\ \bibnamefont {Hilton}}, \bibinfo {author} {\bibfnamefont {S.}~\bibnamefont {Ghosh}}, \bibinfo {author} {\bibfnamefont {H.}~\bibnamefont {Jackson}}, \bibinfo {author} {\bibfnamefont {S.~K.}\ \bibnamefont {Lamoreaux}}, \bibinfo {author} {\bibfnamefont {A.~F.}\ \bibnamefont {Leder}}, \bibinfo {author} {\bibfnamefont {K.~W.}\ \bibnamefont {Lehnert}}, \bibinfo {author} {\bibfnamefont {S.~M.}\ \bibnamefont {Lewis}}, \bibinfo {author} {\bibfnamefont {M.}~\bibnamefont {Malnou}}, \bibinfo {author} {\bibfnamefont {R.~H.}\ \bibnamefont {Maruyama}}, \bibinfo {author}
  {\bibfnamefont {N.~M.}\ \bibnamefont {Rapidis}}, \bibinfo {author} {\bibfnamefont {M.}~\bibnamefont {Simanovskaia}}, \bibinfo {author} {\bibfnamefont {S.}~\bibnamefont {Singh}}, \bibinfo {author} {\bibfnamefont {D.~H.}\ \bibnamefont {Speller}}, \bibinfo {author} {\bibfnamefont {I.}~\bibnamefont {Urdinaran}}, \bibinfo {author} {\bibfnamefont {L.~R.}\ \bibnamefont {Vale}}, \bibinfo {author} {\bibfnamefont {E.~C.}\ \bibnamefont {van Assendelft}}, \bibinfo {author} {\bibfnamefont {K.}~\bibnamefont {van Bibber}},\ and\ \bibinfo {author} {\bibfnamefont {H.}~\bibnamefont {Wang}},\ }\bibfield  {title} {\bibinfo {title} {A quantum enhanced search for dark matter axions},\ }\href {https://doi.org/10.1038/s41586-021-03226-7} {\bibfield  {journal} {\bibinfo  {journal} {Nature}\ }\textbf {\bibinfo {volume} {590}},\ \bibinfo {pages} {238} (\bibinfo {year} {2021})}\BibitemShut {NoStop}%
\bibitem [{\citenamefont {Wallraff}\ \emph {et~al.}(2004)\citenamefont {Wallraff}, \citenamefont {Schuster}, \citenamefont {Blais}, \citenamefont {Frunzio}, \citenamefont {Huang}, \citenamefont {Majer}, \citenamefont {Kumar}, \citenamefont {Girvin},\ and\ \citenamefont {Schoelkopf}}]{OriginalcQEDExpt2004}%
  \BibitemOpen
  \bibfield  {author} {\bibinfo {author} {\bibfnamefont {A.}~\bibnamefont {Wallraff}}, \bibinfo {author} {\bibfnamefont {D.~I.}\ \bibnamefont {Schuster}}, \bibinfo {author} {\bibfnamefont {A.}~\bibnamefont {Blais}}, \bibinfo {author} {\bibfnamefont {L.}~\bibnamefont {Frunzio}}, \bibinfo {author} {\bibfnamefont {R.-.~S.}\ \bibnamefont {Huang}}, \bibinfo {author} {\bibfnamefont {J.}~\bibnamefont {Majer}}, \bibinfo {author} {\bibfnamefont {S.}~\bibnamefont {Kumar}}, \bibinfo {author} {\bibfnamefont {S.~M.}\ \bibnamefont {Girvin}},\ and\ \bibinfo {author} {\bibfnamefont {R.~J.}\ \bibnamefont {Schoelkopf}},\ }\bibfield  {title} {\bibinfo {title} {Strong coupling of a single photon to a superconducting qubit using circuit quantum electrodynamics},\ }\href {https://doi.org/10.1038/nature02851} {\bibfield  {journal} {\bibinfo  {journal} {Nature}\ }\textbf {\bibinfo {volume} {431}},\ \bibinfo {pages} {162} (\bibinfo {year} {2004})}\BibitemShut {NoStop}%
\bibitem [{\citenamefont {Blais}\ \emph {et~al.}(2021)\citenamefont {Blais}, \citenamefont {Grimsmo}, \citenamefont {Girvin},\ and\ \citenamefont {Wallraff}}]{CircuitQEDReview}%
  \BibitemOpen
  \bibfield  {author} {\bibinfo {author} {\bibfnamefont {A.}~\bibnamefont {Blais}}, \bibinfo {author} {\bibfnamefont {A.~L.}\ \bibnamefont {Grimsmo}}, \bibinfo {author} {\bibfnamefont {S.~M.}\ \bibnamefont {Girvin}},\ and\ \bibinfo {author} {\bibfnamefont {A.}~\bibnamefont {Wallraff}},\ }\bibfield  {title} {\bibinfo {title} {Circuit quantum electrodynamics},\ }\href {https://doi.org/10.1103/RevModPhys.93.025005} {\bibfield  {journal} {\bibinfo  {journal} {Rev. Mod. Phys.}\ }\textbf {\bibinfo {volume} {93}},\ \bibinfo {pages} {025005} (\bibinfo {year} {2021})}\BibitemShut {NoStop}%
\bibitem [{\citenamefont {Gao}\ \emph {et~al.}(2018)\citenamefont {Gao}, \citenamefont {Lester}, \citenamefont {Zhang}, \citenamefont {Wang}, \citenamefont {Rosenblum}, \citenamefont {Frunzio}, \citenamefont {Jiang}, \citenamefont {Girvin},\ and\ \citenamefont {Schoelkopf}}]{Yvonne2018_PRX_ProgrammableInterference}%
  \BibitemOpen
  \bibfield  {author} {\bibinfo {author} {\bibfnamefont {Y.~Y.}\ \bibnamefont {Gao}}, \bibinfo {author} {\bibfnamefont {B.~J.}\ \bibnamefont {Lester}}, \bibinfo {author} {\bibfnamefont {Y.}~\bibnamefont {Zhang}}, \bibinfo {author} {\bibfnamefont {C.}~\bibnamefont {Wang}}, \bibinfo {author} {\bibfnamefont {S.}~\bibnamefont {Rosenblum}}, \bibinfo {author} {\bibfnamefont {L.}~\bibnamefont {Frunzio}}, \bibinfo {author} {\bibfnamefont {L.}~\bibnamefont {Jiang}}, \bibinfo {author} {\bibfnamefont {S.~M.}\ \bibnamefont {Girvin}},\ and\ \bibinfo {author} {\bibfnamefont {R.~J.}\ \bibnamefont {Schoelkopf}},\ }\bibfield  {title} {\bibinfo {title} {Programmable interference between two microwave quantum memories},\ }\href {https://doi.org/10.1103/PhysRevX.8.021073} {\bibfield  {journal} {\bibinfo  {journal} {Phys. Rev. X}\ }\textbf {\bibinfo {volume} {8}},\ \bibinfo {pages} {021073} (\bibinfo {year} {2018})}\BibitemShut {NoStop}%
\bibitem [{\citenamefont {Lescanne}\ \emph {et~al.}(2020)\citenamefont {Lescanne}, \citenamefont {Villiers}, \citenamefont {Peronnin}, \citenamefont {Sarlette}, \citenamefont {Delbecq}, \citenamefont {Huard}, \citenamefont {Kontos}, \citenamefont {Mirrahimi},\ and\ \citenamefont {Leghtas}}]{ATS_Paper}%
  \BibitemOpen
  \bibfield  {author} {\bibinfo {author} {\bibfnamefont {R.}~\bibnamefont {Lescanne}}, \bibinfo {author} {\bibfnamefont {M.}~\bibnamefont {Villiers}}, \bibinfo {author} {\bibfnamefont {T.}~\bibnamefont {Peronnin}}, \bibinfo {author} {\bibfnamefont {A.}~\bibnamefont {Sarlette}}, \bibinfo {author} {\bibfnamefont {M.}~\bibnamefont {Delbecq}}, \bibinfo {author} {\bibfnamefont {B.}~\bibnamefont {Huard}}, \bibinfo {author} {\bibfnamefont {T.}~\bibnamefont {Kontos}}, \bibinfo {author} {\bibfnamefont {M.}~\bibnamefont {Mirrahimi}},\ and\ \bibinfo {author} {\bibfnamefont {Z.}~\bibnamefont {Leghtas}},\ }\bibfield  {title} {\bibinfo {title} {Exponential suppression of bit-flips in a qubit encoded in an oscillator},\ }\href {https://doi.org/10.1038/s41567-020-0824-x} {\bibfield  {journal} {\bibinfo  {journal} {Nature Physics}\ }\textbf {\bibinfo {volume} {16}},\ \bibinfo {pages} {509} (\bibinfo {year} {2020})}\BibitemShut {NoStop}%
\bibitem [{\citenamefont {Lu}\ \emph {et~al.}(2023)\citenamefont {Lu}, \citenamefont {Maiti}, \citenamefont {Garmon}, \citenamefont {Ganjam}, \citenamefont {Zhang}, \citenamefont {Claes}, \citenamefont {Frunzio}, \citenamefont {Girvin},\ and\ \citenamefont {Schoelkopf}}]{SQUIDBeamsplitterPaper}%
  \BibitemOpen
  \bibfield  {author} {\bibinfo {author} {\bibfnamefont {Y.}~\bibnamefont {Lu}}, \bibinfo {author} {\bibfnamefont {A.}~\bibnamefont {Maiti}}, \bibinfo {author} {\bibfnamefont {J.~W.~O.}\ \bibnamefont {Garmon}}, \bibinfo {author} {\bibfnamefont {S.}~\bibnamefont {Ganjam}}, \bibinfo {author} {\bibfnamefont {Y.}~\bibnamefont {Zhang}}, \bibinfo {author} {\bibfnamefont {J.}~\bibnamefont {Claes}}, \bibinfo {author} {\bibfnamefont {L.}~\bibnamefont {Frunzio}}, \bibinfo {author} {\bibfnamefont {S.~M.}\ \bibnamefont {Girvin}},\ and\ \bibinfo {author} {\bibfnamefont {R.~J.}\ \bibnamefont {Schoelkopf}},\ }\bibfield  {title} {\bibinfo {title} {High-fidelity parametric beamsplitting with a parity-protected converter},\ }\href {https://doi.org/10.1038/s41467-023-41104-0} {\bibfield  {journal} {\bibinfo  {journal} {Nature Communications}\ }\textbf {\bibinfo {volume} {14}},\ \bibinfo {pages} {5767} (\bibinfo {year} {2023})}\BibitemShut {NoStop}%
\bibitem [{\citenamefont {Chapman}\ \emph {et~al.}(2023)\citenamefont {Chapman}, \citenamefont {de~Graaf}, \citenamefont {Xue}, \citenamefont {Zhang}, \citenamefont {Teoh}, \citenamefont {Curtis}, \citenamefont {Tsunoda}, \citenamefont {Eickbusch}, \citenamefont {Read}, \citenamefont {Koottandavida}, \citenamefont {Mundhada}, \citenamefont {Frunzio}, \citenamefont {Devoret}, \citenamefont {Girvin},\ and\ \citenamefont {Schoelkopf}}]{ChapmanSNAILBeamsplitterPaper}%
  \BibitemOpen
  \bibfield  {author} {\bibinfo {author} {\bibfnamefont {B.~J.}\ \bibnamefont {Chapman}}, \bibinfo {author} {\bibfnamefont {S.~J.}\ \bibnamefont {de~Graaf}}, \bibinfo {author} {\bibfnamefont {S.~H.}\ \bibnamefont {Xue}}, \bibinfo {author} {\bibfnamefont {Y.}~\bibnamefont {Zhang}}, \bibinfo {author} {\bibfnamefont {J.}~\bibnamefont {Teoh}}, \bibinfo {author} {\bibfnamefont {J.~C.}\ \bibnamefont {Curtis}}, \bibinfo {author} {\bibfnamefont {T.}~\bibnamefont {Tsunoda}}, \bibinfo {author} {\bibfnamefont {A.}~\bibnamefont {Eickbusch}}, \bibinfo {author} {\bibfnamefont {A.~P.}\ \bibnamefont {Read}}, \bibinfo {author} {\bibfnamefont {A.}~\bibnamefont {Koottandavida}}, \bibinfo {author} {\bibfnamefont {S.~O.}\ \bibnamefont {Mundhada}}, \bibinfo {author} {\bibfnamefont {L.}~\bibnamefont {Frunzio}}, \bibinfo {author} {\bibfnamefont {M.}~\bibnamefont {Devoret}}, \bibinfo {author} {\bibfnamefont {S.}~\bibnamefont {Girvin}},\ and\ \bibinfo {author} {\bibfnamefont {R.}~\bibnamefont {Schoelkopf}},\ }\bibfield  {title}
  {\bibinfo {title} {High-on-off-ratio beam-splitter interaction for gates on bosonically encoded qubits},\ }\href {https://doi.org/10.1103/PRXQuantum.4.020355} {\bibfield  {journal} {\bibinfo  {journal} {PRX Quantum}\ }\textbf {\bibinfo {volume} {4}},\ \bibinfo {pages} {020355} (\bibinfo {year} {2023})}\BibitemShut {NoStop}%
\bibitem [{\citenamefont {Grimm}\ \emph {et~al.}(2020)\citenamefont {Grimm}, \citenamefont {Frattini}, \citenamefont {Puri}, \citenamefont {Mundhada}, \citenamefont {Touzard}, \citenamefont {Mirrahimi}, \citenamefont {Girvin}, \citenamefont {Shankar},\ and\ \citenamefont {Devoret}}]{GrimmFrattiniKerrCat}%
  \BibitemOpen
  \bibfield  {author} {\bibinfo {author} {\bibfnamefont {A.}~\bibnamefont {Grimm}}, \bibinfo {author} {\bibfnamefont {N.~E.}\ \bibnamefont {Frattini}}, \bibinfo {author} {\bibfnamefont {S.}~\bibnamefont {Puri}}, \bibinfo {author} {\bibfnamefont {S.~O.}\ \bibnamefont {Mundhada}}, \bibinfo {author} {\bibfnamefont {S.}~\bibnamefont {Touzard}}, \bibinfo {author} {\bibfnamefont {M.}~\bibnamefont {Mirrahimi}}, \bibinfo {author} {\bibfnamefont {S.~M.}\ \bibnamefont {Girvin}}, \bibinfo {author} {\bibfnamefont {S.}~\bibnamefont {Shankar}},\ and\ \bibinfo {author} {\bibfnamefont {M.~H.}\ \bibnamefont {Devoret}},\ }\bibfield  {title} {\bibinfo {title} {Stabilization and operation of a kerr-cat qubit},\ }\href {https://doi.org/10.1038/s41586-020-2587-z} {\bibfield  {journal} {\bibinfo  {journal} {Nature}\ }\textbf {\bibinfo {volume} {584}},\ \bibinfo {pages} {205} (\bibinfo {year} {2020})}\BibitemShut {NoStop}%
\bibitem [{\citenamefont {Hajr}\ \emph {et~al.}(2024)\citenamefont {Hajr}, \citenamefont {Qing}, \citenamefont {Wang}, \citenamefont {Koolstra}, \citenamefont {Pedramrazi}, \citenamefont {Kang}, \citenamefont {Chen}, \citenamefont {Nguyen}, \citenamefont {J\"unger}, \citenamefont {Goss}, \citenamefont {Huang}, \citenamefont {Bhandari}, \citenamefont {Frattini}, \citenamefont {Puri}, \citenamefont {Dressel}, \citenamefont {Jordan}, \citenamefont {Santiago},\ and\ \citenamefont {Siddiqi}}]{BerkeleyCat}%
  \BibitemOpen
  \bibfield  {author} {\bibinfo {author} {\bibfnamefont {A.}~\bibnamefont {Hajr}}, \bibinfo {author} {\bibfnamefont {B.}~\bibnamefont {Qing}}, \bibinfo {author} {\bibfnamefont {K.}~\bibnamefont {Wang}}, \bibinfo {author} {\bibfnamefont {G.}~\bibnamefont {Koolstra}}, \bibinfo {author} {\bibfnamefont {Z.}~\bibnamefont {Pedramrazi}}, \bibinfo {author} {\bibfnamefont {Z.}~\bibnamefont {Kang}}, \bibinfo {author} {\bibfnamefont {L.}~\bibnamefont {Chen}}, \bibinfo {author} {\bibfnamefont {L.~B.}\ \bibnamefont {Nguyen}}, \bibinfo {author} {\bibfnamefont {C.}~\bibnamefont {J\"unger}}, \bibinfo {author} {\bibfnamefont {N.}~\bibnamefont {Goss}}, \bibinfo {author} {\bibfnamefont {I.}~\bibnamefont {Huang}}, \bibinfo {author} {\bibfnamefont {B.}~\bibnamefont {Bhandari}}, \bibinfo {author} {\bibfnamefont {N.~E.}\ \bibnamefont {Frattini}}, \bibinfo {author} {\bibfnamefont {S.}~\bibnamefont {Puri}}, \bibinfo {author} {\bibfnamefont {J.}~\bibnamefont {Dressel}}, \bibinfo {author} {\bibfnamefont {A.~N.}\ \bibnamefont {Jordan}},
  \bibinfo {author} {\bibfnamefont {D.~I.}\ \bibnamefont {Santiago}},\ and\ \bibinfo {author} {\bibfnamefont {I.}~\bibnamefont {Siddiqi}},\ }\bibfield  {title} {\bibinfo {title} {High-coherence kerr-cat qubit in 2d architecture},\ }\href {https://doi.org/10.1103/PhysRevX.14.041049} {\bibfield  {journal} {\bibinfo  {journal} {Phys. Rev. X}\ }\textbf {\bibinfo {volume} {14}},\ \bibinfo {pages} {041049} (\bibinfo {year} {2024})}\BibitemShut {NoStop}%
\bibitem [{\citenamefont {Eriksson}\ \emph {et~al.}(2024)\citenamefont {Eriksson}, \citenamefont {Sépulcre}, \citenamefont {Kervinen}, \citenamefont {Hillmann}, \citenamefont {Kudra}, \citenamefont {Dupouy}, \citenamefont {Lu}, \citenamefont {Khanahmadi}, \citenamefont {Yang}, \citenamefont {Castillo-Moreno}, \citenamefont {Delsing},\ and\ \citenamefont {Gasparinetti}}]{CubicPhaseExperiment}%
  \BibitemOpen
  \bibfield  {author} {\bibinfo {author} {\bibfnamefont {A.~M.}\ \bibnamefont {Eriksson}}, \bibinfo {author} {\bibfnamefont {T.}~\bibnamefont {Sépulcre}}, \bibinfo {author} {\bibfnamefont {M.}~\bibnamefont {Kervinen}}, \bibinfo {author} {\bibfnamefont {T.}~\bibnamefont {Hillmann}}, \bibinfo {author} {\bibfnamefont {M.}~\bibnamefont {Kudra}}, \bibinfo {author} {\bibfnamefont {S.}~\bibnamefont {Dupouy}}, \bibinfo {author} {\bibfnamefont {Y.}~\bibnamefont {Lu}}, \bibinfo {author} {\bibfnamefont {M.}~\bibnamefont {Khanahmadi}}, \bibinfo {author} {\bibfnamefont {J.}~\bibnamefont {Yang}}, \bibinfo {author} {\bibfnamefont {C.}~\bibnamefont {Castillo-Moreno}}, \bibinfo {author} {\bibfnamefont {P.}~\bibnamefont {Delsing}},\ and\ \bibinfo {author} {\bibfnamefont {S.}~\bibnamefont {Gasparinetti}},\ }\bibfield  {title} {\bibinfo {title} {Universal control of a bosonic mode via drive-activated native cubic interactions},\ }\href {https://doi.org/10.1038/s41467-024-46507-1} {\bibfield  {journal} {\bibinfo  {journal}
  {Nature Communications}\ }\textbf {\bibinfo {volume} {15}},\ \bibinfo {pages} {2512} (\bibinfo {year} {2024})}\BibitemShut {NoStop}%
\bibitem [{\citenamefont {Zhang}\ \emph {et~al.}(2019)\citenamefont {Zhang}, \citenamefont {Lester}, \citenamefont {Gao}, \citenamefont {Jiang}, \citenamefont {Schoelkopf},\ and\ \citenamefont {Girvin}}]{Yaxing_PRA_BilinearModeCoupling}%
  \BibitemOpen
  \bibfield  {author} {\bibinfo {author} {\bibfnamefont {Y.}~\bibnamefont {Zhang}}, \bibinfo {author} {\bibfnamefont {B.~J.}\ \bibnamefont {Lester}}, \bibinfo {author} {\bibfnamefont {Y.~Y.}\ \bibnamefont {Gao}}, \bibinfo {author} {\bibfnamefont {L.}~\bibnamefont {Jiang}}, \bibinfo {author} {\bibfnamefont {R.~J.}\ \bibnamefont {Schoelkopf}},\ and\ \bibinfo {author} {\bibfnamefont {S.~M.}\ \bibnamefont {Girvin}},\ }\bibfield  {title} {\bibinfo {title} {Engineering bilinear mode coupling in circuit qed: Theory and experiment},\ }\href {https://doi.org/10.1103/PhysRevA.99.012314} {\bibfield  {journal} {\bibinfo  {journal} {Phys. Rev. A}\ }\textbf {\bibinfo {volume} {99}},\ \bibinfo {pages} {012314} (\bibinfo {year} {2019})}\BibitemShut {NoStop}%
\bibitem [{\citenamefont {Roch}\ \emph {et~al.}(2012)\citenamefont {Roch}, \citenamefont {Flurin}, \citenamefont {Nguyen}, \citenamefont {Morfin}, \citenamefont {Campagne-Ibarcq}, \citenamefont {Devoret},\ and\ \citenamefont {Huard}}]{FlurinJPC}%
  \BibitemOpen
  \bibfield  {author} {\bibinfo {author} {\bibfnamefont {N.}~\bibnamefont {Roch}}, \bibinfo {author} {\bibfnamefont {E.}~\bibnamefont {Flurin}}, \bibinfo {author} {\bibfnamefont {F.}~\bibnamefont {Nguyen}}, \bibinfo {author} {\bibfnamefont {P.}~\bibnamefont {Morfin}}, \bibinfo {author} {\bibfnamefont {P.}~\bibnamefont {Campagne-Ibarcq}}, \bibinfo {author} {\bibfnamefont {M.~H.}\ \bibnamefont {Devoret}},\ and\ \bibinfo {author} {\bibfnamefont {B.}~\bibnamefont {Huard}},\ }\bibfield  {title} {\bibinfo {title} {Widely tunable, nondegenerate three-wave mixing microwave device operating near the quantum limit},\ }\href {https://doi.org/10.1103/PhysRevLett.108.147701} {\bibfield  {journal} {\bibinfo  {journal} {Phys. Rev. Lett.}\ }\textbf {\bibinfo {volume} {108}},\ \bibinfo {pages} {147701} (\bibinfo {year} {2012})}\BibitemShut {NoStop}%
\bibitem [{\citenamefont {Chien}\ \emph {et~al.}(2020)\citenamefont {Chien}, \citenamefont {Lanes}, \citenamefont {Liu}, \citenamefont {Cao}, \citenamefont {Lu}, \citenamefont {Motz}, \citenamefont {Liu}, \citenamefont {Pekker},\ and\ \citenamefont {Hatridge}}]{KerrFreeJRMHatridge}%
  \BibitemOpen
  \bibfield  {author} {\bibinfo {author} {\bibfnamefont {T.-C.}\ \bibnamefont {Chien}}, \bibinfo {author} {\bibfnamefont {O.}~\bibnamefont {Lanes}}, \bibinfo {author} {\bibfnamefont {C.}~\bibnamefont {Liu}}, \bibinfo {author} {\bibfnamefont {X.}~\bibnamefont {Cao}}, \bibinfo {author} {\bibfnamefont {P.}~\bibnamefont {Lu}}, \bibinfo {author} {\bibfnamefont {S.}~\bibnamefont {Motz}}, \bibinfo {author} {\bibfnamefont {G.}~\bibnamefont {Liu}}, \bibinfo {author} {\bibfnamefont {D.}~\bibnamefont {Pekker}},\ and\ \bibinfo {author} {\bibfnamefont {M.}~\bibnamefont {Hatridge}},\ }\bibfield  {title} {\bibinfo {title} {Multiparametric amplification and qubit measurement with a kerr-free josephson ring modulator},\ }\href {https://doi.org/10.1103/PhysRevA.101.042336} {\bibfield  {journal} {\bibinfo  {journal} {Phys. Rev. A}\ }\textbf {\bibinfo {volume} {101}},\ \bibinfo {pages} {042336} (\bibinfo {year} {2020})}\BibitemShut {NoStop}%
\bibitem [{\citenamefont {Sivak}\ \emph {et~al.}(2019)\citenamefont {Sivak}, \citenamefont {Frattini}, \citenamefont {Joshi}, \citenamefont {Lingenfelter}, \citenamefont {Shankar},\ and\ \citenamefont {Devoret}}]{SNAIL_Analysis_Sivak}%
  \BibitemOpen
  \bibfield  {author} {\bibinfo {author} {\bibfnamefont {V.}~\bibnamefont {Sivak}}, \bibinfo {author} {\bibfnamefont {N.}~\bibnamefont {Frattini}}, \bibinfo {author} {\bibfnamefont {V.}~\bibnamefont {Joshi}}, \bibinfo {author} {\bibfnamefont {A.}~\bibnamefont {Lingenfelter}}, \bibinfo {author} {\bibfnamefont {S.}~\bibnamefont {Shankar}},\ and\ \bibinfo {author} {\bibfnamefont {M.}~\bibnamefont {Devoret}},\ }\bibfield  {title} {\bibinfo {title} {Kerr-free three-wave mixing in superconducting quantum circuits},\ }\href {https://doi.org/10.1103/PhysRevApplied.11.054060} {\bibfield  {journal} {\bibinfo  {journal} {Phys. Rev. Appl.}\ }\textbf {\bibinfo {volume} {11}},\ \bibinfo {pages} {054060} (\bibinfo {year} {2019})}\BibitemShut {NoStop}%
\bibitem [{\citenamefont {Khabipov}\ \emph {et~al.}(2022)\citenamefont {Khabipov}, \citenamefont {Gaydamachenko}, \citenamefont {Kissling}, \citenamefont {Dolata},\ and\ \citenamefont {Zorin}}]{KerrFreeRFSQUID}%
  \BibitemOpen
  \bibfield  {author} {\bibinfo {author} {\bibfnamefont {M.}~\bibnamefont {Khabipov}}, \bibinfo {author} {\bibfnamefont {V.}~\bibnamefont {Gaydamachenko}}, \bibinfo {author} {\bibfnamefont {C.}~\bibnamefont {Kissling}}, \bibinfo {author} {\bibfnamefont {R.}~\bibnamefont {Dolata}},\ and\ \bibinfo {author} {\bibfnamefont {A.~B.}\ \bibnamefont {Zorin}},\ }\bibfield  {title} {\bibinfo {title} {Superconducting microwave resonators with non-centrosymmetric nonlinearity},\ }\href {https://doi.org/10.1088/1361-6668/ac6989} {\bibfield  {journal} {\bibinfo  {journal} {Superconductor Science and Technology}\ }\textbf {\bibinfo {volume} {35}},\ \bibinfo {pages} {065020} (\bibinfo {year} {2022})}\BibitemShut {NoStop}%
\bibitem [{\citenamefont {Liu}\ \emph {et~al.}(2017)\citenamefont {Liu}, \citenamefont {Chien}, \citenamefont {Cao}, \citenamefont {Lanes}, \citenamefont {Alpern}, \citenamefont {Pekker},\ and\ \citenamefont {Hatridge}}]{GangJPCSaturation}%
  \BibitemOpen
  \bibfield  {author} {\bibinfo {author} {\bibfnamefont {G.}~\bibnamefont {Liu}}, \bibinfo {author} {\bibfnamefont {T.-C.}\ \bibnamefont {Chien}}, \bibinfo {author} {\bibfnamefont {X.}~\bibnamefont {Cao}}, \bibinfo {author} {\bibfnamefont {O.}~\bibnamefont {Lanes}}, \bibinfo {author} {\bibfnamefont {E.}~\bibnamefont {Alpern}}, \bibinfo {author} {\bibfnamefont {D.}~\bibnamefont {Pekker}},\ and\ \bibinfo {author} {\bibfnamefont {M.}~\bibnamefont {Hatridge}},\ }\bibfield  {title} {\bibinfo {title} {{Josephson parametric converter saturation and higher order effects}},\ }\href {https://doi.org/10.1063/1.5003032} {\bibfield  {journal} {\bibinfo  {journal} {Applied Physics Letters}\ }\textbf {\bibinfo {volume} {111}},\ \bibinfo {pages} {202603} (\bibinfo {year} {2017})}\BibitemShut {NoStop}%
\bibitem [{\citenamefont {Frattini}\ \emph {et~al.}(2018)\citenamefont {Frattini}, \citenamefont {Sivak}, \citenamefont {Lingenfelter}, \citenamefont {Shankar},\ and\ \citenamefont {Devoret}}]{SNAIL_Analysis_Frattini}%
  \BibitemOpen
  \bibfield  {author} {\bibinfo {author} {\bibfnamefont {N.~E.}\ \bibnamefont {Frattini}}, \bibinfo {author} {\bibfnamefont {V.~V.}\ \bibnamefont {Sivak}}, \bibinfo {author} {\bibfnamefont {A.}~\bibnamefont {Lingenfelter}}, \bibinfo {author} {\bibfnamefont {S.}~\bibnamefont {Shankar}},\ and\ \bibinfo {author} {\bibfnamefont {M.~H.}\ \bibnamefont {Devoret}},\ }\bibfield  {title} {\bibinfo {title} {Optimizing the nonlinearity and dissipation of a snail parametric amplifier for dynamic range},\ }\href {https://doi.org/10.1103/PhysRevApplied.10.054020} {\bibfield  {journal} {\bibinfo  {journal} {Phys. Rev. Appl.}\ }\textbf {\bibinfo {volume} {10}},\ \bibinfo {pages} {054020} (\bibinfo {year} {2018})}\BibitemShut {NoStop}%
\bibitem [{\citenamefont {Cohen}\ \emph {et~al.}(2023)\citenamefont {Cohen}, \citenamefont {Petrescu}, \citenamefont {Shillito},\ and\ \citenamefont {Blais}}]{BlaisReminiscenceChaos}%
  \BibitemOpen
  \bibfield  {author} {\bibinfo {author} {\bibfnamefont {J.}~\bibnamefont {Cohen}}, \bibinfo {author} {\bibfnamefont {A.}~\bibnamefont {Petrescu}}, \bibinfo {author} {\bibfnamefont {R.}~\bibnamefont {Shillito}},\ and\ \bibinfo {author} {\bibfnamefont {A.}~\bibnamefont {Blais}},\ }\bibfield  {title} {\bibinfo {title} {Reminiscence of classical chaos in driven transmons},\ }\href {https://doi.org/10.1103/PRXQuantum.4.020312} {\bibfield  {journal} {\bibinfo  {journal} {PRX Quantum}\ }\textbf {\bibinfo {volume} {4}},\ \bibinfo {pages} {020312} (\bibinfo {year} {2023})}\BibitemShut {NoStop}%
\bibitem [{\citenamefont {Shillito}\ \emph {et~al.}(2022)\citenamefont {Shillito}, \citenamefont {Petrescu}, \citenamefont {Cohen}, \citenamefont {Beall}, \citenamefont {Hauru}, \citenamefont {Ganahl}, \citenamefont {Lewis}, \citenamefont {Vidal},\ and\ \citenamefont {Blais}}]{BlaisTransmonIonization}%
  \BibitemOpen
  \bibfield  {author} {\bibinfo {author} {\bibfnamefont {R.}~\bibnamefont {Shillito}}, \bibinfo {author} {\bibfnamefont {A.}~\bibnamefont {Petrescu}}, \bibinfo {author} {\bibfnamefont {J.}~\bibnamefont {Cohen}}, \bibinfo {author} {\bibfnamefont {J.}~\bibnamefont {Beall}}, \bibinfo {author} {\bibfnamefont {M.}~\bibnamefont {Hauru}}, \bibinfo {author} {\bibfnamefont {M.}~\bibnamefont {Ganahl}}, \bibinfo {author} {\bibfnamefont {A.~G.}\ \bibnamefont {Lewis}}, \bibinfo {author} {\bibfnamefont {G.}~\bibnamefont {Vidal}},\ and\ \bibinfo {author} {\bibfnamefont {A.}~\bibnamefont {Blais}},\ }\bibfield  {title} {\bibinfo {title} {Dynamics of transmon ionization},\ }\href {https://doi.org/10.1103/PhysRevApplied.18.034031} {\bibfield  {journal} {\bibinfo  {journal} {Phys. Rev. Appl.}\ }\textbf {\bibinfo {volume} {18}},\ \bibinfo {pages} {034031} (\bibinfo {year} {2022})}\BibitemShut {NoStop}%
\bibitem [{\citenamefont {Dumas}\ \emph {et~al.}(2024)\citenamefont {Dumas}, \citenamefont {Groleau-Par\'e}, \citenamefont {McDonald}, \citenamefont {Mu\~noz Arias}, \citenamefont {Lled\'o}, \citenamefont {D'Anjou},\ and\ \citenamefont {Blais}}]{BlaisUnifiedIonization2024}%
  \BibitemOpen
  \bibfield  {author} {\bibinfo {author} {\bibfnamefont {M.~F.}\ \bibnamefont {Dumas}}, \bibinfo {author} {\bibfnamefont {B.}~\bibnamefont {Groleau-Par\'e}}, \bibinfo {author} {\bibfnamefont {A.}~\bibnamefont {McDonald}}, \bibinfo {author} {\bibfnamefont {M.~H.}\ \bibnamefont {Mu\~noz Arias}}, \bibinfo {author} {\bibfnamefont {C.}~\bibnamefont {Lled\'o}}, \bibinfo {author} {\bibfnamefont {B.}~\bibnamefont {D'Anjou}},\ and\ \bibinfo {author} {\bibfnamefont {A.}~\bibnamefont {Blais}},\ }\bibfield  {title} {\bibinfo {title} {Measurement-induced transmon ionization},\ }\href {https://doi.org/10.1103/PhysRevX.14.041023} {\bibfield  {journal} {\bibinfo  {journal} {Phys. Rev. X}\ }\textbf {\bibinfo {volume} {14}},\ \bibinfo {pages} {041023} (\bibinfo {year} {2024})}\BibitemShut {NoStop}%
\bibitem [{\citenamefont {Dykman}(2012)}]{QuantumHeatingYaxingRecommendation}%
  \BibitemOpen
  \bibfield  {author} {\bibinfo {author} {\bibfnamefont {M.~I.}\ \bibnamefont {Dykman}},\ }\bibinfo {title} {Fluctuating nonlinear oscillators: From nanomechanics to quantum superconducting circuits}\ (\bibinfo  {publisher} {Oxford University Press},\ \bibinfo {address} {Oxford,UK},\ \bibinfo {year} {2012})\ pp.\ \bibinfo {pages} {165--197}\BibitemShut {NoStop}%
\bibitem [{\citenamefont {Miano}\ \emph {et~al.}(2022)\citenamefont {Miano}, \citenamefont {Liu}, \citenamefont {Sivak}, \citenamefont {Frattini}, \citenamefont {Joshi}, \citenamefont {Dai}, \citenamefont {Frunzio},\ and\ \citenamefont {Devoret}}]{gSNAIL}%
  \BibitemOpen
  \bibfield  {author} {\bibinfo {author} {\bibfnamefont {A.}~\bibnamefont {Miano}}, \bibinfo {author} {\bibfnamefont {G.}~\bibnamefont {Liu}}, \bibinfo {author} {\bibfnamefont {V.~V.}\ \bibnamefont {Sivak}}, \bibinfo {author} {\bibfnamefont {N.~E.}\ \bibnamefont {Frattini}}, \bibinfo {author} {\bibfnamefont {V.~R.}\ \bibnamefont {Joshi}}, \bibinfo {author} {\bibfnamefont {W.}~\bibnamefont {Dai}}, \bibinfo {author} {\bibfnamefont {L.}~\bibnamefont {Frunzio}},\ and\ \bibinfo {author} {\bibfnamefont {M.~H.}\ \bibnamefont {Devoret}},\ }\bibfield  {title} {\bibinfo {title} {{Frequency-tunable Kerr-free three-wave mixing with a gradiometric SNAIL}},\ }\href {https://doi.org/10.1063/5.0083350} {\bibfield  {journal} {\bibinfo  {journal} {Applied Physics Letters}\ }\textbf {\bibinfo {volume} {120}},\ \bibinfo {pages} {184002} (\bibinfo {year} {2022})}\BibitemShut {NoStop}%
\bibitem [{\citenamefont {Ye}\ \emph {et~al.}(2025)\citenamefont {Ye}, \citenamefont {Kline}, \citenamefont {Yen}, \citenamefont {Cunningham}, \citenamefont {Tan}, \citenamefont {Zang}, \citenamefont {Gingras}, \citenamefont {Niedzielski}, \citenamefont {Stickler}, \citenamefont {Serniak}, \citenamefont {Schwartz},\ and\ \citenamefont {O'Brien}}]{gQuarton}%
  \BibitemOpen
  \bibfield  {author} {\bibinfo {author} {\bibfnamefont {Y.}~\bibnamefont {Ye}}, \bibinfo {author} {\bibfnamefont {J.~B.}\ \bibnamefont {Kline}}, \bibinfo {author} {\bibfnamefont {A.}~\bibnamefont {Yen}}, \bibinfo {author} {\bibfnamefont {G.}~\bibnamefont {Cunningham}}, \bibinfo {author} {\bibfnamefont {M.}~\bibnamefont {Tan}}, \bibinfo {author} {\bibfnamefont {A.}~\bibnamefont {Zang}}, \bibinfo {author} {\bibfnamefont {M.}~\bibnamefont {Gingras}}, \bibinfo {author} {\bibfnamefont {B.~M.}\ \bibnamefont {Niedzielski}}, \bibinfo {author} {\bibfnamefont {H.}~\bibnamefont {Stickler}}, \bibinfo {author} {\bibfnamefont {K.}~\bibnamefont {Serniak}}, \bibinfo {author} {\bibfnamefont {M.~E.}\ \bibnamefont {Schwartz}},\ and\ \bibinfo {author} {\bibfnamefont {K.~P.}\ \bibnamefont {O'Brien}},\ }\bibfield  {title} {\bibinfo {title} {Near-ultrastrong nonlinear light-matter coupling in superconducting circuits},\ }\href {https://doi.org/10.1038/s41467-025-59152-z} {\bibfield  {journal} {\bibinfo  {journal} {Nature
  Communications}\ }\textbf {\bibinfo {volume} {16}},\ \bibinfo {pages} {3799} (\bibinfo {year} {2025})}\BibitemShut {NoStop}%
\bibitem [{\citenamefont {Bhandari}\ \emph {et~al.}(2025)\citenamefont {Bhandari}, \citenamefont {Huang}, \citenamefont {Hajr}, \citenamefont {Yanik}, \citenamefont {Qing}, \citenamefont {Wang}, \citenamefont {Santiago}, \citenamefont {Dressel}, \citenamefont {Siddiqi},\ and\ \citenamefont {Jordan}}]{BerkeleySTS_theory}%
  \BibitemOpen
  \bibfield  {author} {\bibinfo {author} {\bibfnamefont {B.}~\bibnamefont {Bhandari}}, \bibinfo {author} {\bibfnamefont {I.}~\bibnamefont {Huang}}, \bibinfo {author} {\bibfnamefont {A.}~\bibnamefont {Hajr}}, \bibinfo {author} {\bibfnamefont {K.}~\bibnamefont {Yanik}}, \bibinfo {author} {\bibfnamefont {B.}~\bibnamefont {Qing}}, \bibinfo {author} {\bibfnamefont {K.}~\bibnamefont {Wang}}, \bibinfo {author} {\bibfnamefont {D.~I.}\ \bibnamefont {Santiago}}, \bibinfo {author} {\bibfnamefont {J.}~\bibnamefont {Dressel}}, \bibinfo {author} {\bibfnamefont {I.}~\bibnamefont {Siddiqi}},\ and\ \bibinfo {author} {\bibfnamefont {A.~N.}\ \bibnamefont {Jordan}},\ }\bibfield  {title} {\bibinfo {title} {Symmetrically threaded superconducting quantum interference devices as next-generation kerr-cat qubits},\ }\href {https://doi.org/10.1103/c661-yr2z} {\bibfield  {journal} {\bibinfo  {journal} {PRX Quantum}\ }\textbf {\bibinfo {volume} {6}},\ \bibinfo {pages} {030338} (\bibinfo {year} {2025})}\BibitemShut {NoStop}%
\bibitem [{\citenamefont {Hosoya}\ \emph {et~al.}(1991)\citenamefont {Hosoya}, \citenamefont {Hioe}, \citenamefont {Casas}, \citenamefont {Kamikawai}, \citenamefont {Harada}, \citenamefont {Wada}, \citenamefont {Nakane}, \citenamefont {Suda},\ and\ \citenamefont {Goto}}]{QuantumFluxParametron}%
  \BibitemOpen
  \bibfield  {author} {\bibinfo {author} {\bibfnamefont {M.}~\bibnamefont {Hosoya}}, \bibinfo {author} {\bibfnamefont {W.}~\bibnamefont {Hioe}}, \bibinfo {author} {\bibfnamefont {J.}~\bibnamefont {Casas}}, \bibinfo {author} {\bibfnamefont {R.}~\bibnamefont {Kamikawai}}, \bibinfo {author} {\bibfnamefont {Y.}~\bibnamefont {Harada}}, \bibinfo {author} {\bibfnamefont {Y.}~\bibnamefont {Wada}}, \bibinfo {author} {\bibfnamefont {H.}~\bibnamefont {Nakane}}, \bibinfo {author} {\bibfnamefont {R.}~\bibnamefont {Suda}},\ and\ \bibinfo {author} {\bibfnamefont {E.}~\bibnamefont {Goto}},\ }\bibfield  {title} {\bibinfo {title} {Quantum flux parametron: a single quantum flux device for josephson supercomputer},\ }\href {https://doi.org/10.1109/77.84613} {\bibfield  {journal} {\bibinfo  {journal} {IEEE Transactions on Applied Superconductivity}\ }\textbf {\bibinfo {volume} {1}},\ \bibinfo {pages} {77} (\bibinfo {year} {1991})}\BibitemShut {NoStop}%
\bibitem [{\citenamefont {Minev}\ \emph {et~al.}(2021)\citenamefont {Minev}, \citenamefont {Leghtas}, \citenamefont {Mundhada}, \citenamefont {Christakis}, \citenamefont {Pop},\ and\ \citenamefont {Devoret}}]{EPRPaper}%
  \BibitemOpen
  \bibfield  {author} {\bibinfo {author} {\bibfnamefont {Z.~K.}\ \bibnamefont {Minev}}, \bibinfo {author} {\bibfnamefont {Z.}~\bibnamefont {Leghtas}}, \bibinfo {author} {\bibfnamefont {S.~O.}\ \bibnamefont {Mundhada}}, \bibinfo {author} {\bibfnamefont {L.}~\bibnamefont {Christakis}}, \bibinfo {author} {\bibfnamefont {I.~M.}\ \bibnamefont {Pop}},\ and\ \bibinfo {author} {\bibfnamefont {M.~H.}\ \bibnamefont {Devoret}},\ }\bibfield  {title} {\bibinfo {title} {Energy-participation quantization of josephson circuits},\ }\href {https://doi.org/10.1038/s41534-021-00461-8} {\bibfield  {journal} {\bibinfo  {journal} {npj Quantum Information}\ }\textbf {\bibinfo {volume} {7}},\ \bibinfo {pages} {131} (\bibinfo {year} {2021})}\BibitemShut {NoStop}%
\bibitem [{\citenamefont {Bylander}\ \emph {et~al.}(2011)\citenamefont {Bylander}, \citenamefont {Gustavsson}, \citenamefont {Yan}, \citenamefont {Yoshihara}, \citenamefont {Harrabi}, \citenamefont {Fitch}, \citenamefont {Cory}, \citenamefont {Nakamura}, \citenamefont {Tsai},\ and\ \citenamefont {Oliver}}]{BylanderCPMG}%
  \BibitemOpen
  \bibfield  {author} {\bibinfo {author} {\bibfnamefont {J.}~\bibnamefont {Bylander}}, \bibinfo {author} {\bibfnamefont {S.}~\bibnamefont {Gustavsson}}, \bibinfo {author} {\bibfnamefont {F.}~\bibnamefont {Yan}}, \bibinfo {author} {\bibfnamefont {F.}~\bibnamefont {Yoshihara}}, \bibinfo {author} {\bibfnamefont {K.}~\bibnamefont {Harrabi}}, \bibinfo {author} {\bibfnamefont {G.}~\bibnamefont {Fitch}}, \bibinfo {author} {\bibfnamefont {D.~G.}\ \bibnamefont {Cory}}, \bibinfo {author} {\bibfnamefont {Y.}~\bibnamefont {Nakamura}}, \bibinfo {author} {\bibfnamefont {J.-S.}\ \bibnamefont {Tsai}},\ and\ \bibinfo {author} {\bibfnamefont {W.~D.}\ \bibnamefont {Oliver}},\ }\bibfield  {title} {\bibinfo {title} {Noise spectroscopy through dynamical decoupling with a superconducting flux qubit},\ }\href {https://doi.org/10.1038/nphys1994} {\bibfield  {journal} {\bibinfo  {journal} {Nature Physics}\ }\textbf {\bibinfo {volume} {7}},\ \bibinfo {pages} {565} (\bibinfo {year} {2011})}\BibitemShut {NoStop}%
\bibitem [{\citenamefont {Yan}\ \emph {et~al.}(2013)\citenamefont {Yan}, \citenamefont {Gustavsson}, \citenamefont {Bylander}, \citenamefont {Jin}, \citenamefont {Yoshihara}, \citenamefont {Cory}, \citenamefont {Nakamura}, \citenamefont {Orlando},\ and\ \citenamefont {Oliver}}]{FeiYanSpinLocking}%
  \BibitemOpen
  \bibfield  {author} {\bibinfo {author} {\bibfnamefont {F.}~\bibnamefont {Yan}}, \bibinfo {author} {\bibfnamefont {S.}~\bibnamefont {Gustavsson}}, \bibinfo {author} {\bibfnamefont {J.}~\bibnamefont {Bylander}}, \bibinfo {author} {\bibfnamefont {X.}~\bibnamefont {Jin}}, \bibinfo {author} {\bibfnamefont {F.}~\bibnamefont {Yoshihara}}, \bibinfo {author} {\bibfnamefont {D.~G.}\ \bibnamefont {Cory}}, \bibinfo {author} {\bibfnamefont {Y.}~\bibnamefont {Nakamura}}, \bibinfo {author} {\bibfnamefont {T.~P.}\ \bibnamefont {Orlando}},\ and\ \bibinfo {author} {\bibfnamefont {W.~D.}\ \bibnamefont {Oliver}},\ }\bibfield  {title} {\bibinfo {title} {Rotating-frame relaxation as a noise spectrum analyser of a superconducting qubit undergoing driven evolution},\ }\href@noop {} {\bibfield  {journal} {\bibinfo  {journal} {Nature Communications}\ }\textbf {\bibinfo {volume} {4}},\ \bibinfo {pages} {2337} (\bibinfo {year} {2013})}\BibitemShut {NoStop}%
\bibitem [{\citenamefont {Braum\"uller}\ \emph {et~al.}(2020)\citenamefont {Braum\"uller}, \citenamefont {Ding}, \citenamefont {Veps\"al\"ainen}, \citenamefont {Sung}, \citenamefont {Kjaergaard}, \citenamefont {Menke}, \citenamefont {Winik}, \citenamefont {Kim}, \citenamefont {Niedzielski}, \citenamefont {Melville}, \citenamefont {Yoder}, \citenamefont {Hirjibehedin}, \citenamefont {Orlando}, \citenamefont {Gustavsson},\ and\ \citenamefont {Oliver}}]{MIT_SQUID_flux_noise_2020}%
  \BibitemOpen
  \bibfield  {author} {\bibinfo {author} {\bibfnamefont {J.}~\bibnamefont {Braum\"uller}}, \bibinfo {author} {\bibfnamefont {L.}~\bibnamefont {Ding}}, \bibinfo {author} {\bibfnamefont {A.~P.}\ \bibnamefont {Veps\"al\"ainen}}, \bibinfo {author} {\bibfnamefont {Y.}~\bibnamefont {Sung}}, \bibinfo {author} {\bibfnamefont {M.}~\bibnamefont {Kjaergaard}}, \bibinfo {author} {\bibfnamefont {T.}~\bibnamefont {Menke}}, \bibinfo {author} {\bibfnamefont {R.}~\bibnamefont {Winik}}, \bibinfo {author} {\bibfnamefont {D.}~\bibnamefont {Kim}}, \bibinfo {author} {\bibfnamefont {B.~M.}\ \bibnamefont {Niedzielski}}, \bibinfo {author} {\bibfnamefont {A.}~\bibnamefont {Melville}}, \bibinfo {author} {\bibfnamefont {J.~L.}\ \bibnamefont {Yoder}}, \bibinfo {author} {\bibfnamefont {C.~F.}\ \bibnamefont {Hirjibehedin}}, \bibinfo {author} {\bibfnamefont {T.~P.}\ \bibnamefont {Orlando}}, \bibinfo {author} {\bibfnamefont {S.}~\bibnamefont {Gustavsson}},\ and\ \bibinfo {author} {\bibfnamefont {W.~D.}\ \bibnamefont {Oliver}},\ }\bibfield
  {title} {\bibinfo {title} {Characterizing and optimizing qubit coherence based on squid geometry},\ }\href {https://doi.org/10.1103/PhysRevApplied.13.054079} {\bibfield  {journal} {\bibinfo  {journal} {Phys. Rev. Appl.}\ }\textbf {\bibinfo {volume} {13}},\ \bibinfo {pages} {054079} (\bibinfo {year} {2020})}\BibitemShut {NoStop}%
\bibitem [{\citenamefont {Martinis}\ \emph {et~al.}(2003)\citenamefont {Martinis}, \citenamefont {Nam}, \citenamefont {Aumentado}, \citenamefont {Lang},\ and\ \citenamefont {Urbina}}]{Martinis_dephasing}%
  \BibitemOpen
  \bibfield  {author} {\bibinfo {author} {\bibfnamefont {J.~M.}\ \bibnamefont {Martinis}}, \bibinfo {author} {\bibfnamefont {S.}~\bibnamefont {Nam}}, \bibinfo {author} {\bibfnamefont {J.}~\bibnamefont {Aumentado}}, \bibinfo {author} {\bibfnamefont {K.~M.}\ \bibnamefont {Lang}},\ and\ \bibinfo {author} {\bibfnamefont {C.}~\bibnamefont {Urbina}},\ }\bibfield  {title} {\bibinfo {title} {Decoherence of a superconducting qubit due to bias noise},\ }\href {https://doi.org/10.1103/PhysRevB.67.094510} {\bibfield  {journal} {\bibinfo  {journal} {Phys. Rev. B}\ }\textbf {\bibinfo {volume} {67}},\ \bibinfo {pages} {094510} (\bibinfo {year} {2003})}\BibitemShut {NoStop}%
\bibitem [{\citenamefont {Groszkowski}\ \emph {et~al.}(2018)\citenamefont {Groszkowski}, \citenamefont {Paolo}, \citenamefont {Grimsmo}, \citenamefont {Blais}, \citenamefont {Schuster}, \citenamefont {Houck},\ and\ \citenamefont {Koch}}]{Peter_dephasing}%
  \BibitemOpen
  \bibfield  {author} {\bibinfo {author} {\bibfnamefont {P.}~\bibnamefont {Groszkowski}}, \bibinfo {author} {\bibfnamefont {A.~D.}\ \bibnamefont {Paolo}}, \bibinfo {author} {\bibfnamefont {A.~L.}\ \bibnamefont {Grimsmo}}, \bibinfo {author} {\bibfnamefont {A.}~\bibnamefont {Blais}}, \bibinfo {author} {\bibfnamefont {D.~I.}\ \bibnamefont {Schuster}}, \bibinfo {author} {\bibfnamefont {A.~A.}\ \bibnamefont {Houck}},\ and\ \bibinfo {author} {\bibfnamefont {J.}~\bibnamefont {Koch}},\ }\bibfield  {title} {\bibinfo {title} {Coherence properties of the 0-$\pi$qubit},\ }\href {https://doi.org/10.1088/1367-2630/aab7cd} {\bibfield  {journal} {\bibinfo  {journal} {New Journal of Physics}\ }\textbf {\bibinfo {volume} {20}},\ \bibinfo {pages} {043053} (\bibinfo {year} {2018})}\BibitemShut {NoStop}%
\bibitem [{\citenamefont {Valadares}\ \emph {et~al.}(2024)\citenamefont {Valadares}, \citenamefont {Huang}, \citenamefont {Chu}, \citenamefont {Dorogov}, \citenamefont {Chua}, \citenamefont {Kong}, \citenamefont {Song},\ and\ \citenamefont {Gao}}]{YvonneMagneticHose}%
  \BibitemOpen
  \bibfield  {author} {\bibinfo {author} {\bibfnamefont {F.}~\bibnamefont {Valadares}}, \bibinfo {author} {\bibfnamefont {N.-N.}\ \bibnamefont {Huang}}, \bibinfo {author} {\bibfnamefont {K.~T.~N.}\ \bibnamefont {Chu}}, \bibinfo {author} {\bibfnamefont {A.}~\bibnamefont {Dorogov}}, \bibinfo {author} {\bibfnamefont {W.}~\bibnamefont {Chua}}, \bibinfo {author} {\bibfnamefont {L.}~\bibnamefont {Kong}}, \bibinfo {author} {\bibfnamefont {P.}~\bibnamefont {Song}},\ and\ \bibinfo {author} {\bibfnamefont {Y.~Y.}\ \bibnamefont {Gao}},\ }\bibfield  {title} {\bibinfo {title} {On-demand transposition across light-matter interaction regimes in bosonic cqed},\ }\href {https://doi.org/10.1038/s41467-024-50201-7} {\bibfield  {journal} {\bibinfo  {journal} {Nature Communications}\ }\textbf {\bibinfo {volume} {15}},\ \bibinfo {pages} {5816} (\bibinfo {year} {2024})}\BibitemShut {NoStop}%
\bibitem [{\citenamefont {Goldblatt}\ \emph {et~al.}(2024)\citenamefont {Goldblatt}, \citenamefont {Kahn}, \citenamefont {Hazanov}, \citenamefont {Milul}, \citenamefont {Guttel}, \citenamefont {Joshi}, \citenamefont {Chausovsky}, \citenamefont {Lafont},\ and\ \citenamefont {Rosenblum}}]{SergeShotNoiseFeedback2024}%
  \BibitemOpen
  \bibfield  {author} {\bibinfo {author} {\bibfnamefont {U.}~\bibnamefont {Goldblatt}}, \bibinfo {author} {\bibfnamefont {N.}~\bibnamefont {Kahn}}, \bibinfo {author} {\bibfnamefont {S.}~\bibnamefont {Hazanov}}, \bibinfo {author} {\bibfnamefont {O.}~\bibnamefont {Milul}}, \bibinfo {author} {\bibfnamefont {B.}~\bibnamefont {Guttel}}, \bibinfo {author} {\bibfnamefont {L.~M.}\ \bibnamefont {Joshi}}, \bibinfo {author} {\bibfnamefont {D.}~\bibnamefont {Chausovsky}}, \bibinfo {author} {\bibfnamefont {F.}~\bibnamefont {Lafont}},\ and\ \bibinfo {author} {\bibfnamefont {S.}~\bibnamefont {Rosenblum}},\ }\bibfield  {title} {\bibinfo {title} {Recovering quantum coherence of a cavity qubit coupled to a noisy ancilla through real-time feedback},\ }\href {https://doi.org/10.1103/PhysRevX.14.041056} {\bibfield  {journal} {\bibinfo  {journal} {Phys. Rev. X}\ }\textbf {\bibinfo {volume} {14}},\ \bibinfo {pages} {041056} (\bibinfo {year} {2024})}\BibitemShut {NoStop}%
\bibitem [{\citenamefont {Atanasova}\ \emph {et~al.}(2025)\citenamefont {Atanasova}, \citenamefont {Yang}, \citenamefont {H\"onigl-Decrinis}, \citenamefont {Gusenkova}, \citenamefont {Pop},\ and\ \citenamefont {Kirchmair}}]{IoanFluxoniumTunableChi}%
  \BibitemOpen
  \bibfield  {author} {\bibinfo {author} {\bibfnamefont {D.}~\bibnamefont {Atanasova}}, \bibinfo {author} {\bibfnamefont {I.}~\bibnamefont {Yang}}, \bibinfo {author} {\bibfnamefont {T.}~\bibnamefont {H\"onigl-Decrinis}}, \bibinfo {author} {\bibfnamefont {D.}~\bibnamefont {Gusenkova}}, \bibinfo {author} {\bibfnamefont {I.}~\bibnamefont {Pop}},\ and\ \bibinfo {author} {\bibfnamefont {G.}~\bibnamefont {Kirchmair}},\ }\bibfield  {title} {\bibinfo {title} {In situ tunable interaction with an invertible sign between a fluxonium and a post cavity},\ }\href {https://doi.org/10.1103/PRXQuantum.6.020318} {\bibfield  {journal} {\bibinfo  {journal} {PRX Quantum}\ }\textbf {\bibinfo {volume} {6}},\ \bibinfo {pages} {020318} (\bibinfo {year} {2025})}\BibitemShut {NoStop}%
\bibitem [{\citenamefont {Lu}\ \emph {et~al.}()\citenamefont {Lu}, \citenamefont {Zhao}, \citenamefont {Vallières}, \citenamefont {Smith}, \citenamefont {Weiss}, \citenamefont {You}, \citenamefont {Zhang}, \citenamefont {Ganjam}, \citenamefont {Maiti}, \citenamefont {Garmon}, \citenamefont {Mundhada}, \citenamefont {Huang}, , \citenamefont {Mondragon-Shem}, \citenamefont {Grivin}, \citenamefont {Koch},\ and\ \citenamefont {Schoelkopf}}]{Lu2025}%
  \BibitemOpen
  \bibfield  {author} {\bibinfo {author} {\bibfnamefont {Y.}~\bibnamefont {Lu}}, \bibinfo {author} {\bibfnamefont {T.}~\bibnamefont {Zhao}}, \bibinfo {author} {\bibfnamefont {A.}~\bibnamefont {Vallières}}, \bibinfo {author} {\bibfnamefont {K.}~\bibnamefont {Smith}}, \bibinfo {author} {\bibfnamefont {D.}~\bibnamefont {Weiss}}, \bibinfo {author} {\bibfnamefont {X.}~\bibnamefont {You}}, \bibinfo {author} {\bibfnamefont {Y.}~\bibnamefont {Zhang}}, \bibinfo {author} {\bibfnamefont {S.}~\bibnamefont {Ganjam}}, \bibinfo {author} {\bibfnamefont {A.}~\bibnamefont {Maiti}}, \bibinfo {author} {\bibfnamefont {J.}~\bibnamefont {Garmon}}, \bibinfo {author} {\bibfnamefont {S.}~\bibnamefont {Mundhada}}, \bibinfo {author} {\bibfnamefont {Z.}~\bibnamefont {Huang}}, , \bibinfo {author} {\bibfnamefont {I.}~\bibnamefont {Mondragon-Shem}}, \bibinfo {author} {\bibfnamefont {S.}~\bibnamefont {Grivin}}, \bibinfo {author} {\bibfnamefont {J.}~\bibnamefont {Koch}},\ and\ \bibinfo {author} {\bibfnamefont {R.}~\bibnamefont
  {Schoelkopf}},\ }\bibfield  {title} {\bibinfo {title} {Systematic construction of time-dependent {Hamiltonians} for microwave-driven {Josephson} circuits},\ }\bibinfo {note} {in preparation}\BibitemShut {NoStop}%
\bibitem [{\citenamefont {Eickbusch}\ \emph {et~al.}(2022)\citenamefont {Eickbusch}, \citenamefont {Sivak}, \citenamefont {Ding}, \citenamefont {Elder}, \citenamefont {Jha}, \citenamefont {Venkatraman}, \citenamefont {Royer}, \citenamefont {Girvin}, \citenamefont {Schoelkopf},\ and\ \citenamefont {Devoret}}]{ECDControlPaper}%
  \BibitemOpen
  \bibfield  {author} {\bibinfo {author} {\bibfnamefont {A.}~\bibnamefont {Eickbusch}}, \bibinfo {author} {\bibfnamefont {V.}~\bibnamefont {Sivak}}, \bibinfo {author} {\bibfnamefont {A.~Z.}\ \bibnamefont {Ding}}, \bibinfo {author} {\bibfnamefont {S.~S.}\ \bibnamefont {Elder}}, \bibinfo {author} {\bibfnamefont {S.~R.}\ \bibnamefont {Jha}}, \bibinfo {author} {\bibfnamefont {J.}~\bibnamefont {Venkatraman}}, \bibinfo {author} {\bibfnamefont {B.}~\bibnamefont {Royer}}, \bibinfo {author} {\bibfnamefont {S.~M.}\ \bibnamefont {Girvin}}, \bibinfo {author} {\bibfnamefont {R.~J.}\ \bibnamefont {Schoelkopf}},\ and\ \bibinfo {author} {\bibfnamefont {M.~H.}\ \bibnamefont {Devoret}},\ }\bibfield  {title} {\bibinfo {title} {Fast universal control of an oscillator with weak dispersive coupling to a qubit},\ }\href {https://doi.org/10.1038/s41567-022-01776-9} {\bibfield  {journal} {\bibinfo  {journal} {Nature Physics}\ }\textbf {\bibinfo {volume} {18}},\ \bibinfo {pages} {1464} (\bibinfo {year} {2022})}\BibitemShut {NoStop}%
\bibitem [{\citenamefont {Hillmann}\ \emph {et~al.}(2020)\citenamefont {Hillmann}, \citenamefont {Quijandr\'{\i}a}, \citenamefont {Johansson}, \citenamefont {Ferraro}, \citenamefont {Gasparinetti},\ and\ \citenamefont {Ferrini}}]{CubicPhaseTheory}%
  \BibitemOpen
  \bibfield  {author} {\bibinfo {author} {\bibfnamefont {T.}~\bibnamefont {Hillmann}}, \bibinfo {author} {\bibfnamefont {F.}~\bibnamefont {Quijandr\'{\i}a}}, \bibinfo {author} {\bibfnamefont {G.}~\bibnamefont {Johansson}}, \bibinfo {author} {\bibfnamefont {A.}~\bibnamefont {Ferraro}}, \bibinfo {author} {\bibfnamefont {S.}~\bibnamefont {Gasparinetti}},\ and\ \bibinfo {author} {\bibfnamefont {G.}~\bibnamefont {Ferrini}},\ }\bibfield  {title} {\bibinfo {title} {Universal gate set for continuous-variable quantum computation with microwave circuits},\ }\href {https://doi.org/10.1103/PhysRevLett.125.160501} {\bibfield  {journal} {\bibinfo  {journal} {Phys. Rev. Lett.}\ }\textbf {\bibinfo {volume} {125}},\ \bibinfo {pages} {160501} (\bibinfo {year} {2020})}\BibitemShut {NoStop}%
\bibitem [{\citenamefont {Ding}\ \emph {et~al.}(2025)\citenamefont {Ding}, \citenamefont {Brock}, \citenamefont {Eickbusch}, \citenamefont {Koottandavida}, \citenamefont {Frattini}, \citenamefont {Corti{\~n}as}, \citenamefont {Joshi}, \citenamefont {de~Graaf}, \citenamefont {Chapman}, \citenamefont {Ganjam}, \citenamefont {Frunzio}, \citenamefont {Schoelkopf},\ and\ \citenamefont {Devoret}}]{ding2024oscillatorkerrcat}%
  \BibitemOpen
  \bibfield  {author} {\bibinfo {author} {\bibfnamefont {A.~Z.}\ \bibnamefont {Ding}}, \bibinfo {author} {\bibfnamefont {B.~L.}\ \bibnamefont {Brock}}, \bibinfo {author} {\bibfnamefont {A.}~\bibnamefont {Eickbusch}}, \bibinfo {author} {\bibfnamefont {A.}~\bibnamefont {Koottandavida}}, \bibinfo {author} {\bibfnamefont {N.~E.}\ \bibnamefont {Frattini}}, \bibinfo {author} {\bibfnamefont {R.~G.}\ \bibnamefont {Corti{\~n}as}}, \bibinfo {author} {\bibfnamefont {V.~R.}\ \bibnamefont {Joshi}}, \bibinfo {author} {\bibfnamefont {S.~J.}\ \bibnamefont {de~Graaf}}, \bibinfo {author} {\bibfnamefont {B.~J.}\ \bibnamefont {Chapman}}, \bibinfo {author} {\bibfnamefont {S.}~\bibnamefont {Ganjam}}, \bibinfo {author} {\bibfnamefont {L.}~\bibnamefont {Frunzio}}, \bibinfo {author} {\bibfnamefont {R.~J.}\ \bibnamefont {Schoelkopf}},\ and\ \bibinfo {author} {\bibfnamefont {M.~H.}\ \bibnamefont {Devoret}},\ }\bibfield  {title} {\bibinfo {title} {Quantum control of an oscillator with a kerr-cat qubit},\ }\href
  {https://doi.org/10.1038/s41467-025-60352-w} {\bibfield  {journal} {\bibinfo  {journal} {Nature Communications}\ }\textbf {\bibinfo {volume} {16}},\ \bibinfo {pages} {5279} (\bibinfo {year} {2025})}\BibitemShut {NoStop}%
\bibitem [{\citenamefont {Pietik\"ainen}\ \emph {et~al.}(2024)\citenamefont {Pietik\"ainen}, \citenamefont {\ifmmode~\check{C}\else \v{C}\fi{}ernot\'{\i}k}, \citenamefont {Eickbusch}, \citenamefont {Maiti}, \citenamefont {Garmon}, \citenamefont {Filip},\ and\ \citenamefont {Girvin}}]{OndrejHighQMemories}%
  \BibitemOpen
  \bibfield  {author} {\bibinfo {author} {\bibfnamefont {I.}~\bibnamefont {Pietik\"ainen}}, \bibinfo {author} {\bibfnamefont {O.~c.~v.}\ \bibnamefont {\ifmmode~\check{C}\else \v{C}\fi{}ernot\'{\i}k}}, \bibinfo {author} {\bibfnamefont {A.}~\bibnamefont {Eickbusch}}, \bibinfo {author} {\bibfnamefont {A.}~\bibnamefont {Maiti}}, \bibinfo {author} {\bibfnamefont {J.~W.}\ \bibnamefont {Garmon}}, \bibinfo {author} {\bibfnamefont {R.}~\bibnamefont {Filip}},\ and\ \bibinfo {author} {\bibfnamefont {S.~M.}\ \bibnamefont {Girvin}},\ }\bibfield  {title} {\bibinfo {title} {Strategies and trade-offs for controllability and memory time of ultra-high-quality microwave cavities in circuit quantum electrodynamics},\ }\href {https://doi.org/10.1103/PRXQuantum.5.040307} {\bibfield  {journal} {\bibinfo  {journal} {PRX Quantum}\ }\textbf {\bibinfo {volume} {5}},\ \bibinfo {pages} {040307} (\bibinfo {year} {2024})}\BibitemShut {NoStop}%
\bibitem [{\citenamefont {Pozar}(2012)}]{Pozar_2012}%
  \BibitemOpen
  \bibfield  {author} {\bibinfo {author} {\bibfnamefont {D.~M.}\ \bibnamefont {Pozar}},\ }\href@noop {} {\emph {\bibinfo {title} {Microwave Engineering}}},\ \bibinfo {edition} {4th}\ ed.\ (\bibinfo  {publisher} {Wiley \& Sons},\ \bibinfo {year} {2012})\BibitemShut {NoStop}%
\bibitem [{\citenamefont {Remm}\ \emph {et~al.}(2023)\citenamefont {Remm}, \citenamefont {Krinner}, \citenamefont {Lacroix}, \citenamefont {Hellings}, \citenamefont {Swiadek}, \citenamefont {Norris}, \citenamefont {Eichler},\ and\ \citenamefont {Wallraff}}]{IndermodulationDistortionTWPAWalraff}%
  \BibitemOpen
  \bibfield  {author} {\bibinfo {author} {\bibfnamefont {A.}~\bibnamefont {Remm}}, \bibinfo {author} {\bibfnamefont {S.}~\bibnamefont {Krinner}}, \bibinfo {author} {\bibfnamefont {N.}~\bibnamefont {Lacroix}}, \bibinfo {author} {\bibfnamefont {C.}~\bibnamefont {Hellings}}, \bibinfo {author} {\bibfnamefont {F.~m.~c.}\ \bibnamefont {Swiadek}}, \bibinfo {author} {\bibfnamefont {G.~J.}\ \bibnamefont {Norris}}, \bibinfo {author} {\bibfnamefont {C.}~\bibnamefont {Eichler}},\ and\ \bibinfo {author} {\bibfnamefont {A.}~\bibnamefont {Wallraff}},\ }\bibfield  {title} {\bibinfo {title} {Intermodulation distortion in a josephson traveling-wave parametric amplifier},\ }\href {https://doi.org/10.1103/PhysRevApplied.20.034027} {\bibfield  {journal} {\bibinfo  {journal} {Phys. Rev. Appl.}\ }\textbf {\bibinfo {volume} {20}},\ \bibinfo {pages} {034027} (\bibinfo {year} {2023})}\BibitemShut {NoStop}%
\bibitem [{\citenamefont {Spring}\ \emph {et~al.}(2025)\citenamefont {Spring}, \citenamefont {Milanovic}, \citenamefont {Sunada}, \citenamefont {Wang}, \citenamefont {van Loo}, \citenamefont {Tamate},\ and\ \citenamefont {Nakamura}}]{NakamuraMultiplexedReadout}%
  \BibitemOpen
  \bibfield  {author} {\bibinfo {author} {\bibfnamefont {P.~A.}\ \bibnamefont {Spring}}, \bibinfo {author} {\bibfnamefont {L.}~\bibnamefont {Milanovic}}, \bibinfo {author} {\bibfnamefont {Y.}~\bibnamefont {Sunada}}, \bibinfo {author} {\bibfnamefont {S.}~\bibnamefont {Wang}}, \bibinfo {author} {\bibfnamefont {A.~F.}\ \bibnamefont {van Loo}}, \bibinfo {author} {\bibfnamefont {S.}~\bibnamefont {Tamate}},\ and\ \bibinfo {author} {\bibfnamefont {Y.}~\bibnamefont {Nakamura}},\ }\bibfield  {title} {\bibinfo {title} {Fast multiplexed superconducting-qubit readout with intrinsic purcell filtering using a multiconductor transmission line},\ }\href {https://doi.org/10.1103/PRXQuantum.6.020345} {\bibfield  {journal} {\bibinfo  {journal} {PRX Quantum}\ }\textbf {\bibinfo {volume} {6}},\ \bibinfo {pages} {020345} (\bibinfo {year} {2025})}\BibitemShut {NoStop}%
\bibitem [{\citenamefont {Gottesman}\ \emph {et~al.}(2001)\citenamefont {Gottesman}, \citenamefont {Kitaev},\ and\ \citenamefont {Preskill}}]{OriginalGKP}%
  \BibitemOpen
  \bibfield  {author} {\bibinfo {author} {\bibfnamefont {D.}~\bibnamefont {Gottesman}}, \bibinfo {author} {\bibfnamefont {A.}~\bibnamefont {Kitaev}},\ and\ \bibinfo {author} {\bibfnamefont {J.}~\bibnamefont {Preskill}},\ }\bibfield  {title} {\bibinfo {title} {Encoding a qubit in an oscillator},\ }\href {https://doi.org/10.1103/PhysRevA.64.012310} {\bibfield  {journal} {\bibinfo  {journal} {Phys. Rev. A}\ }\textbf {\bibinfo {volume} {64}},\ \bibinfo {pages} {012310} (\bibinfo {year} {2001})}\BibitemShut {NoStop}%
\bibitem [{\citenamefont {Rojkov}\ \emph {et~al.}(2024)\citenamefont {Rojkov}, \citenamefont {R\"oggla}, \citenamefont {Wagener}, \citenamefont {Fontbot\'e-Schmidt}, \citenamefont {Welte}, \citenamefont {Home},\ and\ \citenamefont {Reiter}}]{GKPCiteFromShraddha}%
  \BibitemOpen
  \bibfield  {author} {\bibinfo {author} {\bibfnamefont {I.}~\bibnamefont {Rojkov}}, \bibinfo {author} {\bibfnamefont {P.~M.}\ \bibnamefont {R\"oggla}}, \bibinfo {author} {\bibfnamefont {M.}~\bibnamefont {Wagener}}, \bibinfo {author} {\bibfnamefont {M.}~\bibnamefont {Fontbot\'e-Schmidt}}, \bibinfo {author} {\bibfnamefont {S.}~\bibnamefont {Welte}}, \bibinfo {author} {\bibfnamefont {J.}~\bibnamefont {Home}},\ and\ \bibinfo {author} {\bibfnamefont {F.}~\bibnamefont {Reiter}},\ }\bibfield  {title} {\bibinfo {title} {Two-qubit operations for finite-energy gottesman-kitaev-preskill encodings},\ }\href {https://doi.org/10.1103/PhysRevLett.133.100601} {\bibfield  {journal} {\bibinfo  {journal} {Phys. Rev. Lett.}\ }\textbf {\bibinfo {volume} {133}},\ \bibinfo {pages} {100601} (\bibinfo {year} {2024})}\BibitemShut {NoStop}%
\bibitem [{\citenamefont {Berdou}\ \emph {et~al.}(2023)\citenamefont {Berdou}, \citenamefont {Murani}, \citenamefont {R\'eglade}, \citenamefont {Smith}, \citenamefont {Villiers}, \citenamefont {Palomo}, \citenamefont {Rosticher}, \citenamefont {Denis}, \citenamefont {Morfin}, \citenamefont {Delbecq}, \citenamefont {Kontos}, \citenamefont {Pankratova}, \citenamefont {Rautschke}, \citenamefont {Peronnin}, \citenamefont {Sellem}, \citenamefont {Rouchon}, \citenamefont {Sarlette}, \citenamefont {Mirrahimi}, \citenamefont {Campagne-Ibarcq}, \citenamefont {Jezouin}, \citenamefont {Lescanne},\ and\ \citenamefont {Leghtas}}]{A&B_100_sec_bit_flip_w_ATS}%
  \BibitemOpen
  \bibfield  {author} {\bibinfo {author} {\bibfnamefont {C.}~\bibnamefont {Berdou}}, \bibinfo {author} {\bibfnamefont {A.}~\bibnamefont {Murani}}, \bibinfo {author} {\bibfnamefont {U.}~\bibnamefont {R\'eglade}}, \bibinfo {author} {\bibfnamefont {W.}~\bibnamefont {Smith}}, \bibinfo {author} {\bibfnamefont {M.}~\bibnamefont {Villiers}}, \bibinfo {author} {\bibfnamefont {J.}~\bibnamefont {Palomo}}, \bibinfo {author} {\bibfnamefont {M.}~\bibnamefont {Rosticher}}, \bibinfo {author} {\bibfnamefont {A.}~\bibnamefont {Denis}}, \bibinfo {author} {\bibfnamefont {P.}~\bibnamefont {Morfin}}, \bibinfo {author} {\bibfnamefont {M.}~\bibnamefont {Delbecq}}, \bibinfo {author} {\bibfnamefont {T.}~\bibnamefont {Kontos}}, \bibinfo {author} {\bibfnamefont {N.}~\bibnamefont {Pankratova}}, \bibinfo {author} {\bibfnamefont {F.}~\bibnamefont {Rautschke}}, \bibinfo {author} {\bibfnamefont {T.}~\bibnamefont {Peronnin}}, \bibinfo {author} {\bibfnamefont {L.-A.}\ \bibnamefont {Sellem}}, \bibinfo {author} {\bibfnamefont {P.}~\bibnamefont
  {Rouchon}}, \bibinfo {author} {\bibfnamefont {A.}~\bibnamefont {Sarlette}}, \bibinfo {author} {\bibfnamefont {M.}~\bibnamefont {Mirrahimi}}, \bibinfo {author} {\bibfnamefont {P.}~\bibnamefont {Campagne-Ibarcq}}, \bibinfo {author} {\bibfnamefont {S.}~\bibnamefont {Jezouin}}, \bibinfo {author} {\bibfnamefont {R.}~\bibnamefont {Lescanne}},\ and\ \bibinfo {author} {\bibfnamefont {Z.}~\bibnamefont {Leghtas}},\ }\bibfield  {title} {\bibinfo {title} {One hundred second bit-flip time in a two-photon dissipative oscillator},\ }\href {https://doi.org/10.1103/PRXQuantum.4.020350} {\bibfield  {journal} {\bibinfo  {journal} {PRX Quantum}\ }\textbf {\bibinfo {volume} {4}},\ \bibinfo {pages} {020350} (\bibinfo {year} {2023})}\BibitemShut {NoStop}%
\bibitem [{\citenamefont {Putterman}\ \emph {et~al.}(2025)\citenamefont {Putterman}, \citenamefont {Noh}, \citenamefont {Hann}, \citenamefont {MacCabe}, \citenamefont {Aghaeimeibodi}, \citenamefont {Patel}, \citenamefont {Lee}, \citenamefont {Jones}, \citenamefont {Moradinejad}, \citenamefont {Rodriguez}, \citenamefont {Mahuli}, \citenamefont {Rose}, \citenamefont {Owens}, \citenamefont {Levine}, \citenamefont {Rosenfeld}, \citenamefont {Reinhold}, \citenamefont {Moncelsi}, \citenamefont {Alcid}, \citenamefont {Alidoust}, \citenamefont {Arrangoiz-Arriola}, \citenamefont {Barnett}, \citenamefont {Bienias}, \citenamefont {Carson}, \citenamefont {Chen}, \citenamefont {Chen}, \citenamefont {Chinkezian}, \citenamefont {Chisholm}, \citenamefont {Chou}, \citenamefont {Clerk}, \citenamefont {Clifford}, \citenamefont {Cosmic}, \citenamefont {Curiel}, \citenamefont {Davis}, \citenamefont {DeLorenzo}, \citenamefont {D'Ewart}, \citenamefont {Diky}, \citenamefont {D'Souza}, \citenamefont {Dumitrescu}, \citenamefont
  {Eisenmann}, \citenamefont {Elkhouly}, \citenamefont {Evenbly}, \citenamefont {Fang}, \citenamefont {Fang}, \citenamefont {Fling}, \citenamefont {Fon}, \citenamefont {Garcia}, \citenamefont {Gorshkov}, \citenamefont {Grant}, \citenamefont {Gray}, \citenamefont {Grimberg}, \citenamefont {Grimsmo}, \citenamefont {Haim}, \citenamefont {Hand}, \citenamefont {He}, \citenamefont {Hernandez}, \citenamefont {Hover}, \citenamefont {Hung}, \citenamefont {Hunt}, \citenamefont {Iverson}, \citenamefont {Jarrige}, \citenamefont {Jaskula}, \citenamefont {Jiang}, \citenamefont {Kalaee}, \citenamefont {Karabalin}, \citenamefont {Karalekas}, \citenamefont {Keller}, \citenamefont {Khalajhedayati}, \citenamefont {Kubica}, \citenamefont {Lee}, \citenamefont {Leroux}, \citenamefont {Lieu}, \citenamefont {Ly}, \citenamefont {Madrigal}, \citenamefont {Marcaud}, \citenamefont {McCabe}, \citenamefont {Miles}, \citenamefont {Milsted}, \citenamefont {Minguzzi}, \citenamefont {Mishra}, \citenamefont {Mukherjee}, \citenamefont
  {Naghiloo}, \citenamefont {Oblepias}, \citenamefont {Ortuno}, \citenamefont {Pagdilao}, \citenamefont {Pancotti}, \citenamefont {Panduro}, \citenamefont {Paquette}, \citenamefont {Park}, \citenamefont {Peairs}, \citenamefont {Perello}, \citenamefont {Peterson}, \citenamefont {Ponte}, \citenamefont {Preskill}, \citenamefont {Qiao}, \citenamefont {Refael}, \citenamefont {Resnick}, \citenamefont {Retzker}, \citenamefont {Reyna}, \citenamefont {Runyan}, \citenamefont {Ryan}, \citenamefont {Sahmoud}, \citenamefont {Sanchez}, \citenamefont {Sanil}, \citenamefont {Sankar}, \citenamefont {Sato}, \citenamefont {Scaffidi}, \citenamefont {Siavoshi}, \citenamefont {Sivarajah}, \citenamefont {Skogland}, \citenamefont {Su}, \citenamefont {Swenson}, \citenamefont {Teo}, \citenamefont {Tomada}, \citenamefont {Torlai}, \citenamefont {Wollack}, \citenamefont {Ye}, \citenamefont {Zerrudo}, \citenamefont {Zhang}, \citenamefont {Brand{\~a}o}, \citenamefont {Matheny},\ and\ \citenamefont
  {Painter}}]{AmazonConcatenatedCats_w_ATS}%
  \BibitemOpen
  \bibfield  {author} {\bibinfo {author} {\bibfnamefont {H.}~\bibnamefont {Putterman}}, \bibinfo {author} {\bibfnamefont {K.}~\bibnamefont {Noh}}, \bibinfo {author} {\bibfnamefont {C.~T.}\ \bibnamefont {Hann}}, \bibinfo {author} {\bibfnamefont {G.~S.}\ \bibnamefont {MacCabe}}, \bibinfo {author} {\bibfnamefont {S.}~\bibnamefont {Aghaeimeibodi}}, \bibinfo {author} {\bibfnamefont {R.~N.}\ \bibnamefont {Patel}}, \bibinfo {author} {\bibfnamefont {M.}~\bibnamefont {Lee}}, \bibinfo {author} {\bibfnamefont {W.~M.}\ \bibnamefont {Jones}}, \bibinfo {author} {\bibfnamefont {H.}~\bibnamefont {Moradinejad}}, \bibinfo {author} {\bibfnamefont {R.}~\bibnamefont {Rodriguez}}, \bibinfo {author} {\bibfnamefont {N.}~\bibnamefont {Mahuli}}, \bibinfo {author} {\bibfnamefont {J.}~\bibnamefont {Rose}}, \bibinfo {author} {\bibfnamefont {J.~C.}\ \bibnamefont {Owens}}, \bibinfo {author} {\bibfnamefont {H.}~\bibnamefont {Levine}}, \bibinfo {author} {\bibfnamefont {E.}~\bibnamefont {Rosenfeld}}, \bibinfo {author} {\bibfnamefont
  {P.}~\bibnamefont {Reinhold}}, \bibinfo {author} {\bibfnamefont {L.}~\bibnamefont {Moncelsi}}, \bibinfo {author} {\bibfnamefont {J.~A.}\ \bibnamefont {Alcid}}, \bibinfo {author} {\bibfnamefont {N.}~\bibnamefont {Alidoust}}, \bibinfo {author} {\bibfnamefont {P.}~\bibnamefont {Arrangoiz-Arriola}}, \bibinfo {author} {\bibfnamefont {J.}~\bibnamefont {Barnett}}, \bibinfo {author} {\bibfnamefont {P.}~\bibnamefont {Bienias}}, \bibinfo {author} {\bibfnamefont {H.~A.}\ \bibnamefont {Carson}}, \bibinfo {author} {\bibfnamefont {C.}~\bibnamefont {Chen}}, \bibinfo {author} {\bibfnamefont {L.}~\bibnamefont {Chen}}, \bibinfo {author} {\bibfnamefont {H.}~\bibnamefont {Chinkezian}}, \bibinfo {author} {\bibfnamefont {E.~M.}\ \bibnamefont {Chisholm}}, \bibinfo {author} {\bibfnamefont {M.-H.}\ \bibnamefont {Chou}}, \bibinfo {author} {\bibfnamefont {A.}~\bibnamefont {Clerk}}, \bibinfo {author} {\bibfnamefont {A.}~\bibnamefont {Clifford}}, \bibinfo {author} {\bibfnamefont {R.}~\bibnamefont {Cosmic}}, \bibinfo {author}
  {\bibfnamefont {A.~V.}\ \bibnamefont {Curiel}}, \bibinfo {author} {\bibfnamefont {E.}~\bibnamefont {Davis}}, \bibinfo {author} {\bibfnamefont {L.}~\bibnamefont {DeLorenzo}}, \bibinfo {author} {\bibfnamefont {J.~M.}\ \bibnamefont {D'Ewart}}, \bibinfo {author} {\bibfnamefont {A.}~\bibnamefont {Diky}}, \bibinfo {author} {\bibfnamefont {N.}~\bibnamefont {D'Souza}}, \bibinfo {author} {\bibfnamefont {P.~T.}\ \bibnamefont {Dumitrescu}}, \bibinfo {author} {\bibfnamefont {S.}~\bibnamefont {Eisenmann}}, \bibinfo {author} {\bibfnamefont {E.}~\bibnamefont {Elkhouly}}, \bibinfo {author} {\bibfnamefont {G.}~\bibnamefont {Evenbly}}, \bibinfo {author} {\bibfnamefont {M.~T.}\ \bibnamefont {Fang}}, \bibinfo {author} {\bibfnamefont {Y.}~\bibnamefont {Fang}}, \bibinfo {author} {\bibfnamefont {M.~J.}\ \bibnamefont {Fling}}, \bibinfo {author} {\bibfnamefont {W.}~\bibnamefont {Fon}}, \bibinfo {author} {\bibfnamefont {G.}~\bibnamefont {Garcia}}, \bibinfo {author} {\bibfnamefont {A.~V.}\ \bibnamefont {Gorshkov}}, \bibinfo {author}
  {\bibfnamefont {J.~A.}\ \bibnamefont {Grant}}, \bibinfo {author} {\bibfnamefont {M.~J.}\ \bibnamefont {Gray}}, \bibinfo {author} {\bibfnamefont {S.}~\bibnamefont {Grimberg}}, \bibinfo {author} {\bibfnamefont {A.~L.}\ \bibnamefont {Grimsmo}}, \bibinfo {author} {\bibfnamefont {A.}~\bibnamefont {Haim}}, \bibinfo {author} {\bibfnamefont {J.}~\bibnamefont {Hand}}, \bibinfo {author} {\bibfnamefont {Y.}~\bibnamefont {He}}, \bibinfo {author} {\bibfnamefont {M.}~\bibnamefont {Hernandez}}, \bibinfo {author} {\bibfnamefont {D.}~\bibnamefont {Hover}}, \bibinfo {author} {\bibfnamefont {J.~S.~C.}\ \bibnamefont {Hung}}, \bibinfo {author} {\bibfnamefont {M.}~\bibnamefont {Hunt}}, \bibinfo {author} {\bibfnamefont {J.}~\bibnamefont {Iverson}}, \bibinfo {author} {\bibfnamefont {I.}~\bibnamefont {Jarrige}}, \bibinfo {author} {\bibfnamefont {J.-C.}\ \bibnamefont {Jaskula}}, \bibinfo {author} {\bibfnamefont {L.}~\bibnamefont {Jiang}}, \bibinfo {author} {\bibfnamefont {M.}~\bibnamefont {Kalaee}}, \bibinfo {author} {\bibfnamefont
  {R.}~\bibnamefont {Karabalin}}, \bibinfo {author} {\bibfnamefont {P.~J.}\ \bibnamefont {Karalekas}}, \bibinfo {author} {\bibfnamefont {A.~J.}\ \bibnamefont {Keller}}, \bibinfo {author} {\bibfnamefont {A.}~\bibnamefont {Khalajhedayati}}, \bibinfo {author} {\bibfnamefont {A.}~\bibnamefont {Kubica}}, \bibinfo {author} {\bibfnamefont {H.}~\bibnamefont {Lee}}, \bibinfo {author} {\bibfnamefont {C.}~\bibnamefont {Leroux}}, \bibinfo {author} {\bibfnamefont {S.}~\bibnamefont {Lieu}}, \bibinfo {author} {\bibfnamefont {V.}~\bibnamefont {Ly}}, \bibinfo {author} {\bibfnamefont {K.~V.}\ \bibnamefont {Madrigal}}, \bibinfo {author} {\bibfnamefont {G.}~\bibnamefont {Marcaud}}, \bibinfo {author} {\bibfnamefont {G.}~\bibnamefont {McCabe}}, \bibinfo {author} {\bibfnamefont {C.}~\bibnamefont {Miles}}, \bibinfo {author} {\bibfnamefont {A.}~\bibnamefont {Milsted}}, \bibinfo {author} {\bibfnamefont {J.}~\bibnamefont {Minguzzi}}, \bibinfo {author} {\bibfnamefont {A.}~\bibnamefont {Mishra}}, \bibinfo {author} {\bibfnamefont
  {B.}~\bibnamefont {Mukherjee}}, \bibinfo {author} {\bibfnamefont {M.}~\bibnamefont {Naghiloo}}, \bibinfo {author} {\bibfnamefont {E.}~\bibnamefont {Oblepias}}, \bibinfo {author} {\bibfnamefont {G.}~\bibnamefont {Ortuno}}, \bibinfo {author} {\bibfnamefont {J.}~\bibnamefont {Pagdilao}}, \bibinfo {author} {\bibfnamefont {N.}~\bibnamefont {Pancotti}}, \bibinfo {author} {\bibfnamefont {A.}~\bibnamefont {Panduro}}, \bibinfo {author} {\bibfnamefont {J.}~\bibnamefont {Paquette}}, \bibinfo {author} {\bibfnamefont {M.}~\bibnamefont {Park}}, \bibinfo {author} {\bibfnamefont {G.~A.}\ \bibnamefont {Peairs}}, \bibinfo {author} {\bibfnamefont {D.}~\bibnamefont {Perello}}, \bibinfo {author} {\bibfnamefont {E.~C.}\ \bibnamefont {Peterson}}, \bibinfo {author} {\bibfnamefont {S.}~\bibnamefont {Ponte}}, \bibinfo {author} {\bibfnamefont {J.}~\bibnamefont {Preskill}}, \bibinfo {author} {\bibfnamefont {J.}~\bibnamefont {Qiao}}, \bibinfo {author} {\bibfnamefont {G.}~\bibnamefont {Refael}}, \bibinfo {author} {\bibfnamefont
  {R.}~\bibnamefont {Resnick}}, \bibinfo {author} {\bibfnamefont {A.}~\bibnamefont {Retzker}}, \bibinfo {author} {\bibfnamefont {O.~A.}\ \bibnamefont {Reyna}}, \bibinfo {author} {\bibfnamefont {M.}~\bibnamefont {Runyan}}, \bibinfo {author} {\bibfnamefont {C.~A.}\ \bibnamefont {Ryan}}, \bibinfo {author} {\bibfnamefont {A.}~\bibnamefont {Sahmoud}}, \bibinfo {author} {\bibfnamefont {E.}~\bibnamefont {Sanchez}}, \bibinfo {author} {\bibfnamefont {R.}~\bibnamefont {Sanil}}, \bibinfo {author} {\bibfnamefont {K.}~\bibnamefont {Sankar}}, \bibinfo {author} {\bibfnamefont {Y.}~\bibnamefont {Sato}}, \bibinfo {author} {\bibfnamefont {T.}~\bibnamefont {Scaffidi}}, \bibinfo {author} {\bibfnamefont {S.}~\bibnamefont {Siavoshi}}, \bibinfo {author} {\bibfnamefont {P.}~\bibnamefont {Sivarajah}}, \bibinfo {author} {\bibfnamefont {T.}~\bibnamefont {Skogland}}, \bibinfo {author} {\bibfnamefont {C.-J.}\ \bibnamefont {Su}}, \bibinfo {author} {\bibfnamefont {L.~J.}\ \bibnamefont {Swenson}}, \bibinfo {author} {\bibfnamefont {S.~M.}\
  \bibnamefont {Teo}}, \bibinfo {author} {\bibfnamefont {A.}~\bibnamefont {Tomada}}, \bibinfo {author} {\bibfnamefont {G.}~\bibnamefont {Torlai}}, \bibinfo {author} {\bibfnamefont {E.~A.}\ \bibnamefont {Wollack}}, \bibinfo {author} {\bibfnamefont {Y.}~\bibnamefont {Ye}}, \bibinfo {author} {\bibfnamefont {J.~A.}\ \bibnamefont {Zerrudo}}, \bibinfo {author} {\bibfnamefont {K.}~\bibnamefont {Zhang}}, \bibinfo {author} {\bibfnamefont {F.~G. S.~L.}\ \bibnamefont {Brand{\~a}o}}, \bibinfo {author} {\bibfnamefont {M.~H.}\ \bibnamefont {Matheny}},\ and\ \bibinfo {author} {\bibfnamefont {O.}~\bibnamefont {Painter}},\ }\bibfield  {title} {\bibinfo {title} {Hardware-efficient quantum error correction via concatenated bosonic qubits},\ }\href {https://doi.org/10.1038/s41586-025-08642-7} {\bibfield  {journal} {\bibinfo  {journal} {Nature}\ }\textbf {\bibinfo {volume} {638}},\ \bibinfo {pages} {927} (\bibinfo {year} {2025})}\BibitemShut {NoStop}%
\bibitem [{\citenamefont {Vanselow}\ \emph {et~al.}(2025)\citenamefont {Vanselow}, \citenamefont {Beauseigneur}, \citenamefont {Lattier}, \citenamefont {Villiers}, \citenamefont {Denis}, \citenamefont {Morfin}, \citenamefont {Leghtas},\ and\ \citenamefont {Campagne-Ibarcq}}]{Philippe_four_photon_diss_w_ATS}%
  \BibitemOpen
  \bibfield  {author} {\bibinfo {author} {\bibfnamefont {A.}~\bibnamefont {Vanselow}}, \bibinfo {author} {\bibfnamefont {B.}~\bibnamefont {Beauseigneur}}, \bibinfo {author} {\bibfnamefont {L.}~\bibnamefont {Lattier}}, \bibinfo {author} {\bibfnamefont {M.}~\bibnamefont {Villiers}}, \bibinfo {author} {\bibfnamefont {A.}~\bibnamefont {Denis}}, \bibinfo {author} {\bibfnamefont {P.}~\bibnamefont {Morfin}}, \bibinfo {author} {\bibfnamefont {Z.}~\bibnamefont {Leghtas}},\ and\ \bibinfo {author} {\bibfnamefont {P.}~\bibnamefont {Campagne-Ibarcq}},\ }\href {https://arxiv.org/abs/2501.05960} {\bibinfo {title} {Dissipating quartets of excitations in a superconducting circuit}} (\bibinfo {year} {2025}),\ \Eprint {https://arxiv.org/abs/2501.05960} {arXiv:2501.05960 [quant-ph]} \BibitemShut {NoStop}%
\bibitem [{\citenamefont {Acharya}\ \emph {et~al.}(2025)\citenamefont {Acharya}, \citenamefont {Abanin}, \citenamefont {Aghababaie-Beni}, \citenamefont {Aleiner}, \citenamefont {Andersen}, \citenamefont {Ansmann}, \citenamefont {Arute}, \citenamefont {Arya}, \citenamefont {Asfaw}, \citenamefont {Astrakhantsev}, \citenamefont {Atalaya}, \citenamefont {Babbush}, \citenamefont {Bacon}, \citenamefont {Ballard}, \citenamefont {Bardin}, \citenamefont {Bausch}, \citenamefont {Bengtsson}, \citenamefont {Bilmes}, \citenamefont {Blackwell}, \citenamefont {Boixo}, \citenamefont {Bortoli}, \citenamefont {Bourassa}, \citenamefont {Bovaird}, \citenamefont {Brill}, \citenamefont {Broughton}, \citenamefont {Browne}, \citenamefont {Buchea}, \citenamefont {Buckley}, \citenamefont {Buell}, \citenamefont {Burger}, \citenamefont {Burkett}, \citenamefont {Bushnell}, \citenamefont {Cabrera}, \citenamefont {Campero}, \citenamefont {Chang}, \citenamefont {Chen}, \citenamefont {Chen}, \citenamefont {Chiaro}, \citenamefont {Chik},
  \citenamefont {Chou}, \citenamefont {Claes}, \citenamefont {Cleland}, \citenamefont {Cogan}, \citenamefont {Collins}, \citenamefont {Conner}, \citenamefont {Courtney}, \citenamefont {Crook}, \citenamefont {Curtin}, \citenamefont {Das}, \citenamefont {Davies}, \citenamefont {De~Lorenzo}, \citenamefont {Debroy}, \citenamefont {Demura}, \citenamefont {Devoret}, \citenamefont {Di~Paolo}, \citenamefont {Donohoe}, \citenamefont {Drozdov}, \citenamefont {Dunsworth}, \citenamefont {Earle}, \citenamefont {Edlich}, \citenamefont {Eickbusch}, \citenamefont {Elbag}, \citenamefont {Elzouka}, \citenamefont {Erickson}, \citenamefont {Faoro}, \citenamefont {Farhi}, \citenamefont {Ferreira}, \citenamefont {Burgos}, \citenamefont {Forati}, \citenamefont {Fowler}, \citenamefont {Foxen}, \citenamefont {Ganjam}, \citenamefont {Garcia}, \citenamefont {Gasca}, \citenamefont {Genois}, \citenamefont {Giang}, \citenamefont {Gidney}, \citenamefont {Gilboa}, \citenamefont {Gosula}, \citenamefont {Dau}, \citenamefont {Graumann},
  \citenamefont {Greene}, \citenamefont {Gross}, \citenamefont {Habegger}, \citenamefont {Hall}, \citenamefont {Hamilton}, \citenamefont {Hansen}, \citenamefont {Harrigan}, \citenamefont {Harrington}, \citenamefont {Heras}, \citenamefont {Heslin}, \citenamefont {Heu}, \citenamefont {Higgott}, \citenamefont {Hill}, \citenamefont {Hilton}, \citenamefont {Holland}, \citenamefont {Hong}, \citenamefont {Huang}, \citenamefont {Huff}, \citenamefont {Huggins}, \citenamefont {Ioffe}, \citenamefont {Isakov}, \citenamefont {Iveland}, \citenamefont {Jeffrey}, \citenamefont {Jiang}, \citenamefont {Jones}, \citenamefont {Jordan}, \citenamefont {Joshi}, \citenamefont {Juhas}, \citenamefont {Kafri}, \citenamefont {Kang}, \citenamefont {Karamlou}, \citenamefont {Kechedzhi}, \citenamefont {Kelly}, \citenamefont {Khaire}, \citenamefont {Khattar}, \citenamefont {Khezri}, \citenamefont {Kim}, \citenamefont {Klimov}, \citenamefont {Klots}, \citenamefont {Kobrin}, \citenamefont {Kohli}, \citenamefont {Korotkov}, \citenamefont
  {Kostritsa}, \citenamefont {Kothari}, \citenamefont {Kozlovskii}, \citenamefont {Kreikebaum}, \citenamefont {Kurilovich}, \citenamefont {Lacroix}, \citenamefont {Landhuis}, \citenamefont {Lange-Dei}, \citenamefont {Langley}, \citenamefont {Laptev}, \citenamefont {Lau}, \citenamefont {Le~Guevel}, \citenamefont {Ledford}, \citenamefont {Lee}, \citenamefont {Lee}, \citenamefont {Lensky}, \citenamefont {Leon}, \citenamefont {Lester}, \citenamefont {Li}, \citenamefont {Li}, \citenamefont {Lill}, \citenamefont {Liu}, \citenamefont {Livingston}, \citenamefont {Locharla}, \citenamefont {Lucero}, \citenamefont {Lundahl}, \citenamefont {Lunt}, \citenamefont {Madhuk}, \citenamefont {Malone}, \citenamefont {Maloney}, \citenamefont {Mandr{\`a}}, \citenamefont {Manyika}, \citenamefont {Martin}, \citenamefont {Martin}, \citenamefont {Martin}, \citenamefont {Maxfield}, \citenamefont {McClean}, \citenamefont {McEwen}, \citenamefont {Meeks}, \citenamefont {Megrant}, \citenamefont {Mi}, \citenamefont {Miao}, \citenamefont
  {Mieszala}, \citenamefont {Molavi}, \citenamefont {Molina}, \citenamefont {Montazeri}, \citenamefont {Morvan}, \citenamefont {Movassagh}, \citenamefont {Mruczkiewicz}, \citenamefont {Naaman}, \citenamefont {Neeley}, \citenamefont {Neill}, \citenamefont {Nersisyan}, \citenamefont {Neven}, \citenamefont {Newman}, \citenamefont {Ng}, \citenamefont {Nguyen}, \citenamefont {Nguyen}, \citenamefont {Ni}, \citenamefont {Niu}, \citenamefont {O'Brien}, \citenamefont {Oliver}, \citenamefont {Opremcak}, \citenamefont {Ottosson}, \citenamefont {Petukhov}, \citenamefont {Pizzuto}, \citenamefont {Platt}, \citenamefont {Potter}, \citenamefont {Pritchard}, \citenamefont {Pryadko}, \citenamefont {Quintana}, \citenamefont {Ramachandran}, \citenamefont {Reagor}, \citenamefont {Redding}, \citenamefont {Rhodes}, \citenamefont {Roberts}, \citenamefont {Rosenberg}, \citenamefont {Rosenfeld}, \citenamefont {Roushan}, \citenamefont {Rubin}, \citenamefont {Saei}, \citenamefont {Sank}, \citenamefont {Sankaragomathi}, \citenamefont
  {Satzinger}, \citenamefont {Schurkus}, \citenamefont {Schuster}, \citenamefont {Senior}, \citenamefont {Shearn}, \citenamefont {Shorter}, \citenamefont {Shutty}, \citenamefont {Shvarts}, \citenamefont {Singh}, \citenamefont {Sivak}, \citenamefont {Skruzny}, \citenamefont {Small}, \citenamefont {Smelyanskiy}, \citenamefont {Smith}, \citenamefont {Somma}, \citenamefont {Springer}, \citenamefont {Sterling}, \citenamefont {Strain}, \citenamefont {Suchard}, \citenamefont {Szasz}, \citenamefont {Sztein}, \citenamefont {Thor}, \citenamefont {Torres}, \citenamefont {Torunbalci}, \citenamefont {Vaishnav}, \citenamefont {Vargas}, \citenamefont {Vdovichev}, \citenamefont {Vidal}, \citenamefont {Villalonga}, \citenamefont {Heidweiller}, \citenamefont {Waltman}, \citenamefont {Wang}, \citenamefont {Ware}, \citenamefont {Weber}, \citenamefont {Weidel}, \citenamefont {White}, \citenamefont {Wong}, \citenamefont {Woo}, \citenamefont {Xing}, \citenamefont {Yao}, \citenamefont {Yeh}, \citenamefont {Ying}, \citenamefont
  {Yoo}, \citenamefont {Yosri}, \citenamefont {Young}, \citenamefont {Zalcman}, \citenamefont {Zhang}, \citenamefont {Zhu}, \citenamefont {Zobrist}, \citenamefont {AI},\ and\ \citenamefont {Collaborators}}]{GoogleSurfaceCodeBelowThreshold}%
  \BibitemOpen
  \bibfield  {author} {\bibinfo {author} {\bibfnamefont {R.}~\bibnamefont {Acharya}}, \bibinfo {author} {\bibfnamefont {D.~A.}\ \bibnamefont {Abanin}}, \bibinfo {author} {\bibfnamefont {L.}~\bibnamefont {Aghababaie-Beni}}, \bibinfo {author} {\bibfnamefont {I.}~\bibnamefont {Aleiner}}, \bibinfo {author} {\bibfnamefont {T.~I.}\ \bibnamefont {Andersen}}, \bibinfo {author} {\bibfnamefont {M.}~\bibnamefont {Ansmann}}, \bibinfo {author} {\bibfnamefont {F.}~\bibnamefont {Arute}}, \bibinfo {author} {\bibfnamefont {K.}~\bibnamefont {Arya}}, \bibinfo {author} {\bibfnamefont {A.}~\bibnamefont {Asfaw}}, \bibinfo {author} {\bibfnamefont {N.}~\bibnamefont {Astrakhantsev}}, \bibinfo {author} {\bibfnamefont {J.}~\bibnamefont {Atalaya}}, \bibinfo {author} {\bibfnamefont {R.}~\bibnamefont {Babbush}}, \bibinfo {author} {\bibfnamefont {D.}~\bibnamefont {Bacon}}, \bibinfo {author} {\bibfnamefont {B.}~\bibnamefont {Ballard}}, \bibinfo {author} {\bibfnamefont {J.~C.}\ \bibnamefont {Bardin}}, \bibinfo {author} {\bibfnamefont
  {J.}~\bibnamefont {Bausch}}, \bibinfo {author} {\bibfnamefont {A.}~\bibnamefont {Bengtsson}}, \bibinfo {author} {\bibfnamefont {A.}~\bibnamefont {Bilmes}}, \bibinfo {author} {\bibfnamefont {S.}~\bibnamefont {Blackwell}}, \bibinfo {author} {\bibfnamefont {S.}~\bibnamefont {Boixo}}, \bibinfo {author} {\bibfnamefont {G.}~\bibnamefont {Bortoli}}, \bibinfo {author} {\bibfnamefont {A.}~\bibnamefont {Bourassa}}, \bibinfo {author} {\bibfnamefont {J.}~\bibnamefont {Bovaird}}, \bibinfo {author} {\bibfnamefont {L.}~\bibnamefont {Brill}}, \bibinfo {author} {\bibfnamefont {M.}~\bibnamefont {Broughton}}, \bibinfo {author} {\bibfnamefont {D.~A.}\ \bibnamefont {Browne}}, \bibinfo {author} {\bibfnamefont {B.}~\bibnamefont {Buchea}}, \bibinfo {author} {\bibfnamefont {B.~B.}\ \bibnamefont {Buckley}}, \bibinfo {author} {\bibfnamefont {D.~A.}\ \bibnamefont {Buell}}, \bibinfo {author} {\bibfnamefont {T.}~\bibnamefont {Burger}}, \bibinfo {author} {\bibfnamefont {B.}~\bibnamefont {Burkett}}, \bibinfo {author} {\bibfnamefont
  {N.}~\bibnamefont {Bushnell}}, \bibinfo {author} {\bibfnamefont {A.}~\bibnamefont {Cabrera}}, \bibinfo {author} {\bibfnamefont {J.}~\bibnamefont {Campero}}, \bibinfo {author} {\bibfnamefont {H.-S.}\ \bibnamefont {Chang}}, \bibinfo {author} {\bibfnamefont {Y.}~\bibnamefont {Chen}}, \bibinfo {author} {\bibfnamefont {Z.}~\bibnamefont {Chen}}, \bibinfo {author} {\bibfnamefont {B.}~\bibnamefont {Chiaro}}, \bibinfo {author} {\bibfnamefont {D.}~\bibnamefont {Chik}}, \bibinfo {author} {\bibfnamefont {C.}~\bibnamefont {Chou}}, \bibinfo {author} {\bibfnamefont {J.}~\bibnamefont {Claes}}, \bibinfo {author} {\bibfnamefont {A.~Y.}\ \bibnamefont {Cleland}}, \bibinfo {author} {\bibfnamefont {J.}~\bibnamefont {Cogan}}, \bibinfo {author} {\bibfnamefont {R.}~\bibnamefont {Collins}}, \bibinfo {author} {\bibfnamefont {P.}~\bibnamefont {Conner}}, \bibinfo {author} {\bibfnamefont {W.}~\bibnamefont {Courtney}}, \bibinfo {author} {\bibfnamefont {A.~L.}\ \bibnamefont {Crook}}, \bibinfo {author} {\bibfnamefont {B.}~\bibnamefont
  {Curtin}}, \bibinfo {author} {\bibfnamefont {S.}~\bibnamefont {Das}}, \bibinfo {author} {\bibfnamefont {A.}~\bibnamefont {Davies}}, \bibinfo {author} {\bibfnamefont {L.}~\bibnamefont {De~Lorenzo}}, \bibinfo {author} {\bibfnamefont {D.~M.}\ \bibnamefont {Debroy}}, \bibinfo {author} {\bibfnamefont {S.}~\bibnamefont {Demura}}, \bibinfo {author} {\bibfnamefont {M.}~\bibnamefont {Devoret}}, \bibinfo {author} {\bibfnamefont {A.}~\bibnamefont {Di~Paolo}}, \bibinfo {author} {\bibfnamefont {P.}~\bibnamefont {Donohoe}}, \bibinfo {author} {\bibfnamefont {I.}~\bibnamefont {Drozdov}}, \bibinfo {author} {\bibfnamefont {A.}~\bibnamefont {Dunsworth}}, \bibinfo {author} {\bibfnamefont {C.}~\bibnamefont {Earle}}, \bibinfo {author} {\bibfnamefont {T.}~\bibnamefont {Edlich}}, \bibinfo {author} {\bibfnamefont {A.}~\bibnamefont {Eickbusch}}, \bibinfo {author} {\bibfnamefont {A.~M.}\ \bibnamefont {Elbag}}, \bibinfo {author} {\bibfnamefont {M.}~\bibnamefont {Elzouka}}, \bibinfo {author} {\bibfnamefont {C.}~\bibnamefont
  {Erickson}}, \bibinfo {author} {\bibfnamefont {L.}~\bibnamefont {Faoro}}, \bibinfo {author} {\bibfnamefont {E.}~\bibnamefont {Farhi}}, \bibinfo {author} {\bibfnamefont {V.~S.}\ \bibnamefont {Ferreira}}, \bibinfo {author} {\bibfnamefont {L.~F.}\ \bibnamefont {Burgos}}, \bibinfo {author} {\bibfnamefont {E.}~\bibnamefont {Forati}}, \bibinfo {author} {\bibfnamefont {A.~G.}\ \bibnamefont {Fowler}}, \bibinfo {author} {\bibfnamefont {B.}~\bibnamefont {Foxen}}, \bibinfo {author} {\bibfnamefont {S.}~\bibnamefont {Ganjam}}, \bibinfo {author} {\bibfnamefont {G.}~\bibnamefont {Garcia}}, \bibinfo {author} {\bibfnamefont {R.}~\bibnamefont {Gasca}}, \bibinfo {author} {\bibfnamefont {{\'E}.}~\bibnamefont {Genois}}, \bibinfo {author} {\bibfnamefont {W.}~\bibnamefont {Giang}}, \bibinfo {author} {\bibfnamefont {C.}~\bibnamefont {Gidney}}, \bibinfo {author} {\bibfnamefont {D.}~\bibnamefont {Gilboa}}, \bibinfo {author} {\bibfnamefont {R.}~\bibnamefont {Gosula}}, \bibinfo {author} {\bibfnamefont {A.~G.}\ \bibnamefont {Dau}},
  \bibinfo {author} {\bibfnamefont {D.}~\bibnamefont {Graumann}}, \bibinfo {author} {\bibfnamefont {A.}~\bibnamefont {Greene}}, \bibinfo {author} {\bibfnamefont {J.~A.}\ \bibnamefont {Gross}}, \bibinfo {author} {\bibfnamefont {S.}~\bibnamefont {Habegger}}, \bibinfo {author} {\bibfnamefont {J.}~\bibnamefont {Hall}}, \bibinfo {author} {\bibfnamefont {M.~C.}\ \bibnamefont {Hamilton}}, \bibinfo {author} {\bibfnamefont {M.}~\bibnamefont {Hansen}}, \bibinfo {author} {\bibfnamefont {M.~P.}\ \bibnamefont {Harrigan}}, \bibinfo {author} {\bibfnamefont {S.~D.}\ \bibnamefont {Harrington}}, \bibinfo {author} {\bibfnamefont {F.~J.~H.}\ \bibnamefont {Heras}}, \bibinfo {author} {\bibfnamefont {S.}~\bibnamefont {Heslin}}, \bibinfo {author} {\bibfnamefont {P.}~\bibnamefont {Heu}}, \bibinfo {author} {\bibfnamefont {O.}~\bibnamefont {Higgott}}, \bibinfo {author} {\bibfnamefont {G.}~\bibnamefont {Hill}}, \bibinfo {author} {\bibfnamefont {J.}~\bibnamefont {Hilton}}, \bibinfo {author} {\bibfnamefont {G.}~\bibnamefont {Holland}},
  \bibinfo {author} {\bibfnamefont {S.}~\bibnamefont {Hong}}, \bibinfo {author} {\bibfnamefont {H.-Y.}\ \bibnamefont {Huang}}, \bibinfo {author} {\bibfnamefont {A.}~\bibnamefont {Huff}}, \bibinfo {author} {\bibfnamefont {W.~J.}\ \bibnamefont {Huggins}}, \bibinfo {author} {\bibfnamefont {L.~B.}\ \bibnamefont {Ioffe}}, \bibinfo {author} {\bibfnamefont {S.~V.}\ \bibnamefont {Isakov}}, \bibinfo {author} {\bibfnamefont {J.}~\bibnamefont {Iveland}}, \bibinfo {author} {\bibfnamefont {E.}~\bibnamefont {Jeffrey}}, \bibinfo {author} {\bibfnamefont {Z.}~\bibnamefont {Jiang}}, \bibinfo {author} {\bibfnamefont {C.}~\bibnamefont {Jones}}, \bibinfo {author} {\bibfnamefont {S.}~\bibnamefont {Jordan}}, \bibinfo {author} {\bibfnamefont {C.}~\bibnamefont {Joshi}}, \bibinfo {author} {\bibfnamefont {P.}~\bibnamefont {Juhas}}, \bibinfo {author} {\bibfnamefont {D.}~\bibnamefont {Kafri}}, \bibinfo {author} {\bibfnamefont {H.}~\bibnamefont {Kang}}, \bibinfo {author} {\bibfnamefont {A.~H.}\ \bibnamefont {Karamlou}}, \bibinfo {author}
  {\bibfnamefont {K.}~\bibnamefont {Kechedzhi}}, \bibinfo {author} {\bibfnamefont {J.}~\bibnamefont {Kelly}}, \bibinfo {author} {\bibfnamefont {T.}~\bibnamefont {Khaire}}, \bibinfo {author} {\bibfnamefont {T.}~\bibnamefont {Khattar}}, \bibinfo {author} {\bibfnamefont {M.}~\bibnamefont {Khezri}}, \bibinfo {author} {\bibfnamefont {S.}~\bibnamefont {Kim}}, \bibinfo {author} {\bibfnamefont {P.~V.}\ \bibnamefont {Klimov}}, \bibinfo {author} {\bibfnamefont {A.~R.}\ \bibnamefont {Klots}}, \bibinfo {author} {\bibfnamefont {B.}~\bibnamefont {Kobrin}}, \bibinfo {author} {\bibfnamefont {P.}~\bibnamefont {Kohli}}, \bibinfo {author} {\bibfnamefont {A.~N.}\ \bibnamefont {Korotkov}}, \bibinfo {author} {\bibfnamefont {F.}~\bibnamefont {Kostritsa}}, \bibinfo {author} {\bibfnamefont {R.}~\bibnamefont {Kothari}}, \bibinfo {author} {\bibfnamefont {B.}~\bibnamefont {Kozlovskii}}, \bibinfo {author} {\bibfnamefont {J.~M.}\ \bibnamefont {Kreikebaum}}, \bibinfo {author} {\bibfnamefont {V.~D.}\ \bibnamefont {Kurilovich}}, \bibinfo
  {author} {\bibfnamefont {N.}~\bibnamefont {Lacroix}}, \bibinfo {author} {\bibfnamefont {D.}~\bibnamefont {Landhuis}}, \bibinfo {author} {\bibfnamefont {T.}~\bibnamefont {Lange-Dei}}, \bibinfo {author} {\bibfnamefont {B.~W.}\ \bibnamefont {Langley}}, \bibinfo {author} {\bibfnamefont {P.}~\bibnamefont {Laptev}}, \bibinfo {author} {\bibfnamefont {K.-M.}\ \bibnamefont {Lau}}, \bibinfo {author} {\bibfnamefont {L.}~\bibnamefont {Le~Guevel}}, \bibinfo {author} {\bibfnamefont {J.}~\bibnamefont {Ledford}}, \bibinfo {author} {\bibfnamefont {J.}~\bibnamefont {Lee}}, \bibinfo {author} {\bibfnamefont {K.}~\bibnamefont {Lee}}, \bibinfo {author} {\bibfnamefont {Y.~D.}\ \bibnamefont {Lensky}}, \bibinfo {author} {\bibfnamefont {S.}~\bibnamefont {Leon}}, \bibinfo {author} {\bibfnamefont {B.~J.}\ \bibnamefont {Lester}}, \bibinfo {author} {\bibfnamefont {W.~Y.}\ \bibnamefont {Li}}, \bibinfo {author} {\bibfnamefont {Y.}~\bibnamefont {Li}}, \bibinfo {author} {\bibfnamefont {A.~T.}\ \bibnamefont {Lill}}, \bibinfo {author}
  {\bibfnamefont {W.}~\bibnamefont {Liu}}, \bibinfo {author} {\bibfnamefont {W.~P.}\ \bibnamefont {Livingston}}, \bibinfo {author} {\bibfnamefont {A.}~\bibnamefont {Locharla}}, \bibinfo {author} {\bibfnamefont {E.}~\bibnamefont {Lucero}}, \bibinfo {author} {\bibfnamefont {D.}~\bibnamefont {Lundahl}}, \bibinfo {author} {\bibfnamefont {A.}~\bibnamefont {Lunt}}, \bibinfo {author} {\bibfnamefont {S.}~\bibnamefont {Madhuk}}, \bibinfo {author} {\bibfnamefont {F.~D.}\ \bibnamefont {Malone}}, \bibinfo {author} {\bibfnamefont {A.}~\bibnamefont {Maloney}}, \bibinfo {author} {\bibfnamefont {S.}~\bibnamefont {Mandr{\`a}}}, \bibinfo {author} {\bibfnamefont {J.}~\bibnamefont {Manyika}}, \bibinfo {author} {\bibfnamefont {L.~S.}\ \bibnamefont {Martin}}, \bibinfo {author} {\bibfnamefont {O.}~\bibnamefont {Martin}}, \bibinfo {author} {\bibfnamefont {S.}~\bibnamefont {Martin}}, \bibinfo {author} {\bibfnamefont {C.}~\bibnamefont {Maxfield}}, \bibinfo {author} {\bibfnamefont {J.~R.}\ \bibnamefont {McClean}}, \bibinfo {author}
  {\bibfnamefont {M.}~\bibnamefont {McEwen}}, \bibinfo {author} {\bibfnamefont {S.}~\bibnamefont {Meeks}}, \bibinfo {author} {\bibfnamefont {A.}~\bibnamefont {Megrant}}, \bibinfo {author} {\bibfnamefont {X.}~\bibnamefont {Mi}}, \bibinfo {author} {\bibfnamefont {K.~C.}\ \bibnamefont {Miao}}, \bibinfo {author} {\bibfnamefont {A.}~\bibnamefont {Mieszala}}, \bibinfo {author} {\bibfnamefont {R.}~\bibnamefont {Molavi}}, \bibinfo {author} {\bibfnamefont {S.}~\bibnamefont {Molina}}, \bibinfo {author} {\bibfnamefont {S.}~\bibnamefont {Montazeri}}, \bibinfo {author} {\bibfnamefont {A.}~\bibnamefont {Morvan}}, \bibinfo {author} {\bibfnamefont {R.}~\bibnamefont {Movassagh}}, \bibinfo {author} {\bibfnamefont {W.}~\bibnamefont {Mruczkiewicz}}, \bibinfo {author} {\bibfnamefont {O.}~\bibnamefont {Naaman}}, \bibinfo {author} {\bibfnamefont {M.}~\bibnamefont {Neeley}}, \bibinfo {author} {\bibfnamefont {C.}~\bibnamefont {Neill}}, \bibinfo {author} {\bibfnamefont {A.}~\bibnamefont {Nersisyan}}, \bibinfo {author} {\bibfnamefont
  {H.}~\bibnamefont {Neven}}, \bibinfo {author} {\bibfnamefont {M.}~\bibnamefont {Newman}}, \bibinfo {author} {\bibfnamefont {J.~H.}\ \bibnamefont {Ng}}, \bibinfo {author} {\bibfnamefont {A.}~\bibnamefont {Nguyen}}, \bibinfo {author} {\bibfnamefont {M.}~\bibnamefont {Nguyen}}, \bibinfo {author} {\bibfnamefont {C.-H.}\ \bibnamefont {Ni}}, \bibinfo {author} {\bibfnamefont {M.~Y.}\ \bibnamefont {Niu}}, \bibinfo {author} {\bibfnamefont {T.~E.}\ \bibnamefont {O'Brien}}, \bibinfo {author} {\bibfnamefont {W.~D.}\ \bibnamefont {Oliver}}, \bibinfo {author} {\bibfnamefont {A.}~\bibnamefont {Opremcak}}, \bibinfo {author} {\bibfnamefont {K.}~\bibnamefont {Ottosson}}, \bibinfo {author} {\bibfnamefont {A.}~\bibnamefont {Petukhov}}, \bibinfo {author} {\bibfnamefont {A.}~\bibnamefont {Pizzuto}}, \bibinfo {author} {\bibfnamefont {J.}~\bibnamefont {Platt}}, \bibinfo {author} {\bibfnamefont {R.}~\bibnamefont {Potter}}, \bibinfo {author} {\bibfnamefont {O.}~\bibnamefont {Pritchard}}, \bibinfo {author} {\bibfnamefont {L.~P.}\
  \bibnamefont {Pryadko}}, \bibinfo {author} {\bibfnamefont {C.}~\bibnamefont {Quintana}}, \bibinfo {author} {\bibfnamefont {G.}~\bibnamefont {Ramachandran}}, \bibinfo {author} {\bibfnamefont {M.~J.}\ \bibnamefont {Reagor}}, \bibinfo {author} {\bibfnamefont {J.}~\bibnamefont {Redding}}, \bibinfo {author} {\bibfnamefont {D.~M.}\ \bibnamefont {Rhodes}}, \bibinfo {author} {\bibfnamefont {G.}~\bibnamefont {Roberts}}, \bibinfo {author} {\bibfnamefont {E.}~\bibnamefont {Rosenberg}}, \bibinfo {author} {\bibfnamefont {E.}~\bibnamefont {Rosenfeld}}, \bibinfo {author} {\bibfnamefont {P.}~\bibnamefont {Roushan}}, \bibinfo {author} {\bibfnamefont {N.~C.}\ \bibnamefont {Rubin}}, \bibinfo {author} {\bibfnamefont {N.}~\bibnamefont {Saei}}, \bibinfo {author} {\bibfnamefont {D.}~\bibnamefont {Sank}}, \bibinfo {author} {\bibfnamefont {K.}~\bibnamefont {Sankaragomathi}}, \bibinfo {author} {\bibfnamefont {K.~J.}\ \bibnamefont {Satzinger}}, \bibinfo {author} {\bibfnamefont {H.~F.}\ \bibnamefont {Schurkus}}, \bibinfo {author}
  {\bibfnamefont {C.}~\bibnamefont {Schuster}}, \bibinfo {author} {\bibfnamefont {A.~W.}\ \bibnamefont {Senior}}, \bibinfo {author} {\bibfnamefont {M.~J.}\ \bibnamefont {Shearn}}, \bibinfo {author} {\bibfnamefont {A.}~\bibnamefont {Shorter}}, \bibinfo {author} {\bibfnamefont {N.}~\bibnamefont {Shutty}}, \bibinfo {author} {\bibfnamefont {V.}~\bibnamefont {Shvarts}}, \bibinfo {author} {\bibfnamefont {S.}~\bibnamefont {Singh}}, \bibinfo {author} {\bibfnamefont {V.}~\bibnamefont {Sivak}}, \bibinfo {author} {\bibfnamefont {J.}~\bibnamefont {Skruzny}}, \bibinfo {author} {\bibfnamefont {S.}~\bibnamefont {Small}}, \bibinfo {author} {\bibfnamefont {V.}~\bibnamefont {Smelyanskiy}}, \bibinfo {author} {\bibfnamefont {W.~C.}\ \bibnamefont {Smith}}, \bibinfo {author} {\bibfnamefont {R.~D.}\ \bibnamefont {Somma}}, \bibinfo {author} {\bibfnamefont {S.}~\bibnamefont {Springer}}, \bibinfo {author} {\bibfnamefont {G.}~\bibnamefont {Sterling}}, \bibinfo {author} {\bibfnamefont {D.}~\bibnamefont {Strain}}, \bibinfo {author}
  {\bibfnamefont {J.}~\bibnamefont {Suchard}}, \bibinfo {author} {\bibfnamefont {A.}~\bibnamefont {Szasz}}, \bibinfo {author} {\bibfnamefont {A.}~\bibnamefont {Sztein}}, \bibinfo {author} {\bibfnamefont {D.}~\bibnamefont {Thor}}, \bibinfo {author} {\bibfnamefont {A.}~\bibnamefont {Torres}}, \bibinfo {author} {\bibfnamefont {M.~M.}\ \bibnamefont {Torunbalci}}, \bibinfo {author} {\bibfnamefont {A.}~\bibnamefont {Vaishnav}}, \bibinfo {author} {\bibfnamefont {J.}~\bibnamefont {Vargas}}, \bibinfo {author} {\bibfnamefont {S.}~\bibnamefont {Vdovichev}}, \bibinfo {author} {\bibfnamefont {G.}~\bibnamefont {Vidal}}, \bibinfo {author} {\bibfnamefont {B.}~\bibnamefont {Villalonga}}, \bibinfo {author} {\bibfnamefont {C.~V.}\ \bibnamefont {Heidweiller}}, \bibinfo {author} {\bibfnamefont {S.}~\bibnamefont {Waltman}}, \bibinfo {author} {\bibfnamefont {S.~X.}\ \bibnamefont {Wang}}, \bibinfo {author} {\bibfnamefont {B.}~\bibnamefont {Ware}}, \bibinfo {author} {\bibfnamefont {K.}~\bibnamefont {Weber}}, \bibinfo {author}
  {\bibfnamefont {T.}~\bibnamefont {Weidel}}, \bibinfo {author} {\bibfnamefont {T.}~\bibnamefont {White}}, \bibinfo {author} {\bibfnamefont {K.}~\bibnamefont {Wong}}, \bibinfo {author} {\bibfnamefont {B.~W.~K.}\ \bibnamefont {Woo}}, \bibinfo {author} {\bibfnamefont {C.}~\bibnamefont {Xing}}, \bibinfo {author} {\bibfnamefont {Z.~J.}\ \bibnamefont {Yao}}, \bibinfo {author} {\bibfnamefont {P.}~\bibnamefont {Yeh}}, \bibinfo {author} {\bibfnamefont {B.}~\bibnamefont {Ying}}, \bibinfo {author} {\bibfnamefont {J.}~\bibnamefont {Yoo}}, \bibinfo {author} {\bibfnamefont {N.}~\bibnamefont {Yosri}}, \bibinfo {author} {\bibfnamefont {G.}~\bibnamefont {Young}}, \bibinfo {author} {\bibfnamefont {A.}~\bibnamefont {Zalcman}}, \bibinfo {author} {\bibfnamefont {Y.}~\bibnamefont {Zhang}}, \bibinfo {author} {\bibfnamefont {N.}~\bibnamefont {Zhu}}, \bibinfo {author} {\bibfnamefont {N.}~\bibnamefont {Zobrist}}, \bibinfo {author} {\bibfnamefont {G.~Q.}\ \bibnamefont {AI}},\ and\ \bibinfo {author} {\bibnamefont {Collaborators}},\
  }\bibfield  {title} {\bibinfo {title} {Quantum error correction below the surface code threshold},\ }\href {https://doi.org/10.1038/s41586-024-08449-y} {\bibfield  {journal} {\bibinfo  {journal} {Nature}\ }\textbf {\bibinfo {volume} {638}},\ \bibinfo {pages} {920} (\bibinfo {year} {2025})}\BibitemShut {NoStop}%
\bibitem [{\citenamefont {Gao}\ \emph {et~al.}(2025)\citenamefont {Gao}, \citenamefont {Fan}, \citenamefont {Zha}, \citenamefont {Bei}, \citenamefont {Cai}, \citenamefont {Cai}, \citenamefont {Cao}, \citenamefont {Chen}, \citenamefont {Chen}, \citenamefont {Chen}, \citenamefont {Chen}, \citenamefont {Chen}, \citenamefont {Chen}, \citenamefont {Chen}, \citenamefont {Chen}, \citenamefont {Chu}, \citenamefont {Deng}, \citenamefont {Deng}, \citenamefont {Ding}, \citenamefont {Ding}, \citenamefont {Ding}, \citenamefont {Dong}, \citenamefont {Dong}, \citenamefont {Fan}, \citenamefont {Fu}, \citenamefont {Gao}, \citenamefont {Ge}, \citenamefont {Gong}, \citenamefont {Gui}, \citenamefont {Guo}, \citenamefont {Guo}, \citenamefont {Guo}, \citenamefont {Han}, \citenamefont {He}, \citenamefont {Hong}, \citenamefont {Hu}, \citenamefont {Huang}, \citenamefont {Huo}, \citenamefont {Jiang}, \citenamefont {Jiang}, \citenamefont {Jin}, \citenamefont {Leng}, \citenamefont {Li}, \citenamefont {Li}, \citenamefont {Li},
  \citenamefont {Li}, \citenamefont {Li}, \citenamefont {Li}, \citenamefont {Li}, \citenamefont {Li}, \citenamefont {Li}, \citenamefont {Li}, \citenamefont {Li}, \citenamefont {Li}, \citenamefont {Liang}, \citenamefont {Liang}, \citenamefont {Liao}, \citenamefont {Lin}, \citenamefont {Lin}, \citenamefont {Liu}, \citenamefont {Liu}, \citenamefont {Liu}, \citenamefont {Liu}, \citenamefont {Liu}, \citenamefont {Liu}, \citenamefont {Lou}, \citenamefont {Ma}, \citenamefont {Meng}, \citenamefont {Mou}, \citenamefont {Nan}, \citenamefont {Nie}, \citenamefont {Nie}, \citenamefont {Ning}, \citenamefont {Niu}, \citenamefont {Peng}, \citenamefont {Qian}, \citenamefont {Rong}, \citenamefont {Rong}, \citenamefont {Shen}, \citenamefont {Shen}, \citenamefont {Su}, \citenamefont {Su}, \citenamefont {Sun}, \citenamefont {Sun}, \citenamefont {Sun}, \citenamefont {Sun}, \citenamefont {Tan}, \citenamefont {Tan}, \citenamefont {Tang}, \citenamefont {Tu}, \citenamefont {Wan}, \citenamefont {Wang}, \citenamefont {Wang},
  \citenamefont {Wang}, \citenamefont {Wang}, \citenamefont {Wang}, \citenamefont {Wang}, \citenamefont {Wang}, \citenamefont {Wang}, \citenamefont {Wang}, \citenamefont {Wang}, \citenamefont {Wang}, \citenamefont {Wang}, \citenamefont {Wang}, \citenamefont {Wei}, \citenamefont {Wei}, \citenamefont {Wu}, \citenamefont {Wu}, \citenamefont {Wu}, \citenamefont {Wu}, \citenamefont {Wu}, \citenamefont {Xie}, \citenamefont {Xin}, \citenamefont {Xu}, \citenamefont {Xue}, \citenamefont {Yan}, \citenamefont {Yang}, \citenamefont {Yang}, \citenamefont {Yang}, \citenamefont {Ye}, \citenamefont {Ye}, \citenamefont {Ying}, \citenamefont {Yu}, \citenamefont {Yu}, \citenamefont {Yu}, \citenamefont {Zeng}, \citenamefont {Zhan}, \citenamefont {Zhang}, \citenamefont {Zhang}, \citenamefont {Zhang}, \citenamefont {Zhang}, \citenamefont {Zhang}, \citenamefont {Zhang}, \citenamefont {Zhang}, \citenamefont {Zhang}, \citenamefont {Zhao}, \citenamefont {Zhao}, \citenamefont {Zhao}, \citenamefont {Zhao}, \citenamefont {Zhao},
  \citenamefont {Zhao}, \citenamefont {Zheng}, \citenamefont {Zhou}, \citenamefont {Zhou}, \citenamefont {Zhou}, \citenamefont {Zhou}, \citenamefont {Zhou}, \citenamefont {Zhou}, \citenamefont {Zhou}, \citenamefont {Zhu}, \citenamefont {Zhu}, \citenamefont {Zou}, \citenamefont {Zou}, \citenamefont {Zhang}, \citenamefont {Lu}, \citenamefont {Peng}, \citenamefont {Zhu},\ and\ \citenamefont {Pan}}]{ZuchongzhiChinaProcessor}%
  \BibitemOpen
  \bibfield  {author} {\bibinfo {author} {\bibfnamefont {D.}~\bibnamefont {Gao}}, \bibinfo {author} {\bibfnamefont {D.}~\bibnamefont {Fan}}, \bibinfo {author} {\bibfnamefont {C.}~\bibnamefont {Zha}}, \bibinfo {author} {\bibfnamefont {J.}~\bibnamefont {Bei}}, \bibinfo {author} {\bibfnamefont {G.}~\bibnamefont {Cai}}, \bibinfo {author} {\bibfnamefont {J.}~\bibnamefont {Cai}}, \bibinfo {author} {\bibfnamefont {S.}~\bibnamefont {Cao}}, \bibinfo {author} {\bibfnamefont {F.}~\bibnamefont {Chen}}, \bibinfo {author} {\bibfnamefont {J.}~\bibnamefont {Chen}}, \bibinfo {author} {\bibfnamefont {K.}~\bibnamefont {Chen}}, \bibinfo {author} {\bibfnamefont {X.}~\bibnamefont {Chen}}, \bibinfo {author} {\bibfnamefont {X.}~\bibnamefont {Chen}}, \bibinfo {author} {\bibfnamefont {Z.}~\bibnamefont {Chen}}, \bibinfo {author} {\bibfnamefont {Z.}~\bibnamefont {Chen}}, \bibinfo {author} {\bibfnamefont {Z.}~\bibnamefont {Chen}}, \bibinfo {author} {\bibfnamefont {W.}~\bibnamefont {Chu}}, \bibinfo {author} {\bibfnamefont
  {H.}~\bibnamefont {Deng}}, \bibinfo {author} {\bibfnamefont {Z.}~\bibnamefont {Deng}}, \bibinfo {author} {\bibfnamefont {P.}~\bibnamefont {Ding}}, \bibinfo {author} {\bibfnamefont {X.}~\bibnamefont {Ding}}, \bibinfo {author} {\bibfnamefont {Z.}~\bibnamefont {Ding}}, \bibinfo {author} {\bibfnamefont {S.}~\bibnamefont {Dong}}, \bibinfo {author} {\bibfnamefont {Y.}~\bibnamefont {Dong}}, \bibinfo {author} {\bibfnamefont {B.}~\bibnamefont {Fan}}, \bibinfo {author} {\bibfnamefont {Y.}~\bibnamefont {Fu}}, \bibinfo {author} {\bibfnamefont {S.}~\bibnamefont {Gao}}, \bibinfo {author} {\bibfnamefont {L.}~\bibnamefont {Ge}}, \bibinfo {author} {\bibfnamefont {M.}~\bibnamefont {Gong}}, \bibinfo {author} {\bibfnamefont {J.}~\bibnamefont {Gui}}, \bibinfo {author} {\bibfnamefont {C.}~\bibnamefont {Guo}}, \bibinfo {author} {\bibfnamefont {S.}~\bibnamefont {Guo}}, \bibinfo {author} {\bibfnamefont {X.}~\bibnamefont {Guo}}, \bibinfo {author} {\bibfnamefont {L.}~\bibnamefont {Han}}, \bibinfo {author} {\bibfnamefont
  {T.}~\bibnamefont {He}}, \bibinfo {author} {\bibfnamefont {L.}~\bibnamefont {Hong}}, \bibinfo {author} {\bibfnamefont {Y.}~\bibnamefont {Hu}}, \bibinfo {author} {\bibfnamefont {H.-L.}\ \bibnamefont {Huang}}, \bibinfo {author} {\bibfnamefont {Y.-H.}\ \bibnamefont {Huo}}, \bibinfo {author} {\bibfnamefont {T.}~\bibnamefont {Jiang}}, \bibinfo {author} {\bibfnamefont {Z.}~\bibnamefont {Jiang}}, \bibinfo {author} {\bibfnamefont {H.}~\bibnamefont {Jin}}, \bibinfo {author} {\bibfnamefont {Y.}~\bibnamefont {Leng}}, \bibinfo {author} {\bibfnamefont {D.}~\bibnamefont {Li}}, \bibinfo {author} {\bibfnamefont {D.}~\bibnamefont {Li}}, \bibinfo {author} {\bibfnamefont {F.}~\bibnamefont {Li}}, \bibinfo {author} {\bibfnamefont {J.}~\bibnamefont {Li}}, \bibinfo {author} {\bibfnamefont {J.}~\bibnamefont {Li}}, \bibinfo {author} {\bibfnamefont {J.}~\bibnamefont {Li}}, \bibinfo {author} {\bibfnamefont {J.}~\bibnamefont {Li}}, \bibinfo {author} {\bibfnamefont {N.}~\bibnamefont {Li}}, \bibinfo {author} {\bibfnamefont
  {S.}~\bibnamefont {Li}}, \bibinfo {author} {\bibfnamefont {W.}~\bibnamefont {Li}}, \bibinfo {author} {\bibfnamefont {Y.}~\bibnamefont {Li}}, \bibinfo {author} {\bibfnamefont {Y.}~\bibnamefont {Li}}, \bibinfo {author} {\bibfnamefont {F.}~\bibnamefont {Liang}}, \bibinfo {author} {\bibfnamefont {X.}~\bibnamefont {Liang}}, \bibinfo {author} {\bibfnamefont {N.}~\bibnamefont {Liao}}, \bibinfo {author} {\bibfnamefont {J.}~\bibnamefont {Lin}}, \bibinfo {author} {\bibfnamefont {W.}~\bibnamefont {Lin}}, \bibinfo {author} {\bibfnamefont {D.}~\bibnamefont {Liu}}, \bibinfo {author} {\bibfnamefont {H.}~\bibnamefont {Liu}}, \bibinfo {author} {\bibfnamefont {M.}~\bibnamefont {Liu}}, \bibinfo {author} {\bibfnamefont {X.}~\bibnamefont {Liu}}, \bibinfo {author} {\bibfnamefont {X.}~\bibnamefont {Liu}}, \bibinfo {author} {\bibfnamefont {Y.}~\bibnamefont {Liu}}, \bibinfo {author} {\bibfnamefont {H.}~\bibnamefont {Lou}}, \bibinfo {author} {\bibfnamefont {Y.}~\bibnamefont {Ma}}, \bibinfo {author} {\bibfnamefont {L.}~\bibnamefont
  {Meng}}, \bibinfo {author} {\bibfnamefont {H.}~\bibnamefont {Mou}}, \bibinfo {author} {\bibfnamefont {K.}~\bibnamefont {Nan}}, \bibinfo {author} {\bibfnamefont {B.}~\bibnamefont {Nie}}, \bibinfo {author} {\bibfnamefont {M.}~\bibnamefont {Nie}}, \bibinfo {author} {\bibfnamefont {J.}~\bibnamefont {Ning}}, \bibinfo {author} {\bibfnamefont {L.}~\bibnamefont {Niu}}, \bibinfo {author} {\bibfnamefont {W.}~\bibnamefont {Peng}}, \bibinfo {author} {\bibfnamefont {H.}~\bibnamefont {Qian}}, \bibinfo {author} {\bibfnamefont {H.}~\bibnamefont {Rong}}, \bibinfo {author} {\bibfnamefont {T.}~\bibnamefont {Rong}}, \bibinfo {author} {\bibfnamefont {H.}~\bibnamefont {Shen}}, \bibinfo {author} {\bibfnamefont {Q.}~\bibnamefont {Shen}}, \bibinfo {author} {\bibfnamefont {H.}~\bibnamefont {Su}}, \bibinfo {author} {\bibfnamefont {F.}~\bibnamefont {Su}}, \bibinfo {author} {\bibfnamefont {C.}~\bibnamefont {Sun}}, \bibinfo {author} {\bibfnamefont {L.}~\bibnamefont {Sun}}, \bibinfo {author} {\bibfnamefont {T.}~\bibnamefont {Sun}},
  \bibinfo {author} {\bibfnamefont {Y.}~\bibnamefont {Sun}}, \bibinfo {author} {\bibfnamefont {Y.}~\bibnamefont {Tan}}, \bibinfo {author} {\bibfnamefont {J.}~\bibnamefont {Tan}}, \bibinfo {author} {\bibfnamefont {L.}~\bibnamefont {Tang}}, \bibinfo {author} {\bibfnamefont {W.}~\bibnamefont {Tu}}, \bibinfo {author} {\bibfnamefont {C.}~\bibnamefont {Wan}}, \bibinfo {author} {\bibfnamefont {J.}~\bibnamefont {Wang}}, \bibinfo {author} {\bibfnamefont {B.}~\bibnamefont {Wang}}, \bibinfo {author} {\bibfnamefont {C.}~\bibnamefont {Wang}}, \bibinfo {author} {\bibfnamefont {C.}~\bibnamefont {Wang}}, \bibinfo {author} {\bibfnamefont {C.}~\bibnamefont {Wang}}, \bibinfo {author} {\bibfnamefont {J.}~\bibnamefont {Wang}}, \bibinfo {author} {\bibfnamefont {L.}~\bibnamefont {Wang}}, \bibinfo {author} {\bibfnamefont {R.}~\bibnamefont {Wang}}, \bibinfo {author} {\bibfnamefont {S.}~\bibnamefont {Wang}}, \bibinfo {author} {\bibfnamefont {X.}~\bibnamefont {Wang}}, \bibinfo {author} {\bibfnamefont {X.}~\bibnamefont {Wang}}, \bibinfo
  {author} {\bibfnamefont {X.}~\bibnamefont {Wang}}, \bibinfo {author} {\bibfnamefont {Y.}~\bibnamefont {Wang}}, \bibinfo {author} {\bibfnamefont {Z.}~\bibnamefont {Wei}}, \bibinfo {author} {\bibfnamefont {J.}~\bibnamefont {Wei}}, \bibinfo {author} {\bibfnamefont {D.}~\bibnamefont {Wu}}, \bibinfo {author} {\bibfnamefont {G.}~\bibnamefont {Wu}}, \bibinfo {author} {\bibfnamefont {J.}~\bibnamefont {Wu}}, \bibinfo {author} {\bibfnamefont {S.}~\bibnamefont {Wu}}, \bibinfo {author} {\bibfnamefont {Y.}~\bibnamefont {Wu}}, \bibinfo {author} {\bibfnamefont {S.}~\bibnamefont {Xie}}, \bibinfo {author} {\bibfnamefont {L.}~\bibnamefont {Xin}}, \bibinfo {author} {\bibfnamefont {Y.}~\bibnamefont {Xu}}, \bibinfo {author} {\bibfnamefont {C.}~\bibnamefont {Xue}}, \bibinfo {author} {\bibfnamefont {K.}~\bibnamefont {Yan}}, \bibinfo {author} {\bibfnamefont {W.}~\bibnamefont {Yang}}, \bibinfo {author} {\bibfnamefont {X.}~\bibnamefont {Yang}}, \bibinfo {author} {\bibfnamefont {Y.}~\bibnamefont {Yang}}, \bibinfo {author}
  {\bibfnamefont {Y.}~\bibnamefont {Ye}}, \bibinfo {author} {\bibfnamefont {Z.}~\bibnamefont {Ye}}, \bibinfo {author} {\bibfnamefont {C.}~\bibnamefont {Ying}}, \bibinfo {author} {\bibfnamefont {J.}~\bibnamefont {Yu}}, \bibinfo {author} {\bibfnamefont {Q.}~\bibnamefont {Yu}}, \bibinfo {author} {\bibfnamefont {W.}~\bibnamefont {Yu}}, \bibinfo {author} {\bibfnamefont {X.}~\bibnamefont {Zeng}}, \bibinfo {author} {\bibfnamefont {S.}~\bibnamefont {Zhan}}, \bibinfo {author} {\bibfnamefont {F.}~\bibnamefont {Zhang}}, \bibinfo {author} {\bibfnamefont {H.}~\bibnamefont {Zhang}}, \bibinfo {author} {\bibfnamefont {K.}~\bibnamefont {Zhang}}, \bibinfo {author} {\bibfnamefont {P.}~\bibnamefont {Zhang}}, \bibinfo {author} {\bibfnamefont {W.}~\bibnamefont {Zhang}}, \bibinfo {author} {\bibfnamefont {Y.}~\bibnamefont {Zhang}}, \bibinfo {author} {\bibfnamefont {Y.}~\bibnamefont {Zhang}}, \bibinfo {author} {\bibfnamefont {L.}~\bibnamefont {Zhang}}, \bibinfo {author} {\bibfnamefont {G.}~\bibnamefont {Zhao}}, \bibinfo {author}
  {\bibfnamefont {P.}~\bibnamefont {Zhao}}, \bibinfo {author} {\bibfnamefont {X.}~\bibnamefont {Zhao}}, \bibinfo {author} {\bibfnamefont {X.}~\bibnamefont {Zhao}}, \bibinfo {author} {\bibfnamefont {Y.}~\bibnamefont {Zhao}}, \bibinfo {author} {\bibfnamefont {Z.}~\bibnamefont {Zhao}}, \bibinfo {author} {\bibfnamefont {L.}~\bibnamefont {Zheng}}, \bibinfo {author} {\bibfnamefont {F.}~\bibnamefont {Zhou}}, \bibinfo {author} {\bibfnamefont {L.}~\bibnamefont {Zhou}}, \bibinfo {author} {\bibfnamefont {N.}~\bibnamefont {Zhou}}, \bibinfo {author} {\bibfnamefont {N.}~\bibnamefont {Zhou}}, \bibinfo {author} {\bibfnamefont {S.}~\bibnamefont {Zhou}}, \bibinfo {author} {\bibfnamefont {S.}~\bibnamefont {Zhou}}, \bibinfo {author} {\bibfnamefont {Z.}~\bibnamefont {Zhou}}, \bibinfo {author} {\bibfnamefont {C.}~\bibnamefont {Zhu}}, \bibinfo {author} {\bibfnamefont {Q.}~\bibnamefont {Zhu}}, \bibinfo {author} {\bibfnamefont {G.}~\bibnamefont {Zou}}, \bibinfo {author} {\bibfnamefont {H.}~\bibnamefont {Zou}}, \bibinfo {author}
  {\bibfnamefont {Q.}~\bibnamefont {Zhang}}, \bibinfo {author} {\bibfnamefont {C.-Y.}\ \bibnamefont {Lu}}, \bibinfo {author} {\bibfnamefont {C.-Z.}\ \bibnamefont {Peng}}, \bibinfo {author} {\bibfnamefont {X.}~\bibnamefont {Zhu}},\ and\ \bibinfo {author} {\bibfnamefont {J.-W.}\ \bibnamefont {Pan}},\ }\bibfield  {title} {\bibinfo {title} {Establishing a new benchmark in quantum computational advantage with 105-qubit zuchongzhi 3.0 processor},\ }\href {https://doi.org/10.1103/PhysRevLett.134.090601} {\bibfield  {journal} {\bibinfo  {journal} {Phys. Rev. Lett.}\ }\textbf {\bibinfo {volume} {134}},\ \bibinfo {pages} {090601} (\bibinfo {year} {2025})}\BibitemShut {NoStop}%
\bibitem [{\citenamefont {Smith}\ \emph {et~al.}(2016)\citenamefont {Smith}, \citenamefont {Kou}, \citenamefont {Vool}, \citenamefont {Pop}, \citenamefont {Frunzio}, \citenamefont {Schoelkopf},\ and\ \citenamefont {Devoret}}]{ClarkeISTQuantization}%
  \BibitemOpen
  \bibfield  {author} {\bibinfo {author} {\bibfnamefont {W.~C.}\ \bibnamefont {Smith}}, \bibinfo {author} {\bibfnamefont {A.}~\bibnamefont {Kou}}, \bibinfo {author} {\bibfnamefont {U.}~\bibnamefont {Vool}}, \bibinfo {author} {\bibfnamefont {I.~M.}\ \bibnamefont {Pop}}, \bibinfo {author} {\bibfnamefont {L.}~\bibnamefont {Frunzio}}, \bibinfo {author} {\bibfnamefont {R.~J.}\ \bibnamefont {Schoelkopf}},\ and\ \bibinfo {author} {\bibfnamefont {M.~H.}\ \bibnamefont {Devoret}},\ }\bibfield  {title} {\bibinfo {title} {Quantization of inductively shunted superconducting circuits},\ }\href {https://doi.org/10.1103/PhysRevB.94.144507} {\bibfield  {journal} {\bibinfo  {journal} {Phys. Rev. B}\ }\textbf {\bibinfo {volume} {94}},\ \bibinfo {pages} {144507} (\bibinfo {year} {2016})}\BibitemShut {NoStop}%
\bibitem [{\citenamefont {Zakka-Bajjani}\ \emph {et~al.}(2011)\citenamefont {Zakka-Bajjani}, \citenamefont {Nguyen}, \citenamefont {Lee}, \citenamefont {Vale}, \citenamefont {Simmonds},\ and\ \citenamefont {Aumentado}}]{Zakka2011}%
  \BibitemOpen
  \bibfield  {author} {\bibinfo {author} {\bibfnamefont {E.}~\bibnamefont {Zakka-Bajjani}}, \bibinfo {author} {\bibfnamefont {F.}~\bibnamefont {Nguyen}}, \bibinfo {author} {\bibfnamefont {M.}~\bibnamefont {Lee}}, \bibinfo {author} {\bibfnamefont {L.~R.}\ \bibnamefont {Vale}}, \bibinfo {author} {\bibfnamefont {R.~W.}\ \bibnamefont {Simmonds}},\ and\ \bibinfo {author} {\bibfnamefont {J.}~\bibnamefont {Aumentado}},\ }\bibfield  {title} {\bibinfo {title} {Quantum superposition of a single microwave photon in two different 'colour'states},\ }\href@noop {} {\bibfield  {journal} {\bibinfo  {journal} {Nature Physics}\ }\textbf {\bibinfo {volume} {7}},\ \bibinfo {pages} {599} (\bibinfo {year} {2011})}\BibitemShut {NoStop}%
\bibitem [{\citenamefont {Noh}\ \emph {et~al.}(2023)\citenamefont {Noh}, \citenamefont {Xiao}, \citenamefont {Jin}, \citenamefont {Cicak}, \citenamefont {Doucet}, \citenamefont {Aumentado}, \citenamefont {Govia}, \citenamefont {Ranzani}, \citenamefont {Kamal},\ and\ \citenamefont {Simmonds}}]{ParametricCQEDPaper}%
  \BibitemOpen
  \bibfield  {author} {\bibinfo {author} {\bibfnamefont {T.}~\bibnamefont {Noh}}, \bibinfo {author} {\bibfnamefont {Z.}~\bibnamefont {Xiao}}, \bibinfo {author} {\bibfnamefont {X.~Y.}\ \bibnamefont {Jin}}, \bibinfo {author} {\bibfnamefont {K.}~\bibnamefont {Cicak}}, \bibinfo {author} {\bibfnamefont {E.}~\bibnamefont {Doucet}}, \bibinfo {author} {\bibfnamefont {J.}~\bibnamefont {Aumentado}}, \bibinfo {author} {\bibfnamefont {L.~C.~G.}\ \bibnamefont {Govia}}, \bibinfo {author} {\bibfnamefont {L.}~\bibnamefont {Ranzani}}, \bibinfo {author} {\bibfnamefont {A.}~\bibnamefont {Kamal}},\ and\ \bibinfo {author} {\bibfnamefont {R.~W.}\ \bibnamefont {Simmonds}},\ }\bibfield  {title} {\bibinfo {title} {Strong parametric dispersive shifts in a statically decoupled two-qubit cavity qed system},\ }\href {https://doi.org/10.1038/s41567-023-02107-2} {\bibfield  {journal} {\bibinfo  {journal} {Nature Physics}\ }\textbf {\bibinfo {volume} {19}},\ \bibinfo {pages} {1445} (\bibinfo {year} {2023})}\BibitemShut {NoStop}%
\bibitem [{\citenamefont {de~Graaf}(2024)}]{StijnThesis}%
  \BibitemOpen
  \bibfield  {author} {\bibinfo {author} {\bibfnamefont {S.~J.}\ \bibnamefont {de~Graaf}},\ }\emph {\bibinfo {title} {Microwave beamsplitters for oscillator-based quantum information processing}},\ \href@noop {} {\bibinfo {type} {Phd thesis}},\ \bibinfo  {school} {Yale University} (\bibinfo {year} {2024}),\ \bibinfo {note} {pgs. 55-56}\BibitemShut {NoStop}%
\bibitem [{\citenamefont {Miano}\ \emph {et~al.}(2023)\citenamefont {Miano}, \citenamefont {Joshi}, \citenamefont {Liu}, \citenamefont {Dai}, \citenamefont {Parakh}, \citenamefont {Frunzio},\ and\ \citenamefont {Devoret}}]{NinaPaper}%
  \BibitemOpen
  \bibfield  {author} {\bibinfo {author} {\bibfnamefont {A.}~\bibnamefont {Miano}}, \bibinfo {author} {\bibfnamefont {V.}~\bibnamefont {Joshi}}, \bibinfo {author} {\bibfnamefont {G.}~\bibnamefont {Liu}}, \bibinfo {author} {\bibfnamefont {W.}~\bibnamefont {Dai}}, \bibinfo {author} {\bibfnamefont {P.}~\bibnamefont {Parakh}}, \bibinfo {author} {\bibfnamefont {L.}~\bibnamefont {Frunzio}},\ and\ \bibinfo {author} {\bibfnamefont {M.}~\bibnamefont {Devoret}},\ }\bibfield  {title} {\bibinfo {title} {Hamiltonian extrema of an arbitrary flux-biased josephson circuit},\ }\href {https://doi.org/10.1103/PRXQuantum.4.030324} {\bibfield  {journal} {\bibinfo  {journal} {PRX Quantum}\ }\textbf {\bibinfo {volume} {4}},\ \bibinfo {pages} {030324} (\bibinfo {year} {2023})}\BibitemShut {NoStop}%
\bibitem [{\citenamefont {Smith}\ \emph {et~al.}(2020)\citenamefont {Smith}, \citenamefont {Kou}, \citenamefont {Xiao}, \citenamefont {Vool},\ and\ \citenamefont {Devoret}}]{Clarkecos2phiTheory}%
  \BibitemOpen
  \bibfield  {author} {\bibinfo {author} {\bibfnamefont {W.~C.}\ \bibnamefont {Smith}}, \bibinfo {author} {\bibfnamefont {A.}~\bibnamefont {Kou}}, \bibinfo {author} {\bibfnamefont {X.}~\bibnamefont {Xiao}}, \bibinfo {author} {\bibfnamefont {U.}~\bibnamefont {Vool}},\ and\ \bibinfo {author} {\bibfnamefont {M.~H.}\ \bibnamefont {Devoret}},\ }\bibfield  {title} {\bibinfo {title} {Superconducting circuit protected by two-{Cooper}-pair tunneling},\ }\href {https://doi.org/10.1038/s41534-019-0231-2} {\bibfield  {journal} {\bibinfo  {journal} {npj Quantum Information}\ }\textbf {\bibinfo {volume} {6}},\ \bibinfo {pages} {8} (\bibinfo {year} {2020})}\BibitemShut {NoStop}%
\bibitem [{\citenamefont {Smith}\ \emph {et~al.}(2022)\citenamefont {Smith}, \citenamefont {Villiers}, \citenamefont {Marquet}, \citenamefont {Palomo}, \citenamefont {Delbecq}, \citenamefont {Kontos}, \citenamefont {Campagne-Ibarcq}, \citenamefont {Dou\ifmmode~\mbox{\c{c}}\else \c{c}\fi{}ot},\ and\ \citenamefont {Leghtas}}]{ClarkeKITEcos2phiExpt}%
  \BibitemOpen
  \bibfield  {author} {\bibinfo {author} {\bibfnamefont {W.~C.}\ \bibnamefont {Smith}}, \bibinfo {author} {\bibfnamefont {M.}~\bibnamefont {Villiers}}, \bibinfo {author} {\bibfnamefont {A.}~\bibnamefont {Marquet}}, \bibinfo {author} {\bibfnamefont {J.}~\bibnamefont {Palomo}}, \bibinfo {author} {\bibfnamefont {M.~R.}\ \bibnamefont {Delbecq}}, \bibinfo {author} {\bibfnamefont {T.}~\bibnamefont {Kontos}}, \bibinfo {author} {\bibfnamefont {P.}~\bibnamefont {Campagne-Ibarcq}}, \bibinfo {author} {\bibfnamefont {B.}~\bibnamefont {Dou\ifmmode~\mbox{\c{c}}\else \c{c}\fi{}ot}},\ and\ \bibinfo {author} {\bibfnamefont {Z.}~\bibnamefont {Leghtas}},\ }\bibfield  {title} {\bibinfo {title} {Magnifying quantum phase fluctuations with cooper-pair pairing},\ }\href {https://doi.org/10.1103/PhysRevX.12.021002} {\bibfield  {journal} {\bibinfo  {journal} {Phys. Rev. X}\ }\textbf {\bibinfo {volume} {12}},\ \bibinfo {pages} {021002} (\bibinfo {year} {2022})}\BibitemShut {NoStop}%
\end{thebibliography}%

\end{document}